\documentclass[twocolumn,dvipsnames,prx,nofootinbib,floatfix,superscriptaddress]{revtex4-2}
\usepackage{xcolor}
\usepackage{amsmath,amssymb}
\usepackage{appendix}
\usepackage{makecell}
\usepackage{relsize}
\usepackage{graphicx}
\usepackage{stmaryrd}
\setcellgapes{4pt}
\usepackage[caption=false, position=top]{subfig}
\usepackage{ulem}

\usepackage[colorlinks,plainpages]{hyperref}
\hypersetup{urlcolor=blue,citecolor=blue, linkcolor=blue,hypertexnames=false}

\usepackage{xspace}
\usepackage{setspace}
\usepackage{bm}
\usepackage{algorithm}
\usepackage{algpseudocode}

\definecolor{newblue}{RGB}{8,81,156}

\newcommand{\CCA}{\affiliation{Center for Computational Astrophysics, Flatiron Institute, 162 Fifth Avenue, New York, NY 10010, USA}}

\newcommand{\KICP}{\affiliation{Kavli Institute for Cosmological Physics, University of Chicago, 5640 S. Ellis Ave., Chicago, IL 60615, USA}}

\newcommand{\SBU}{\affiliation{Department of Physics and Astronomy, Stony Brook University, Stony Brook NY 11794, USA}}

\newcommand{\MassFitFpeakOne}{0.75}
\newcommand{\MassFitFpeakTwo}{0.07}
\newcommand{\MassFitMuOne}{10.0}
\newcommand{\MassFitSigOne}{1.1}
\newcommand{\MassFitMuTwo}{33.4}
\newcommand{\MassFitSigTwo}{4.2}
\newcommand{\MassFitSlopeOne}{-2.7}
\newcommand{\MassFitSlopeTwo}{-4.2}
\newcommand{\MassFitMbreak}{42.7}
\newcommand{\MassFitMmin}{12.9}
\newcommand{\MassFitMmax}{100}
\newcommand{\MassFitDeltaMmin}{0.6}

\newcommand{\MassSignificancePercentageLowPeak}{96\%}
\newcommand{\MassSignificancePercentageHighPeak}{94\%}
\newcommand{\MassSignificanceLowSlope}{-1.1^{+2.7}_{-2.6}}
\newcommand{\MassSignificanceHighSlope}{-3.8^{+2.6}_{-2.7}}
\newcommand{\MassSignificancePercentageSteepening}{89\%}

\newcommand{\RedshiftFitSigmoidZCenter}{0.65}
\newcommand{\RedshiftFitSigmoidDeltaZ}{0.08}
\newcommand{\RedshiftFitSigmoidRateLow}{19.1}
\newcommand{\RedshiftFitSigmoidDeltaRate}{27.0}
\newcommand{\RedshiftFitBplZBreak}{0.44}
\newcommand{\RedshiftFitBplKappaLow}{0.3}
\newcommand{\RedshiftFitBplKappaHigh}{3.4}
\newcommand{\RedshiftFitBplRateBreak}{20.4}

\newcommand{\MarginalRedshiftFitSigmoidZCenter}{0.64}
\newcommand{\MarginalRedshiftFitSigmoidDeltaZ}{0.08}
\newcommand{\MarginalRedshiftFitSigmoidRateLow}{25.0}
\newcommand{\MarginalRedshiftFitSigmoidDeltaRate}{33.9}
\newcommand{\MarginalRedshiftFitBplZBreak}{0.44}
\newcommand{\MarginalRedshiftFitBplKappaLow}{0.2}
\newcommand{\MarginalRedshiftFitBplKappaHigh}{3.5}
\newcommand{\MarginalRedshiftFitBplRateBreak}{26.4}

\newcommand{\RedshiftSignificancePercentageLowRise}{53\%}
\newcommand{\RedshiftSignificancePercentageHighRise}{93\%}

\newcommand{\SpinMagnitudeFifty}{0.21^{+0.07}_{-0.07}}
\newcommand{\SpinMagnitudeNinetyFive}{0.72^{+0.17}_{-0.25}}
\newcommand{\SpinTiltFifty}{0.16^{+0.28}_{-0.17}}
\newcommand{\SpinTiltMedianPercentPositive}{88\%}
\newcommand{\SpinTiltPercentNegative}{41^{+9}_{-17}\%}
\newcommand{\SpinTiltRateOneOrBothNegative}{17.0^{+10.7}_{-7.4}}

\newcommand{\TiltSignificancePercentagePeak}{37\%}
\newcommand{\MagnitudeSignificancePercentagePeak}{66\%}

\newcommand{\MagFitLorentzianMean}{0.15}
\newcommand{\MagFitLorentzianGamma}{0.18}
\newcommand{\CosTiltFitFractionIso}{0.67}
\newcommand{\CosTiltFitMean}{0.59}
\newcommand{\CosTiltFitStd}{0.58}

\newcommand{\ChiEffPercentNegative}{27^{+17}_{-14}\%}
\newcommand{\ChiEffRateNegative}{7.7^{+8.2}_{-4.3}}
\newcommand{\ChiPPercentLarge}{42^{+35}_{-32}\%}
\newcommand{\ChiEffAsymmetricPercentage}{98.3\%}
\newcommand{\ChiEffMedianPercentPositive}{98.2\%}

\newcommand{\ChiEffMean}{0.04^{+0.05}_{-0.09}}

\newcommand{\ChiEffFitGaussianMean}{0.07}
\newcommand{\ChiEffFitGaussianStd}{0.09}
\newcommand{\ChiEffFitLorentzianMean}{0.07}
\newcommand{\ChiEffFitLorentzianGamma}{0.07}
\newcommand{\ChiPFitGaussianMean}{0.0}
\newcommand{\ChiPFitGaussianStd}{0.23}
\newcommand{\ChiPFitLorentzianMean}{0.0}
\newcommand{\ChiPFitLorentzianGamma}{0.18}

\newcommand{\MassRatioSignifanceRising}{90\%}

\newcommand{\ComponentMagTailHighTwo}{0.90^{+0.08}_{-0.26}}
\newcommand{\ComponentMagTailHighFour}{0.82^{+0.13}_{-0.29}}
\newcommand{\ComponentTiltTailLowTwo}{-0.96^{+0.12}_{-0.03}}
\newcommand{\ComponentTiltTailLowFour}{-0.91^{+0.19}_{-0.06}}

\newcommand{\EffectiveSpinTailLowTwo}{-0.29^{+0.22}_{-0.45}}

\newcommand{\EffectiveSpinTailHighTwo}{0.49^{+0.31}_{-0.29}}

\newcommand{\PrecessingSpinTailHighTwo}{0.84^{+0.14}_{-0.44}}

\graphicspath{{./}{figures/}}
\begin{document}

\title{A Parameter-Free Tour of the Binary Black Hole Population}


\author{Thomas A. Callister}
\email{tcallister@uchicago.edu}
\KICP{}
\CCA{}

\author{Will M. Farr}
\CCA{}
\SBU{}

\begin{abstract}
  The continued operation of the Advanced LIGO and Advanced Virgo gravitational-wave detectors is enabling the first detailed measurements of the mass, spin, and redshift distributions of the merging binary black hole population.
  Our present knowledge of these distributions, however, is based largely on strongly parameteric models.
  Such models typically assume the distributions of binary parameters to be superpositions of ``building block'' features like power laws, peaks, dips, and breaks.
  Although this approach has yielded great progress in the initial characterization of the compact binary population, the strong assumptions entailed often leave it unclear which physical conclusions are driven by observation and which by the specific choice of model.
  In this paper, we instead model the merger rate of binary black holes as an unknown \textit{autoregressive process} over the space of binary parameters, allowing us to measure the distributions of binary black hole masses, redshifts, component spins, and effective spins with near-complete agnosticism.
  We find the primary mass spectrum of binary black holes to be doubly peaked, with a fairly flat continuum that steepens at high masses.
  We identify signs of unexpected structure in the redshift distribution of binary black holes: a uniform-in-comoving volume merger rate at low redshift followed by an increase in the merger rate beyond redshift $z\approx 0.5$.
  Finally, we find that the distribution of black hole spin magnitudes is unimodal and concentrated at small but nonzero values, and that spin orientations span a wide range of spin-orbit misalignment angles but are also moderately unlikely to be truly isotropic.
\end{abstract}

\maketitle

\section{Background}

The recent release of the third gravitational-wave transient catalog (GWTC-3)~\cite{GWTC3} by the LIGO Scientific Collaboration~\cite{Aasi2015}, Virgo Collaboration~\cite{Acernese2015}, and KAGRA Collaboration~\cite{Akutsu2021} has increased the number of confident gravitational-wave detections to 76\footnote{Counting those events with false alarm rates $\mathrm{FAR}<1\,\mathrm{yr}^{-1}$~\cite{O3b-pop}.}, with yet more candidates identified in independent reanalyses of LIGO-Virgo data~\cite{Nitz2022,Olsen2022}.
This growing body of detections has pushed gravitational-wave astronomy firmly into the catalog era; we can move beyond interrogating the properties of individual binary mergers to instead exploring the ensemble properties of the complete compact binary population~\cite{O3a-pop,O3b-pop}.

Most present-day analyses of the compact binary population adopt a \textit{strongly modeled} approach, in which the distributions of binary masses, spins, and redshifts are assumed to follow specific parametric forms.
The binary black hole mass spectrum, for example, is commonly assumed to be the superposition of power laws and/or Gaussians~\cite{Talbot2018,Wysocki2019,O3a-pop,Edelman2021,Tiwari2022,O3b-pop,Farah2022}.
The hyperparameters describing these functional forms (e.g., power-law slopes and Gaussian means and widths) are then measured using our catalog of gravitational-wave detections.

This strongly modeled ``building-block'' approach has yielded significant insight.
We have learned, for example, that the black hole merger rate is highest at $m\approx 10\,M_\odot$, declining steeply towards larger masses but with a secondary bump near $35\,M_\odot$~\cite{O3a-pop,O3b-pop}.
Black hole spins are small but nonzero, with a wide range of misalignment angles between spins and binary orbital angular momenta~\cite{O3a-pop,O3b-pop,Callister2022,Mould2022}.
And the rate of binary mergers grows as we look to higher redshifts~\cite{O3b-pop}.

At the same time, this approach has some less-desirable downsides.
\begin{itemize}
\item First, our chosen functional form prescribes, from the very outset, the set of possible population features.
Thus, it is not always clear which conclusions come from informative data and which are built, by assumption, into the models themselves.
Parametrized models including sharp features, for example, are prone to ``false alarms,'' favoring the existence of such features even when none exist~\cite{Callister2022}.
\item Second, different models may yield very different or even conflicting conclusions if they prescribe different sets of features, further obscuring which conclusions are robust and which are model-induced.
\item Finally, strongly parametrized models allow us to search for ``known unknowns'' (e.g. \textit{is there a pair instability cut-off in the black hole mass spectrum?}) but do not let us search for the ``unknown unknowns,'' truly unexpected features that might challenge our astrophysical understanding of compact binary formation and evolution.
Several features in the binary black hole population (a peak in the merger rate near $35\,M_\odot$~\cite{O3a-pop}, a correlation between binary mass ratio and spin~\cite{Callister2021,Adamcewicz2022}, etc.), for example, were discovered serendipitously only after a fortuitous choice of model.
\end{itemize}

These concerns have spurred the development of flexible methods that aim to characterize the compact binary population while imposing few \textit{a priori} assumptions regarding the form of the population.
Examples of these flexible approaches include modeling the distribution of binary parameters using splines~\cite{Edelman2022a,Edelman2022b,Golomb2022}, piecewise-constant ``binned'' models~\cite{Mandel2017,Farr2018,O2-pop,Fishbach2020,Veske2021,O3b-pop}, and Gaussian mixture models~\cite{Tiwari2021,Tiwari2022,Rinaldi2022}, as well as non-Bayesian methods that seek to identify clustering in gravitational-wave catalogs~\cite{Powell2019,Sadiq2022}.

In this paper, we will explore an alternative and complementary approach,  treating the merger rate of binary black holes as an unknown \textit{autoregressive process} defined over masses, spins, and/or redshifts.
Whereas all other population models entail the use of hyperparameters to specify the dependence of the merger rate on mass, spin, and redshift, under our approach  the \textit{merger rates at every posterior sample} are themselves the quantities that we directly infer from data.
This allows us to characterize the compact binary distribution with a high degree of agnosticism, assuming only a prior preference that the  merger rate be a continuous function of binary parameters.
This approach will allow us to confirm the robustness of features previously identified using standard strongly parameterized models, as well as identify new features that might otherwise be overlooked.

More specifically, our goals in this work are threefold:
    \begin{enumerate}
    \item First, we present a flexible measurement of the merger rate densities and probability distributions over binary black hole masses, mass ratios, redshifts, and spins.
    \item Inference of the binary merger rate is only a first step. As we discuss further in Sect.~\ref{sec:conclusion}, a conceptually distinct and equally important step is \textit{feature extraction}: the subsequent identification of features and assessment of their significance.
    For strongly parametrized models, rate inference and feature extraction are by definition performed simultaneously.
    This is not the case for flexible approaches like spline methods or our autoregressive process, and so the importance of developing techniques for feature extraction is particularly acute.
    For each binary black hole parameter, we therefore seek to methodically assess feature significance without resorting to hyperparameters, but instead by calculating and comparing merger rates between different regions of interest.
    \item Finally, our goal is to leverage our autoregressive inference to provide new or extended strongly parametrized models that reflect our most up-to-date understanding of the binary black hole population.
    This goal serves two purposes.
    First, these updated models may provide a robust and accessible point of comparison between theory and observation.
    Second, these strongly-parametrized models can, in turn, be adopted in traditional hierarchical analyses, helping to confirm (or reject) possible features that appear in more flexible analyses of the black hole population.
    \end{enumerate}

In Sect.~\ref{sec:method}, we begin by motivating and defining autoregressive processes as a useful tool in the inference of compact binary populations.
In Sects.~\ref{sec:masses}--\ref{sec:effective-spins}, we then apply our method to study the distributions of masses, redshifts, component spins, and effective spins among the binary black hole population.
Along the way, we systematically demonstrate feature extraction and discuss new or expanded strongly parametrized models motivated by our results.
We conclude in Sect.~\ref{sec:conclusion} by commenting on the relationship between flexible and strongly parameterized models and avenues for future work.

\section{The Compact Binary Population as an Autoregressive Process}
\label{sec:method}

\begin{figure*}
    \centering
    \includegraphics[width=0.7\textwidth]{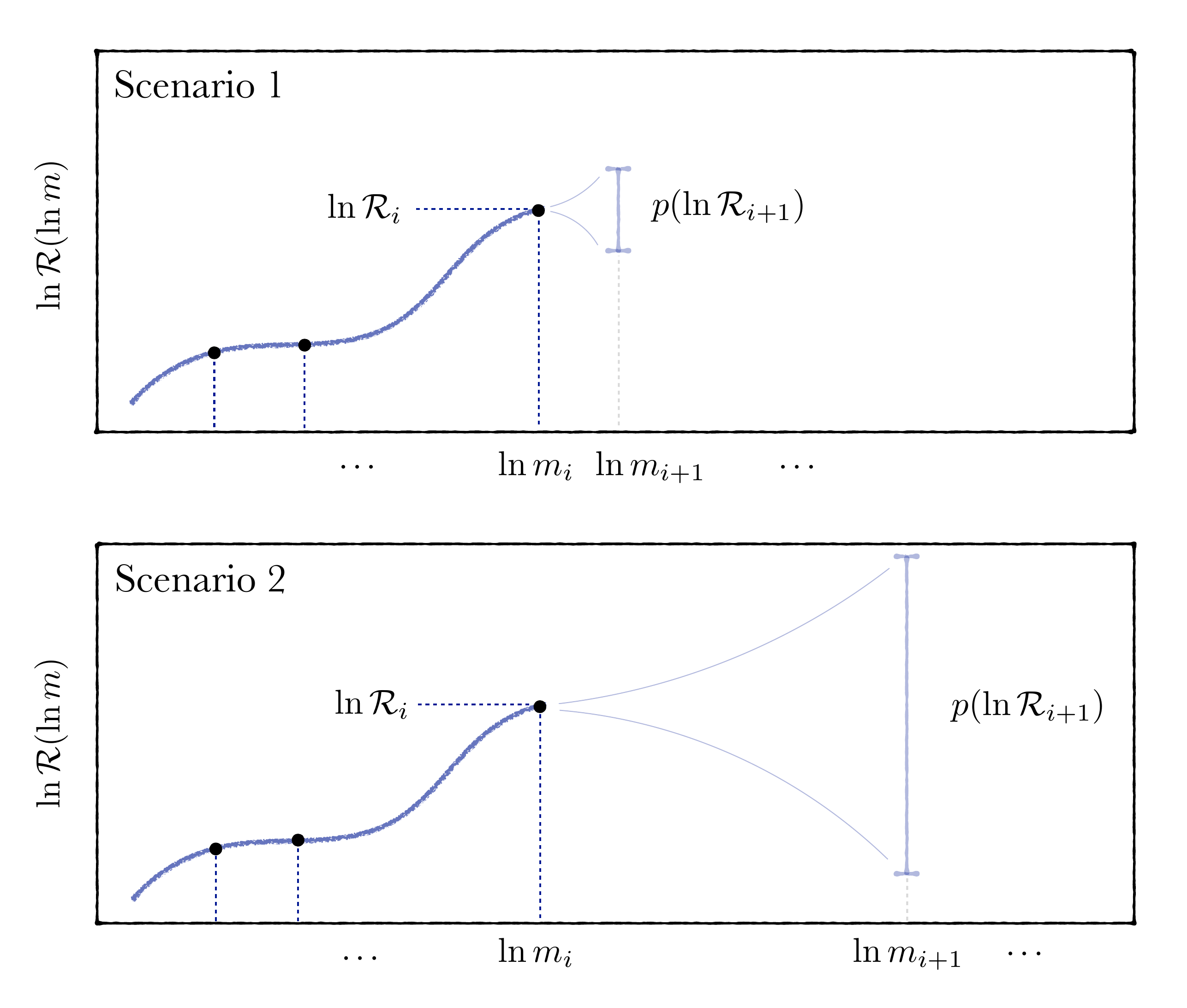}
    \caption{Cartoon demonstrating the physical principle behind our autoregressive inference of the binary hole population.
    Consider a situation in which we know the differential merger rate $\mathcal{R}(\ln m)$ at several different masses, up to some $\ln m_i$.
    Next, take some \textit{new} mass, $\ln m_{i+1}$.
    What prior should we place on the merger rate at this new location?
    If $\ln m_i$ and $\ln m_{i+1}$ are close (Scenario 1), then we expect the merger rates at this location to be reasonably close as well; we might, therefore, place a tight prior on $\mathcal{R}_{i+1}$ about the value $\mathcal{R}_i$.
    If, on the other hand, $\ln m_i$ and $\ln m_{i+1}$ are distant (Scenario 2), the merger rates at each point are unlikely to be related, so we might place a considerably less informative prior on $\mathcal{R}_{i+1}$.
    This intuition is codified by choosing an \textit{autoregressive prior} on $\ln \mathcal{R}(\ln m)$, such that the merger rate at any mass value has a Gaussian prior about the merger rate at the previous value, with variance related to the separation between points.
    }
    \label{fig:ar-demo}
\end{figure*}

\subsection{Autoregressive Processes}

To help make our discussion concrete, consider the problem of measuring the binary black hole primary mass distribution.
This amounts to measuring the differential merger rate
    \begin{equation}
    \frac{dR}{d\ln m} \equiv \frac{dN}{dV_c\,dt_s\,d\ln m},
    \label{eq:volumetric}
    \end{equation}
giving the number $dN$ of mergers per unit comoving volume $dV_c$, per unit source-frame time $dt_s$, and per logarithmic mass interval $d\ln m$.
For notational convenience, we will use the shorthand $\mathcal{R}(\theta) \equiv dR/d\theta$ to denote the merger rate density over parameters $\theta$, e.g.,
    \begin{equation}
    \mathcal{R}(\ln m) \equiv \frac{dR}{d\ln m}.
    \end{equation}
The standard strongly parameterized approach involves assuming some particular functional form for $\mathcal{R}(\ln m)$, such as a superposition of power laws, Gaussians, and/or truncations, and then measuring the parameters of these functions~\cite{Talbot2018,Wysocki2019,O3a-pop,Edelman2021,Tiwari2022,O3b-pop,Farah2022}.
Stepping back, however, we can think more generally about the merger rate density $\mathcal{R}(\ln m)$ that we seek to measure.

In nature, there exists \textit{some} underlying function that describes the true mass spectrum of compact binaries; this is illustrated in cartoon form by the dark blue curve in Fig.~\ref{fig:ar-demo}.
We know nothing \textit{a priori} about the exact shape of this function.
However, we can still attempt to write down prior assumptions about this function's likely behavior.
In Fig.~\ref{fig:ar-demo}, we hypothetically know the merger rate $\mathcal{R}_i$ at some particular value $\ln m_i$.
Given this knowledge, what is our prior expectation on the merger rate at a new point $\ln m_{i+1}$?
A reasonable expectation is that, if $\ln m_i$ and $\ln m_{i+1}$ are close together (Scenario 1 in the top panel), then the rates at these locations are likely similar as well.
In fact, in the limit that $\ln m_i=\ln m_{i+1}$, we should recover $\mathcal{R}_i=\mathcal{R}_{i+1}$.
Conversely, if $\ln m_i$ and $\ln m_{i+1}$ are far apart (Scenario 2), then the rates at each point need not be similar at all.

This intuition forms the basis of an \textit{autoregressive process} prior.
An autoregressive process $\Psi(x)$ is a stochastic function whose value $\Psi_i$ at some new point is related to the values at all previous points by
    \begin{equation}
    \Psi_i = \sum_{j=1}^p c_j \Psi_{i-j} + w_i.
    \end{equation}
Here, the $\{c_j\}$ are deterministic coefficients and $w_i$ is a random variable.
Qualitatively, the coefficients $\{c_j\}$ govern the degree to which $\Psi(x)$ ``remembers'' its past values, while the parent distribution of $\{w_i\}$ governs the degree to which the function is allowed to randomly fluctuate.
The parameter $p$ is called the ``order'' of the process and determines the smoothness of the resulting functions;
an autoregressive process of order $p$ has $p-1$ continuous derivatives.
Choosing order $p=1$ gives us the simplest ``AR(1)'' autoregressive process, which obeys
    \begin{equation}
    \Psi_i = c_i \Psi_{i-1} + w_i.
    \label{eq:general-ar1}
    \end{equation}

We can adopt this language as a framework with which to codify our intuition regarding possible merger rate densities, considering the merger rate as a function of mass to be of the form
    \begin{equation}
    \mathcal{R}(\ln m) = r\,e^{\Psi(\ln m)},
    \end{equation}
where $\Psi(\ln m)$ is an autoregressive process in $\ln m$ of order $p=1$.
This implies that, if we know the merger rate $\mathcal{R}_{i-1}$ at one mass location, then we take the rate at a new location $\ln m_i$ to be probabilistically given by the relation
    \begin{equation}
    \label{eq:mass-ar1}
    \begin{aligned}
    \ln \mathcal{R}_i - \ln r &= c_i \left[\ln \mathcal{R}_{i-1} - \ln r \right] + w_i,
    \end{aligned}
    \end{equation}
for some choice of $c_i$ and $w_i$ (discussed further below).
The quantity $r$ sets the mean merger rate; it is the \textit{departures} from $\ln r$ that are described via an AR(1) model.
Note also that it is the \textit{log} of the merger rate, not the merger rate itself, that is modeled as an AR(1) process.
This choice guarantees that predicted merger rates are everywhere positive, but has the downside that our inferred merger rate can never strictly go to $\mathcal{R} = 0$ (corresponding to $\ln\mathcal{R}\to-\infty$); see Appendix~\ref{app:inference-details}.

We take $c_i$ and $w_i$ to be of the form
    \begin{equation}
    c_i = e^{-\Delta_i/\tau}
    \label{eq:c}
    \end{equation}
and
    \begin{equation}
    w_i = \sigma \left(1 - e^{-2\Delta_i/\tau}\right)^{1/2} n_i,
    \label{eq:w}
    \end{equation}
where $\Delta_i = \ln m_i - \ln m_{i-1}$ is the distance between mass locations and $n_i$ is a random variable drawn from a unit normal distribution: $n_i\sim N(0,1)$.
The parameter $\sigma$ functions to rescale the random variable $n_i$ and thus controls the allowed variance of the merger rate.
The parameter $\tau$, meanwhile, defines the mass scale over which the mass spectrum remains significantly correlated with itself.
In the limit that $\Delta_i \ll \tau$, Eq.~\eqref{eq:mass-ar1} demands that $\ln\mathcal{R}_i \to \ln\mathcal{R}_{i-1}$.
And in the opposite limit that $\Delta_i \gg \tau$, we instead have $\ln \mathcal{R}_i$ drawn randomly from $N(\ln r,\sigma)$, with no memory of earlier merger rate values.
The exact forms of Eqs.~\eqref{eq:c} and \eqref{eq:w} are chosen to ensure that $\sigma^2$ and $\tau$ indeed control the variance and autocorrelation length of the process; see Appendices~\ref{app:ar-appendix} and \ref{app:inference-details} for more information about these expressions.

\begin{figure}
    \centering
    \includegraphics[width=0.48\textwidth]{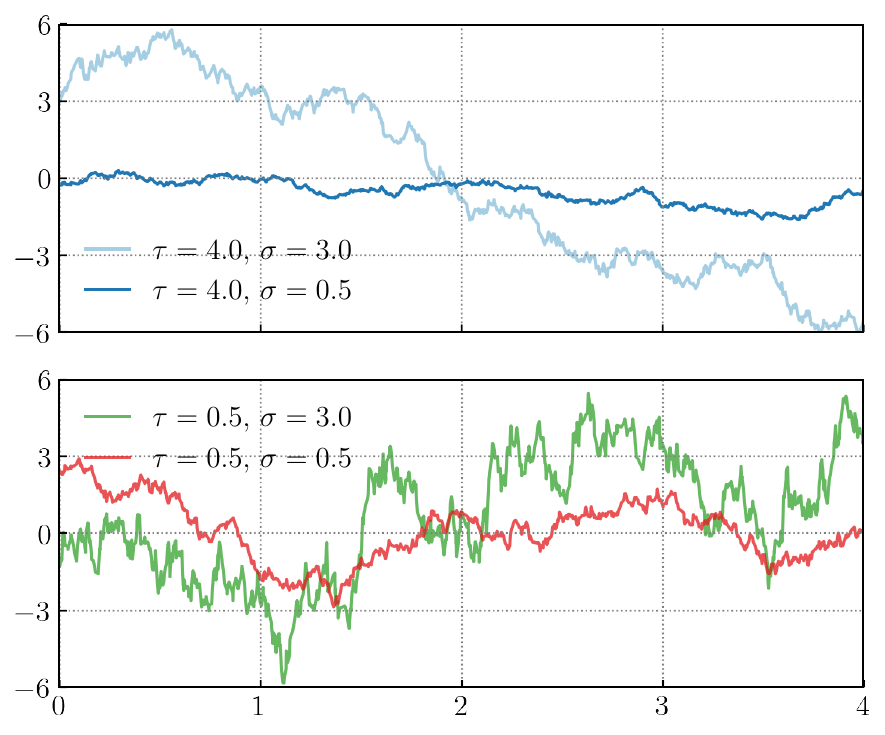}
    \caption{
    Examples of various autoregressive processes $\Psi(x)$.
    Each curve is random draw from an AR(1) process, subject to different autocorrelation lengths $\tau$ and standard deviations $\sigma$; see Eqs.~\eqref{eq:general-ar1}, \eqref{eq:c}, and \eqref{eq:w}.
    The top panel shows example autoregressive processes with large $\tau$, while the bottom panel illustrates two processes with short $\tau$.
    In Secs.~\ref{sec:masses}--\ref{sec:effective-spins} below, we model the mass, redshift, and spin distributions of binary black holes assuming they are each describable as unknown AR(1) processes.
    }
    \label{fig:ar-illustration}
\end{figure}

Figure~\ref{fig:ar-illustration} illustrates several random AR(1) processes $\Psi(x)$, generated with various choices of $\tau$ and $\sigma$.
Processes with large $\tau$ (top panel) exhibit much stronger correlations between adjacent points, yielding larger observed scale lengths than those with smaller $\tau$ (bottom panel).
Processes with large $\sigma$, meanwhile, traverse a much larger vertical range than processes with small $\sigma$.
Note that these functions are continuous but do not have well-defined first derivatives.
If we wanted to consider functions with continuous first derivatives, we could instead adopt ``AR(2)'' processes of order $p=2$.
We continue with an AR(1) process, however, in order to better capture any sharp or nondifferentiable features in the binary black hole population that could be missed by models that require continuous derivatives.

\subsection{Hierarchical Inference with an Autoregressive Prior}

Consider a set of $N_\mathrm{obs}$ gravitational-wave detections with sets of posterior samples $\{\lambda_I\}$ on the properties of each event $I$.
The likelihood that our data, denoted $\{d\}$, arise from an underlying population described by $\Lambda$ is~\cite{Loredo2004,Taylor2018,Mandel2019,Vitale2020}
    \begin{equation}
    \begin{aligned}
    p(\{d\}|\Lambda) &\propto e^{-N_\mathrm{exp}(\Lambda)}
        \prod_{I=1}^{N_\mathrm{obs}} \left\langle \frac{R_d(\lambda_{I,j};\Lambda)}{p_\mathrm{pe}(\lambda_{I,j})} \right\rangle_{\mathrm{samples}\,j}.
    \label{eq:pop-likelihood-mc}
    \end{aligned}
    \end{equation}
Here, the product is taken over detected events $I$, and the expectation value is taken over posterior samples $j$ for each event.
The quantity $p_\mathrm{pe}(\lambda_{I,j})$ is the prior probability assigned to each posterior sample under parameter estimation, while
    \begin{equation}
    R_d(\lambda;\Lambda) = \frac{dN}{dt_d\,d\lambda} (\lambda;\Lambda)
    \label{eq:det-rate}
    \end{equation}
is the detector-frame merger rate density, to be evaluated at each posterior sample.
We use semicolons to indicate that $R_d(\lambda;\Lambda)$ is a function of the population model $\Lambda$ but not a \textit{density} over $\Lambda$.
Note also that $R_d(\lambda;\Lambda)$ is \textit{not} a volumetric density, as in Eq.~\eqref{eq:volumetric}.
If redshift $z$ is a parameter in $\lambda$ such that $R_d$ is a merger rate per unit redshift, then Eqs.~\eqref{eq:volumetric} and \eqref{eq:det-rate} are related by
    \begin{equation}
    R_d(\lambda;\Lambda) = \mathcal{R}(\tilde\lambda;z,\Lambda) \frac{dV_c}{dz} \left(1+z\right)^{-1},
    \label{eq:rate-relation}
    \end{equation}
where $\tilde\lambda$ is the set of all binary parameters \textit{excluding} $z$ and $\mathcal{R}(\tilde\lambda;z,\Lambda)$ is the volumetric merger rate density as evaluated at redshift $z$.
The factor $\frac{dV_c}{dz}$ is the differential comoving volume per unit redshift, while the factor $(1+z)^{-1}$ is needed to convert between source-frame and detector-frame times.

Equation~\eqref{eq:pop-likelihood-mc} additionally depends on $N_\mathrm{exp}(\Lambda)$, the expected number of detections over our observation time $T_\mathrm{obs}$ given the population $\Lambda$.
We evaluate $N_\mathrm{exp}(\Lambda)$ using a set of successfully recovered signals injected into LIGO and Virgo data~\cite{O3b-pop,injections}.
If $p_\mathrm{inj}(\lambda)$ is the reference probability distribution from which these injections were drawn, then \cite{Farr2019}
    \begin{equation}
    N_\mathrm{exp}(\Lambda) \approx \frac{T_\mathrm{obs}}{N_\mathrm{inj}} \sum_\mathrm{found} \frac{R_d(\lambda_{\mathrm{inj},i};\Lambda)}{p_\mathrm{inj}(\lambda_{\mathrm{inj},i})},
    \label{eq:Nexp-mc}
    \end{equation}
where $N_\mathrm{inj}$ is the total number of injections performed, detected or otherwise, and $T_\mathrm{obs}$ is our total search time.
The detector-frame rate $R_d(\lambda_{\mathrm{inj},i};\Lambda)$ at the location of each injection can, once again, be related to the underlying volumetric rate using Eq.~\eqref{eq:rate-relation}.

The critical ingredients underlying Eqs.~\eqref{eq:pop-likelihood-mc} and \eqref{eq:Nexp-mc} are the differential rates $\mathcal{R}(\lambda_{I,j};z,\Lambda)$ and $\mathcal{R}(\lambda_{\mathrm{inj},i};z,\Lambda)$ at the locations of every posterior sample and every found injection.
In the usual strongly parametrized approach, we obtain these quantities by assuming some functional form for $\mathcal{R}(\lambda;z,\Lambda)$.
Here, our goal is to not assume a particular functional form for the differential rate, but to \textit{directly infer} the merger rate at every posterior sample and every found injection using our autoregressive prior.
In adopting this approach, we have rid ourselves of (nearly) all ordinary hyperparameters.
Instead, the \textit{merger rates at every posterior sample and every injection} are themselves the parameters that we directly infer from the data.
This results in a rather large-dimensional parameter space.
If we have $N_\mathrm{obs}$ events (each with $N_\mathrm{samp}$ posterior samples) and $N_\mathrm{inj}$ injections, we are directly inferring the binary merger rate at $N_\mathrm{obs}N_\mathrm{samp} + N_\mathrm{inj}$ discrete locations.
The form of Eq.~\eqref{eq:mass-ar1}, however, imposes an almost equally large number of constraints, ensuring that the inference problem remains tractable.

We have not quite discarded all hyperparameters: We still need to determine the variance $\sigma$ and autocorrelation length $\tau$ associated with our autoregressive rate prior.
Rather than fix $\sigma$ and $\tau$, we hierarchically infer them as well, allowing our data to dictate the characteristic length scale and size of features present in the binary black hole population.
In Appendix~\ref{app:inference-details}, we derive and discuss the priors we place on $\sigma$ and $\tau$.
To obtain physically meaningful priors, we approach the problem indirectly, considering not constraints on $\sigma$ and $\tau$ themselves but instead on allowed variations in the black hole merger rate $\mathcal{R}(\lambda)$; these choices then induce priors on $\sigma$, $\tau$, and the ratio $\sigma/\sqrt{\tau}$.

It is worthwhile to compare our methodology to other flexible approaches appearing in the literature.
One similar approach is the spline-based method appearing in Refs.~\cite{Edelman2022a,Edelman2022b,Golomb2022}; this model proceeds by first defining merger rates over a discrete set of ``knot'' locations and then constructing a spline interpolant between these knots.
The rates at each knot location may themselves be linked via a Gaussian process prior or other regularization schemes~\cite{Edelman2022b}.
The ``binned Gaussian process'' models in Refs.~\cite{Mandel2017,Farr2018,O2-pop,Fishbach2020,Veske2021,O3b-pop} operate similarly.
In this approach, a merger rate is again defined across a discrete grid of points and interpolated, but now assuming that the merger rate is a piecewise constant between grid points rather than a spline.

A primary methodological difference between these approaches and ours is that we do not perform interpolation: the parameters governing our models are the direct merger rates at each point of interest rather than the rates defined over some reference grid.
Our autoregressive model also behaves in ways that make it complementary to these other approaches.
Because spline interpolants are continuous in their derivatives, the spline-based approaches above are suitable for identifying smooth trends in the data but may struggle to resolve sharp features or features at the same scale as the knot separations.
The continuity and differentiability imposed by spline models can additionally sometimes give rise to oscillatory ``ringing'' that depends on the precise choice of knot locations~\cite{Golomb2022,Farah2023}.
Our AR(1) model requires no differentiability, however, avoiding this oscillatory behavior.
The lack of a reference grid also means that we require no \textit{a priori} choice of scale.
This freedom, however, greatly boosts the computational cost of our approach and gives rise to possible instabilities in the hierarchical likelihood; this instability is described in Appendix~\ref{app:inference-details}.
Even with the flexibility afforded by an AR(1) process, there remain limitations on the degree to which our model can recover discontinuously sharp features; see Appendix~\ref{app:demo} for further discussion.

We implement our autoregressive model using \texttt{jax}~\cite{jax} and \texttt{numpyro}~\cite{numpyro1,numpyro2}, which enable compilation and auto-differentiation of our likelihood.
We perform our Bayesian inference using \texttt{numpyro}'s implementation of the \texttt{NUTS} (``No U-Turn Sampler'') algorithm~\cite{Hoffman2011}, a variant of Hamiltonian Monte Carlo (HMC) sampling~\cite{Betancourt2017}.
As noted above, our autoregressive models actually comprise a vast number of latent parameters: one per posterior sample and found injection.
In practice, this amounts to approximately $2.5\times10^5$ parameters for the analyses presented in this paper.
Given this extremely high-dimensional space, the computational acceleration and sampling efficiency afforded by auto-differentiation and HMC methods is critical.
Further details regarding our hierarchical inference, including the exact data and priors used, are given in Appendix~\ref{app:inference-details}.

\section{Stop One: Masses}
\label{sec:masses}

We first use our autoregressive model to investigate the distribution of binary black hole primary masses $m_1$ and mass ratios $q$.
We consider the merger rate to be the combination of two parallel autoregressive processes, $\Psi(\ln m_1)$ and $\Phi(q)$, that capture the dependence of the merger rate on both $\ln m_1$ and $q$:
\begin{equation}
\begin{aligned}
&\mathcal{R}(\ln m_1,q,\chi_1,\chi_2,\cos\theta_1,\cos\theta_2;z) \\
&\quad = r \Big[ e^{\Psi(\ln m_1)} e^{\Phi(q)} \Big] \left(\frac{1+z}{1+0.2}\right)^\kappa p(\chi_1,\chi_2,\cos\theta_1,\cos\theta_2).
\end{aligned}
\label{eq:ar-m1-q}
\end{equation}
We fit for both $\Psi(\ln m_1)$ and $\Phi(q)$ simultaneously, allowing each process to possess its own variance and autocorrelation length.

While our focus in this section is on the mass distribution of binary black holes, when measuring the population distribution of any one parameter it is usually important to simultaneously fit the distributions of other parameters, like spin magnitudes $\chi_i$, spin-orbit misalignment angles $\theta_i$, and redshifts.
There is no fundamental reason why the distributions over \textit{all} binary parameters cannot be fit as simultaneous autoregressive processes.
Since each additional AR(1) process introduces a fairly high computational cost, however, for simplicity we will fit the ``leftover'' redshift and spin distributions by falling back on ordinary parametrized models.
We assume that the merger rate evolves as $(1+z)^\kappa$ for some unknown index $\kappa$~\cite{Fishbach2018} and that component spins are independently and identically distributed with a probability distribution $p(\chi_1,\chi_2,\cos\theta_1,\cos\theta_2)$ of the form given in Appendix~\ref{app:strong-models}; these redshift and spin distributions are hierarchically fit alongside our autoregressive mass model.
Finally, note that our model in Eq.~\eqref{eq:ar-m1-q} is presumed to be separable, with no correlations between the masses, mass ratios, and spins of binary black holes.

\begin{figure*}
    \centering
    \includegraphics[width=0.9\textwidth]{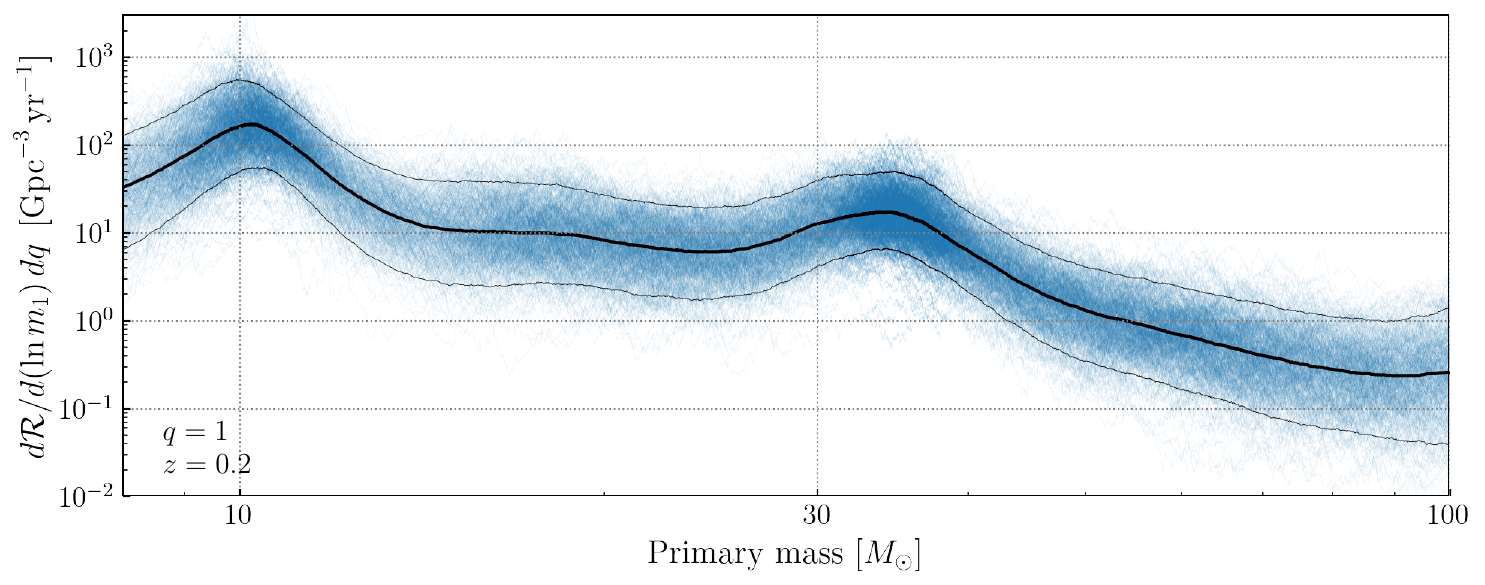} \\
    \includegraphics[width=0.9\textwidth]{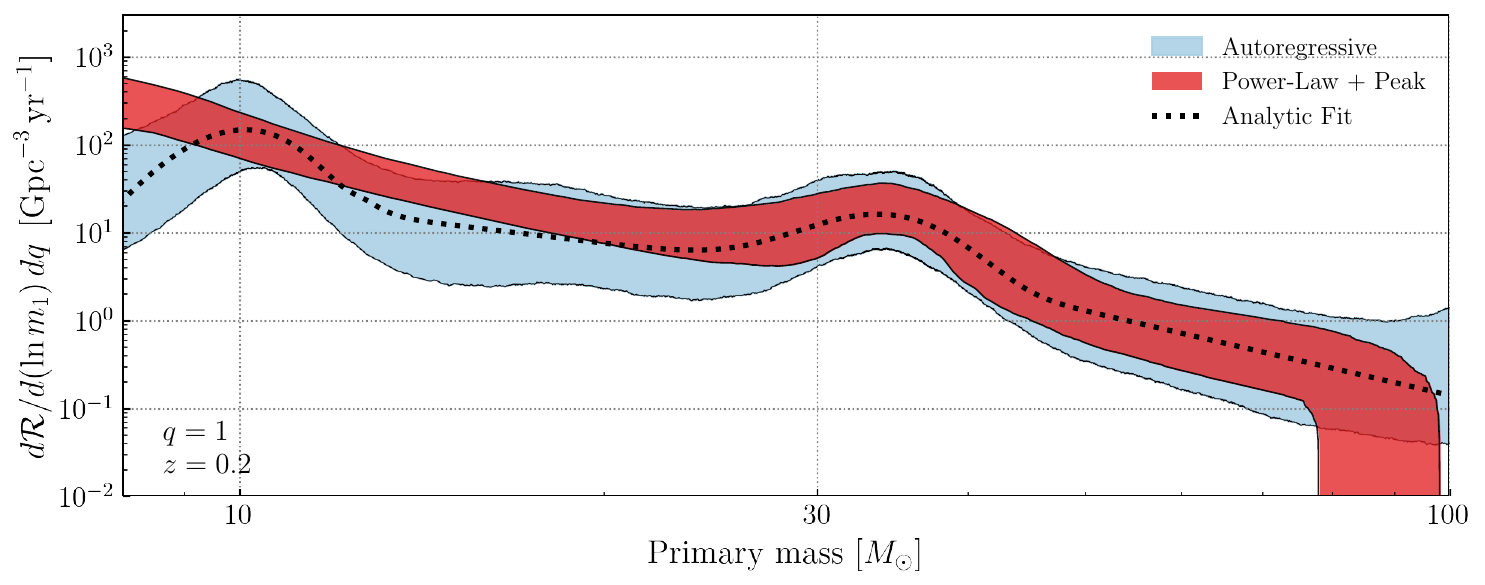}
    \caption{
    \textit{Top}: The binary black hole merger rate as a function of primary mass, inferred non-parametrically under an autoregressive prior.
    The merger rate is evaluated at mass ratio $q=1$ and redshift $z=0.2$, and is marginalized over spins, following the model defined in Eq.~\eqref{eq:ar-m1-q}.
    The solid black trace marks the mean inferred $\mathcal{R}(\ln m_1)$ as a function of $m_1$, while the lighter black traces bound our central 90\% credible bounds.
    Individual posterior draws on $\mathcal{R}(\ln m_1)$ are shown via light blue traces.
    Three features naturally emerge in the inferred mass distribution: a global maximum in the merger rate at $m_1 \approx 10\,M_\odot$, a secondary maximum at $m_1 \approx 35\,M_\odot$, and an otherwise smooth continuum that steepens above $40\,M_\odot$.
    Each of these three features is exhibited by approximately 90\% of posterior draws.
    \textit{Bottom}:
    A comparison between our autoregressive inference (blue band) and results obtained using the strongly parametrized \textsc{PowerLaw+Peak} model in Ref.~\cite{O3b-pop} (red).
    Rates are again evaluated at $q=1$ and $z=0.2$, and marginalized over spin.
    Each band encompasses the central 90\% credible region inferred using the given model.
    Both approaches give consistent estimates of the merger rate at $m_1 \approx 10\,M_\odot$, as well as the merger rate in the $30$-$70\,M_\odot$ interval.
    In order to match these rates, though, we see that the parametrized model is forced to overestimate the merger rate between $15$-$30\,M_\odot$, as well as the merger rate below $10\,M_\odot$.
    Furthermore, our autoregressive model shows no indication of a sharp cutoff in the binary mass distribution at or above $80\,M_\odot$ (this feature is included \textit{a priori} in the strongly parametrized model).
    A simple fit to our median inferred rate, using the parametric form of Eq.~\eqref{eq:mass-fit}, is shown via the black dotted curve.
    }
    \label{fig:lnm1}
\end{figure*}

\begin{figure}
    \centering
    \includegraphics[width=0.45\textwidth]{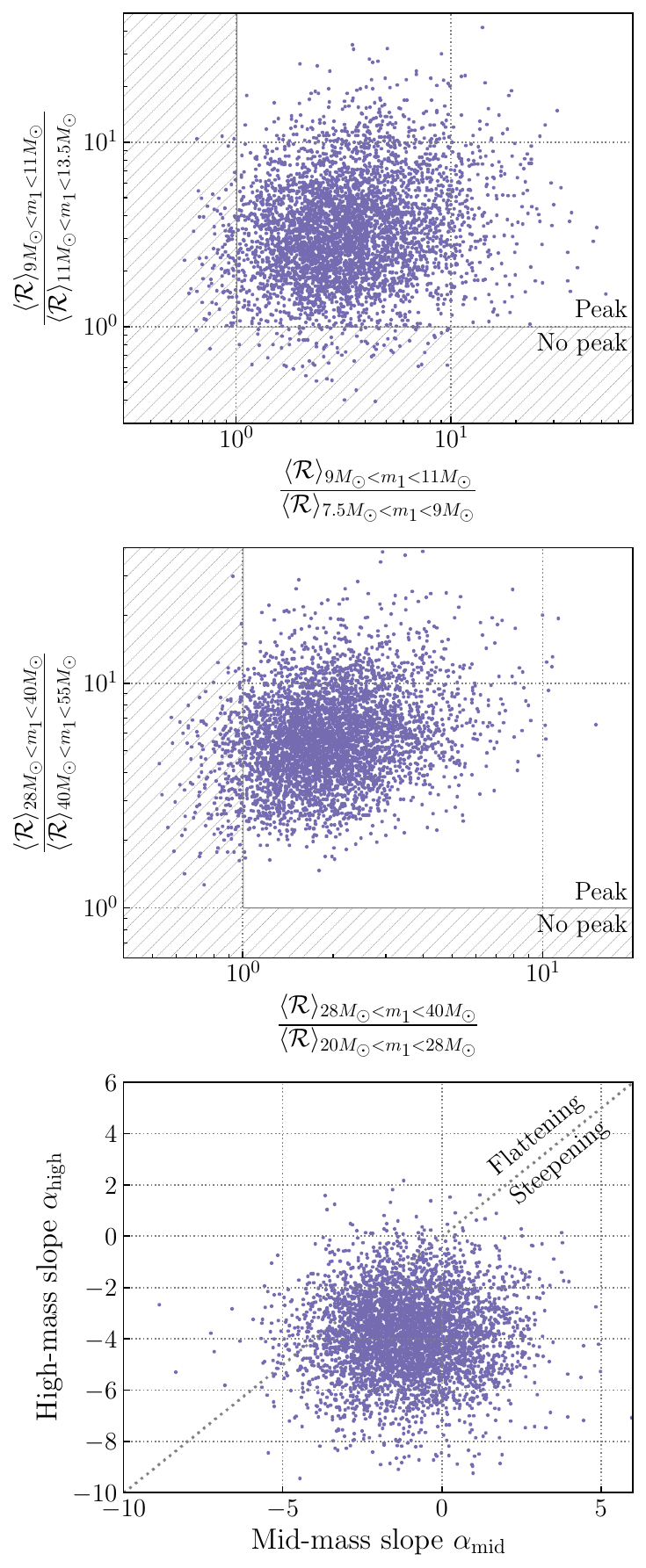}
    \caption{
    Tests quantifying the significances of features identified in Fig.~\ref{fig:lnm1}.
    \textit{Top}: Ratios between the average merger rate across $9\,M_\odot < m_1 < 11\,M_\odot$ and in adjacent lower- and higher-mass intervals.
    If a peak is present near $10\,M_\odot$, both ratios should be greater than one; this is true for $\MassSignificancePercentageLowPeak$ of our samples.
    \textit{Middle}: Similarly, ratios between the average merger rate across $28\,M_\odot <m_1<40\,M_\odot$ and in adjacent bands.
    A peak near $35\,M_\odot$ is present in $\MassSignificancePercentageHighPeak$ of samples.
    \textit{Bottom}: The implied power-law indices characterizing the $15\text{--}25\,M_\odot$ and $45\text{--}75\,M_\odot$ intervals ($\alpha_\mathrm{mid}$ and $\alpha_\mathrm{high}$, respectively).
    We find that $\MassSignificancePercentageSteepening$ of samples show a steepening in the mass spectrum, with $\alpha_\mathrm{high}<\alpha_\mathrm{mid}$.
    }
    \label{fig:mass-features}
\end{figure}

\subsection{Features in the black hole mass distribution}

The top panel of Fig.~\ref{fig:lnm1} shows our autoregressive measurement on the merger density rate of binary black holes as a function of primary mass, evaluated at $q=1$ and $z=0.2$ and marginalized over spin degrees of freedom.
Each blue trace shows a single posterior sample for $\mathcal{R}(\ln m_1)$,\footnote{For convenience we neglect the full functional dependence $\mathcal{R}(\ln m_1,q,\chi_1,\chi_2,\cos\theta_1,\cos\theta_2;z)$ and just abbreviate this quantity as $\mathcal{R}(\ln m_1)$, since we are only concerned with the dependence of the merger rate on mass at the moment.
We will use an analogous shorthand below when focusing on other parameters as well.} while the thick and thin black curves mark a running median and central 90\% credible bounds, respectively.
We note that our presentation of the mass spectrum, conditioned on some particular reference values of mass ratio and redshift, is slightly unusual; it is more common to show a mass distribution that has been fully marginalized over other parameters.
When marginalizing a merger rate over one or more parameters, however, the result can show extreme systematic dependence on the exact model presumed for these marginalized parameters, particularly across regions of parameter space that are not well measured.
An extreme example can be found in Ref.~\cite{O3b-pop}, in which the fully marginalized binary neutron star merger rate can vary by two orders of magnitude depending on the mass model used.
Our approach in this paper is to minimize such systematics by instead quoting differential merger rates at well-measured locations in parameter space (e.g. $q=1$ and $z=0.2$); this approach maximizes precision and best enables comparison to predictions between observation and theory.

Returning to Fig.~\ref{fig:lnm1}, we see three possible features in the black hole primary mass spectrum:

\textit{1. A global maximum at $m_1\approx 10\,M_\odot$.}
The binary merger rate appears to be maximized at $m_1\approx 10\,M_\odot$ primary masses, falling off with both lower and higher primary masses.
We can quantify the significance of this feature by computing the fraction of posterior samples that exhibit a systematic peak in this neighborhood.
To do so, we compute and compare the average merger rates across three bins: $7.5$-$9\,M_\odot$, $9$-$11\,M_\odot$, and  $11$-$13.5\,M_\odot$ (chosen to have roughly equal logarithmic widths).
We regard a ``peak'' as a case when the averaged merger rate in the middle interval is higher than the averaged merger rates in both adjacent bins.
As shown in the top panel of Fig.~\ref{fig:mass-features}, we find that $\MassSignificancePercentageLowPeak$ of our samples meet this criterion and exhibit a systematic peak near $10\,M_\odot$.

\textit{2. A local maximum at $m_1\approx 35\,M_\odot$.}
We can again quantify the significance of this feature by comparing the average rates across three bins: $20$-$28\,M_\odot$, $28$-$40\,M_\odot$, and $40$-$55\,M_\odot$.
As shown in the middle panel of Fig.~\ref{fig:mass-features}, $\MassSignificancePercentageHighPeak$ of our posterior draws yield higher averaged merger rates in the $28$-$40\,M_\odot$ range than in both adjacent bins.
Thus both the $10\,M_\odot$ and $35\,M_\odot$ maxima have roughly equal significance; although neither is unambiguously required by the data, both are favored to exist at greater than $90\%$ credibility.

\textit{3. Steepening of the continuum above $40\,M_\odot$.}
Between the $10\,M_\odot$ and $35\,M_\odot$ maxima is a large, relatively flat continuum.
Above the $35\,M_\odot$ maximum, the continuum appears to steepen, falling off more rapidly with increasing mass.
We quantify the evidence for this steepening by computing and comparing the mean power-law slope of the black hole merger rate above and below the $35\,M_\odot$ maximum.
From each posterior sample, we extract the merger rates $\mathcal{R}(\ln m_1)$ near $15$, $25$, $45$, and $85\,M_\odot$;
these rates are then used to compute the power-law indices characterizing the middle and high end of the mass spectrum:
    \begin{equation}
    \alpha_\mathrm{mid} = \frac{\ln\overline{\mathcal{R}(25\,M_\odot)} - \ln\overline{\mathcal{R}(15\,M_\odot)}}{\ln(25\,M_\odot) - \ln(15\,M_\odot)}
    \end{equation}
and
    \begin{equation}
    \alpha_\mathrm{high} = \frac{\ln\overline{\mathcal{R}(85\,M_\odot)} - \ln\overline{\mathcal{R}(45\,M_\odot)}}{\ln(85\,M_\odot) - \ln(45\,M_\odot)}.
    \end{equation}
We write $\overline{\mathcal{R}(25\,M_\odot)}$, for example, to indicate the average merger rate in a $1\,M_\odot$ window around $25\,M_\odot$.
Using window-averaged rates in this fashion enables more reliable estimates of representative power-law indices due to the rapid oscillations exhibited by individual $\mathcal{R}(\ln m_1)$ traces.
The joint distribution of both power-law slopes is plotted in the lower panel of Fig.~\ref{fig:mass-features}.
In the $15$-$25\,M_\odot$ interval, we find an average power-law index $\alpha_\mathrm{mid} = \MassSignificanceLowSlope$, while in the $45$-$85\,M_\odot$ range we find $\alpha_\mathrm{high} = \MassSignificanceHighSlope$.
We identify a preference for steepening, with $\alpha_\mathrm{high} < \alpha_\mathrm{mid}$, in $\MassSignificancePercentageSteepening$ of samples, although this behavior is not strictly required by the data.

\subsection{Discussion}

The significances of the $10\,M_\odot$ and $35\,M_\odot$ peaks, as computed here, are similar to but more conservative than significance estimates presented elsewhere.
A strongly parameterized analysis presented in Ref.~\cite{O3b-pop} identifies a $35\,M_\odot$ excess at effectively $100\%$ credibility, and an analysis in the same study using splines to measure deviations from an ordinary power law finds upward fluctuations at $10\,M_\odot$ and $35\,M_\odot$ with greater than $99\%$ credibility (see also Refs.~\cite{Edelman2022a,Edelman2022b}).
Reference~\cite{Farah2023} alternatively explores the frequency with which apparent peaks might arise purely from random counting statistics due to our still-moderate number of binary black hole detections.
By repeatedly drawing realizations of 69 events from a peakless power-law population, they find the observed $10\,M_\odot$ and $35\,M_\odot$ peaks to be more statistically significant than over $99\%$ of false peaks arising from random clustering.

The difference between these significance estimates and ours is likely two-fold.
First, these significance estimates test slightly different features; an upward fluctuation relative to a power law does not necessarily indicate a local maximum, but can also be caused by a plateau or change in slope.
Second, by virtue of its extreme flexibility, our autoregressive prior likely maximizes the variance in our $\mathcal{R}(\ln m_1)$ measurements, diminishing slightly our confidence in any given feature.
We note that our assessment of feature significance does not depend on the particular choice of reference mass ratio and redshift adopted in Fig.~\ref{fig:lnm1}; different reference values would rescale each merger rate by a hyperparameter-dependent constant, which cancels when subsequently taking ratios between rates as in Fig.~\ref{fig:mass-features}.

In addition to the $10\,M_\odot$ and $35\,M_\odot$ maxima, other studies have noted the possible existence of other features in the primary mass spectrum, namely additional maxima or minima in the $15$-$25\,M_\odot$ range~\cite{Tiwari2021b,O3b-pop,Tiwari2022,Edelman2022b}.
We do not see evidence for any such features here, however, indicating that any additional features are likely prior dependent and consistent with random clustering of a still-small number of observations.
References~\cite{Edelman2022a,O3b-pop,Edelman2022b,Farah2023} note a somewhat significant \textit{dip} in the mass spectrum, relative to a power law, near $14\,M_\odot$.
We interpret this result not as a local minimum but just as a flattening of the power-law index at lower masses, as seen in Fig.~\ref{fig:lnm1} and discussed further below.
Additionally, various studies have searched for the presence of a high-mass cutoff in the black hole mass spectrum~\cite{Fishbach2017,Talbot2018,O2-pop,Fishbach2020,O3a-pop,Baxter2021,Edelman2021,Fishbach2021,O3b-pop}, due possibly to the occurrence of pair-instability supernova~\cite{Woosley2021}.
In Ref.~\cite{O3b-pop}, for example, it is inferred that if such a cutoff exists, then it must occur at  $m_1>78\,M_\odot$ at 95\% credibility.
Our analysis, however, shows no indication of a cutoff in the black hole mass spectrum, instead recovering a distribution that continues to smoothly decline out to $m_1\approx 100\,M_\odot$.
Note that the slight increase in the variance of $\mathcal{R}$ seen near $100\,M_\odot$ marks a reversion to the prior in the region $m_1\gtrsim 100\,M_\odot$ where we have little data.

It is valuable to compare our autoregressive results to measurements made using standard strongly parameterized models, in order to identify regions where strongly parameterized models may fail to capture features in the data and to guide the iterative development of improved models going forward.
In the bottom panel of Fig.~\ref{fig:lnm1}, we compare our autoregressive model of the binary merger rate (blue) with results obtained under the \textsc{PowerLaw+Peak} model~\cite{Talbot2018} presented in Ref.~\cite{O3b-pop}.
Both models identify an excess of mergers near $35\,M_\odot$, and both measure approximately consistent merger rates near $10\,M_\odot$.
We see two signs of tension, however.
First, the \textsc{PowerLaw+Peak} model otherwise adopts a single unbroken power law; in order to match the merger rate at both low $(10\,M_\odot)$ and high $(\geq 30\,M_\odot)$ masses, it is therefore forced to overestimate the merger rate in the $15$-$30\,M_\odot$ range.
This result is consistent with the \textit{downward} perturbation identified by spline-based methods in this region~\cite{Edelman2022a,O3b-pop,Edelman2022b,Farah2023}; this downward perturbation may not be caused by a local minima in the mass spectrum, but just a flattening of the power-law index at lower masses.

\begin{figure}
    \centering
    \includegraphics[width=0.45\textwidth]{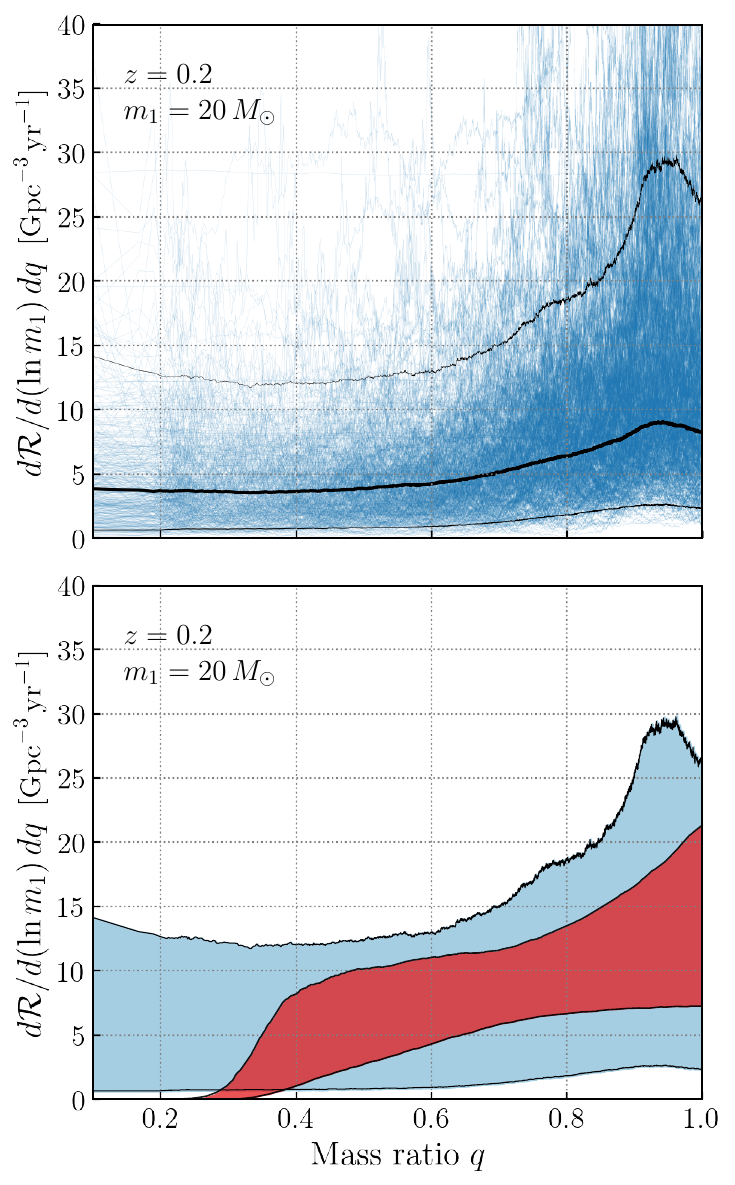}
    \caption{
    \textit{Top}: The merger rate of binary black holes as a function of mass ratio, evaluated at $m_1=20\,M_\odot$ and $z=0.2$, and integrated over possible spins, following Eq.~\eqref{eq:ar-m1-q}.
    The thick and thin black lines mark the mean and central 90\% bounds on $\mathcal{R}(q)$, while thin blue traces show individual draws from our posterior on $\mathcal{R}(q)$.
    We see a preference for an increasing merger rate as a function of $q$, but this behavior is not strictly required.
    \textit{Bottom}: Comparison between $\mathcal{R}(q)$ as inferred by our autoregressive model (blue) and the strongly parametrized analysis of Ref.~\cite{O3b-pop} (red), which assumes a power-law dependence on $q$ with a truncation in the merger rate below $q = m_\mathrm{min}/m_1$ for some minimum mass $m_\mathrm{min}$.
    The two results are broadly consistent, although under our autoregressive model we find reduced evidence for a merger rate that increases with larger $q$.
    }
    \label{fig:q}
\end{figure}

\subsection{Suitable parametric model}

In cases where a strongly parameterized phenomenological model is needed, our autoregressive result suggests that a sufficient choice is a model comprising two Gaussian peaks and a broken power law, with a probability density
    \begin{equation}
    \begin{aligned}
    &p(m_1) = f_{p,1} N(m_1|\mu_{m,1},\sigma_{m,1}) \\
        &\hspace{1.5cm} + f_{p,2} N(m_2|\mu_{m,2},\sigma_{m,2}) \\
        &\hspace{1.5cm} + (1-f_{p,1}-f_{p,2}) \Gamma(m_1).
    \end{aligned}
    \label{eq:mass-fit}
    \end{equation}
Here, we use $N(m_1|\mu,\sigma)$ to signify a normalized Gaussian distribution with mean $\mu$ and standard deviation $\sigma$, and $\Gamma(m_1)$ to denote a broken power law tapered towards zero at low masses:
    \begin{equation}
    \Gamma(m_1) \propto \begin{cases}
        e^{-\frac{(m_1-m_\mathrm{min})^2}{2\delta m^2}} \left(\dfrac{m_1}{m_b}\right)^{\alpha_1} & (m_1<m_\mathrm{min}) \\[8pt]
        \left(\dfrac{m_1}{m_b}\right)^{\alpha_1} & (m_\mathrm{min} \leq m_1 < m_b) \\[8pt]
        \left(\dfrac{m_1}{m_b}\right)^{\alpha_2} & (m_b \leq m_1 < m_\mathrm{max}) \\[8pt]
        0 & (\mathrm{else}),
        \end{cases}
    \end{equation}
with a proportionality constant chosen to enforce $\int \Gamma(m_1) dm_1 = 1$.
A least-squares fit against our mean inferred $\ln\mathcal{R}(\ln m)$ gives best-fit parameters
    \begin{equation}
    \begin{aligned}
    &\mu_1 = \MassFitMuOne\,M_\odot 
        &&\mu_2 = \MassFitMuTwo\,M_\odot \\
    &\sigma_1 = \MassFitSigOne\,M_\odot
        &&\sigma_2 = \MassFitSigTwo\,M_\odot \\
    &f_{p,1} = \MassFitFpeakOne 
        &&f_{p,2} = \MassFitFpeakTwo \\
    &\alpha_1 = \MassFitSlopeOne 
        &&\alpha_2 = \MassFitSlopeTwo \\
    &m_\mathrm{min} = \MassFitMmin\,M_\odot 
        &&\delta_m = \MassFitDeltaMmin\,M_\odot \\
    & m_b = \MassFitMbreak\,M_\odot
        && m_\mathrm{max} = \MassFitMmax\,M_\odot\,.
    \end{aligned}
    \end{equation}
The corresponding distribution $p(\ln m_1) = p(m_1) m_1$ is shown as a dotted line in Fig.~\ref{fig:lnm1}.
Note that this fit approximates the fully marginalized primary mass distribution and is thus valid at any choice of $q$, $z$, etc.

\subsection{Features in the black hole mass ratio distribution}
    
Compared to the primary mass distribution, we resolve relatively little information about the distribution of black hole mass ratios.
The top panel of Fig.~\ref{fig:q} illustrates our constraints on $\mathcal{R}(q)$, evaluated at $z=0.2$ and $m_1=20\,M_\odot$, and integrated over component spins.
The only feature that manifests in Fig.~\ref{fig:q} is a possible preference for larger $q$.
As above, we can compare integrated merger rates in two bands, $0.5\leq q \leq 0.6$ and $0.9\leq q \leq 1$, to quantify the significance of this feature.
We find that the merger rate in the high-$q$ interval is greater than the rate in the low-$q$ interval for $\MassRatioSignifanceRising$ of samples, such that the binary black hole population likely favors equal mass ratios.

In the lower panel of Fig.~\ref{fig:q} we compare our results with the strongly parametrized measurements presented in Ref.~\cite{O3b-pop} using the \textsc{PowerLaw+Peak} model, in which the mass ratio distribution is modeled as a power law with a primary-mass-dependent truncation:
    \begin{equation}
    p(q|m_1) \propto S(q;m_1)\,q^{\beta_q}.
    \end{equation}
Here, $S(q;m_1)$ is a tapering function that sends $p(q|m_1)$ to zero when $q<m_\mathrm{min}/m_1$ for some $m_\mathrm{min}$.
Both results are again evaluated at $z=0.2$ and $m_1=20\,M_\odot$, and integrated over black hole spins.
Other than the truncation below $q\approx 0.2$ (imposed in Ref.~\cite{O3b-pop} as an \textit{a priori} modeling choice), both sets of results are broadly consistent.
In the strongly parameterized analysis of Ref.~\cite{O3b-pop}, it is found that $\beta_q>0$ with 92\% credibility, comparable to our significance estimate above.\footnote{Within Ref.~\cite{O3b-pop}, the right-hand panel of Fig.~10 appears to be in tension with this significance estimate, instead showing a measurement of $\mathcal{R}(q)$ (marginalized over all $m_1$) that unambiguously increases as a function of $q$.
The behavior in Fig.~10 is actually due to the presumed \textit{truncation} in the mass ratio distribution, rather than a confident measurement of positive $\beta_q$.
Since the overall merger rate is highest at small $m_1$, the structure of the marginalized $\mathcal{R}(q)$ must correspondingly be dominated by the mass ratio distribution at small $m_1$.
The truncation in $p(q|m_1)$, however, enforces that $q\approx 1$ when $m_1$ is small.
This combination of effects requires $\mathcal{R}(q)$ to be maximized at $q=1$ after marginalization over $m_1$, nearly independently of $\beta_q$.
}

\section{Stop Two: Redshifts}
\label{sec:redshifts}

\begin{figure*}
    \centering
    \includegraphics[width=0.92\textwidth]{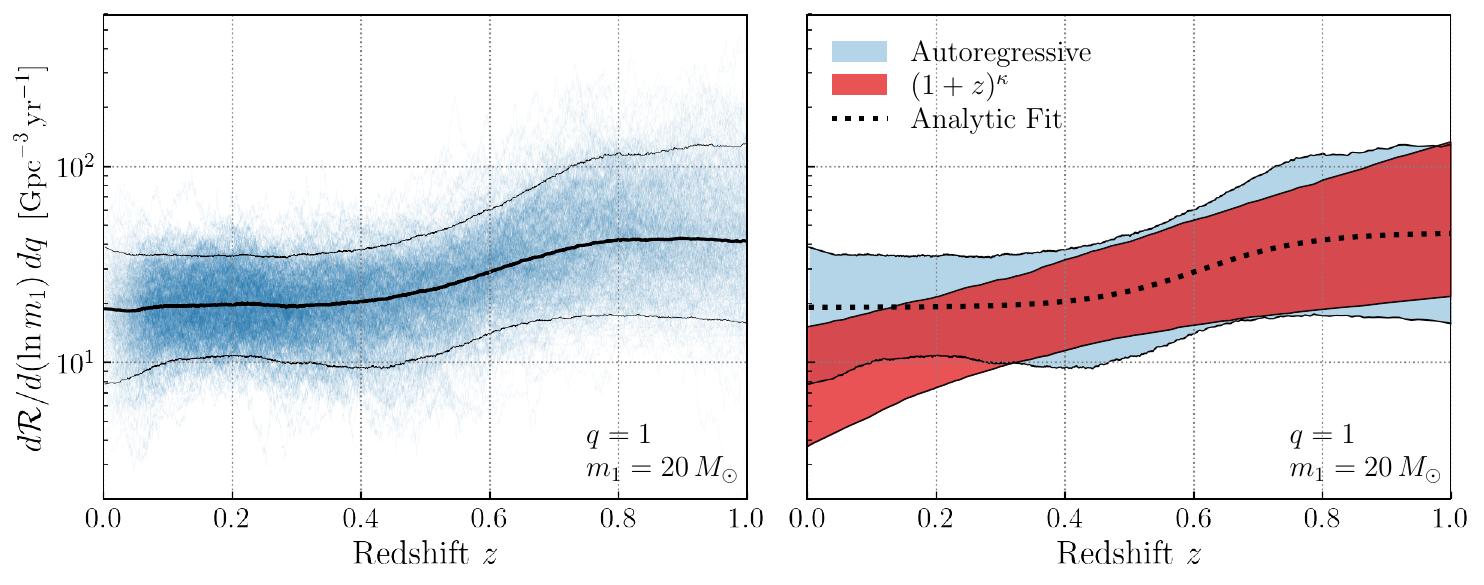}
    \caption{
    \textit{Left}: The binary black hole merger rate as a function of redshift, inferred non-parametrically using an autoregressive process prior.
    The merger rate is evaluated at a primary mass $m_1=20\,M_\odot$ and mass ratio $q=1$, and integrated over black hole spins.
    Light blue traces show individual draws from our posterior, while the black and grey curves denote a running median and central 90\% credible bounds, respectively.
    \textit{Right}: A comparison between our non-parametric result (blue) and the result obtained in Ref.~\cite{O3b-pop} when assuming that the merger rate evolves as $(1+z)^\kappa$ for an unknown index $\kappa$.
    Both bands denote 90\% credible bounds.
    We see that both approaches recover similar merger rates at $z\approx 0.3$ and $z\approx 1$, and both indicate that the black hole merger rate systematically grows with redshift.
    Our autoregressive result, however, suggests that this growth may not be well modeled by a power law but instead by a slowly growing or constant merger rate that begins to evolve more sharply only beyond $z\gtrsim0.4$.
    The dashed black curve, for example, shows the result of a simple least-squares fit to our median inferred merger rate using the sigmoid model defined in Eq.~\eqref{eq:sigmoid}.
    A broken power law, as in Eq.~\eqref{eq:z-bpl}, also yields a good fit at $z\lesssim 0.8$.
    }
    \label{fig:z}
\end{figure*}

\begin{figure*}
    \centering
    \includegraphics[width=0.92\textwidth]{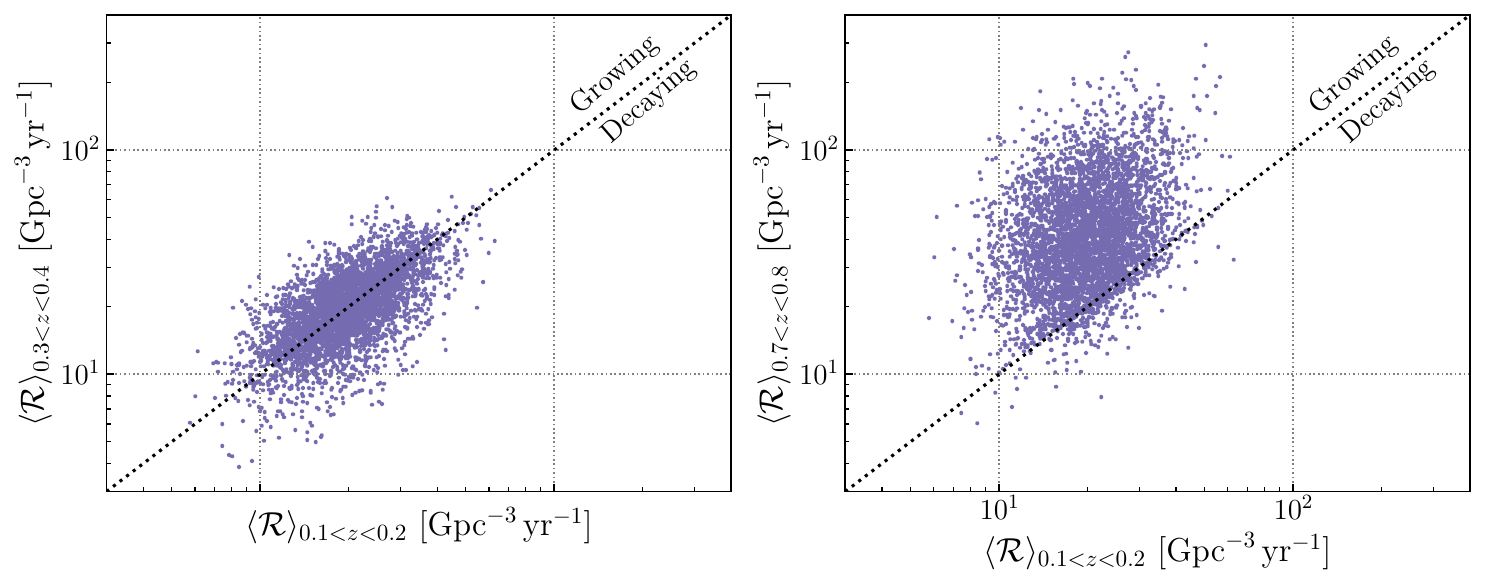}
    \caption{
    \textit{Left}: A comparison between the mean merger rate across the interval $0.3 < z < 0.4$ and the mean rate across $0.1<z<0.2$.
    Each point corresponds to a single posterior draw from Fig.~\ref{fig:z}.
    All estimates cluster around the diagonal, indicating that the merger rates in both intervals are consistent with one another.
    The data are therefore consistent with a non-evolving merger rate below $z\lesssim 0.4$.
    \textit{Right}: An analogous comparison between the mean merger rate in the interval $0.7 < z < 0.8$ and the mean rate within $0.1<z<0.2$.
    The merger rate in the high-redshift interval is greater than that in the low-redshift interval for $96\%$ of samples, indicating a preference for a merger rate that grows at large redshifts.
    }
    \label{fig:rate-diagnostics}
\end{figure*}

Next, we investigate the redshift distribution of binary black holes.
In most analyses, the redshift dependence of the binary black hole merger rate is presumed to follow a power-law form: $\mathcal{R}(z) \propto (1+z)^\kappa$ for some index $\kappa$~\cite{Fishbach2018,O3a-pop,Callister2020,Fishbach2021,O3b-pop,vanSon2022}.
Under this model, it has been concluded that the binary black hole merger rate systematically grows with redshift at a rate consistent with star formation in the local Universe~\cite{O3b-pop}.
Here, we instead model the redshift dependence of the black hole merger rate as an autoregressive process, searching for any features that might be missed under a more strongly parameterized approach.
We simultaneously measure the mass and component spin distributions by falling back on the strongly-parameterized models described in Appendix~\ref{app:strong-models}.
Together, our model is of the form
    \begin{equation}
    \begin{aligned}
    &\mathcal{R}(\ln m_1,q,\chi_1,\chi_2,\cos\theta_1,\cos\theta_2;z) \\
        &\quad = r\frac{f(m_1,q)}{f(20\,M_\odot,1)}  e^\Psi(z)\, p(\chi_1,\chi_2,\cos\theta_1,\cos\theta_2).
    \end{aligned}
    \label{eq:ar-z}
    \end{equation}

\subsection{Features in the black hole redshift distribution}

The left panel of Fig.~\ref{fig:z} shows our resulting inference on the binary merger rate as a function of redshift, evaluated at $m_1=20\,M_\odot$ and $q=1$ and integrated across spins.
Blue traces show individual draws from our posterior, the solid black curve marks the running median rate, and thin grey lines denote central $90\%$ credible bounds on the merger rate at each redshift.
The right panel of Fig.~\ref{fig:z} compares these results (in blue) to the results obtained in Ref.~\cite{O3b-pop} using the strongly parameterized power-law model for the black hole merger rate.
Both approaches yield consistent estimates of the merger rate at $z\approx 0.3$ and $z\approx 1$, but our autoregressive result suggests that the intervening evolution is not necessarily well modeled by a power law.
Instead, our result is consistent with a ``sigmoid'' shape displaying the following features:

\textit{1. A non-evolving, uniform-in-comoving-volume rate below $z\approx0.4$}.
At the lowest redshifts, the data do not require the merger rate to evolve with redshift.
Instead, our autoregressive results are consistent with a rate that remains constant out to $z\approx 0.4$.
To gauge the significance of this feature, we compute and compare the mean merger rates in two intervals: $0.1<z<0.2$ and $0.3<z<0.4$.
As shown in the left panel of Fig.~\ref{fig:rate-diagnostics}, we find these mean rates to be consistent with one another, with the mean rate in the $0.3<z<0.4$ interval exceeding the rate in the $0.1<z<0.2$ interval only $\RedshiftSignificancePercentageLowRise$ of the time.

\textit{2. A rise in the merger rate between $z\approx0.4$ and $0.8$}.
Beyond redshift $z\approx 0.4$, however, the merger rate is required to increase by up to an order of magnitude by $z\approx0.8$.
We quantify the significance of this rise by comparing the mean merger rate in the range $0.1<z<0.2$ to the mean rate in the range $0.7<z<0.8$.
As shown in the right panel of Fig.~\ref{fig:rate-diagnostics}, the mean rates in these high- and low-redshift intervals are confidently unequal, with the $0.7<z<0.8$ merger rate exceeding the $0.1<z<0.2$ rate $\RedshiftSignificancePercentageHighRise$ of the time.
Beyond redshift $z\approx1$, the absence of informative data causes our measurement to asymptote back towards the autoregressive prior, yielding expanding error bars towards higher redshifts.

\subsection{Discussion}

Other studies employing flexible non-parameteric analysis have also obtained results indicating a possible tension with a $(1+z)^\kappa$ power law.
Reference~\cite{Payne2022} explored the use of population models composed of ``Green's-functions''-like delta functions as a tool with which to diagnose the performance of strongly parameterized models.
In that work, the likelihood is found to be maximized when $\mathcal{R}(z)$ is modeled as a sequence of delta functions that initially \textit{decrease} in height below $z\approx 0.13$, followed by an elevated but flat merger rate in the range $0.2\lesssim z \lesssim 0.5$ that then more sharply rises between $ 0.5\lesssim z \lesssim 0.75$; see their Fig.~5.
Other than the initially decreasing merger rate, which we do not recover, these results are consistent with the behavior we see in Fig.~\ref{fig:z}.
Reference~\cite{Edelman2022b}, in turn, measured the redshift-dependent merger rate using a set of basis splines to capture deviations from a $(1+z)^\kappa$ power law.
That work also recovered a largely constant merger rate density below $z\approx 0.4$, followed by a steeper increase in the merger rate out to $z\approx 1$; see their Fig.~8.

If real, the steplike structure in the redshift-dependent merger rate could arise from a variety of effects.
The redshift-dependent merger rate $\mathcal{R}(z)$ is generally modeled by convolving an estimate of the metallicity-dependent cosmic star formation rate with a distribution of time delays between progenitor formation and binary merger; the time delay distribution is itself typically modeled as a power law.
The resulting merger rate is also usually well described by a power law at low redshifts.
If the observed binary black hole population is dominated by a single formation channel, the possible non-power-law behavior in Fig.~\ref{fig:z} could indicate additional nontrivial structure in the birth rate or time delay distribution of binary progenitors.
Alternatively, the observed binary population could comprise a mixture of several distinct formation channels.
A shift from a flat to an evolving merger rate at $z\approx 0.5$ could mark a transition between two formation channels, one of which dominates low-redshift mergers and the other of which takes over at larger redshifts.
If a mixture between formation channels is the correct explanation of Fig.~\ref{fig:z}, then we should also expect to see systematic evolution in \textit{other} intrinsic properties of binary black holes between low and high redshifts.
Although no such evolution has been found in the binary black hole mass spectrum~\cite{Fishbach2021,vanSon2022}, the binary black hole \textit{spin} distribution potentially evolves with redshift, with the effective inspiral spin (further discussed in Sect.~\ref{sec:effective-spins} below) becoming larger and more positive at higher $z$~\cite{Biscoveanu2022}.
Additional observations will be critical in confirming the trends identified in Fig.~\ref{fig:z} and in Ref.~\cite{Biscoveanu2022} and in probing any relationship between these two trends.

\subsection{Suitable parametric models}

When a parametric model is required, our autoregressive results suggest that one might replace the standard power-law model with a broken power law:
    \begin{equation}
    \frac{d\mathcal{R}}{d\ln m_1\,dq}(z) = \begin{cases}
    \mathcal{R}_b \left(\dfrac{1+z}{1+z_b}\right)^{\kappa_1} & (z\leq z_b) \\[5mm]
    \mathcal{R}_b \left(\dfrac{1+z}{1+z_b}\right)^{\kappa_2} & (z > z_b),
    \end{cases}
    \label{eq:z-bpl}
    \end{equation}
with a transition between power-law indices $\kappa_1$ and $\kappa_2$ occurring at $z=z_b$, or a sigmoid,
    \begin{equation}
    \frac{d\mathcal{R}}{d\ln m_1\,dq}(z) = \mathcal{R}_0 + \frac{\delta\mathcal{R}}{1 + e^{-(z-z_b)/\delta z}},
    \label{eq:sigmoid}
    \end{equation}
in which the merger rate density increases from $\mathcal{R}_0$ to $\mathcal{R}_0 + \delta\mathcal{R}$ across an interval of width $\delta z$ around a transition redshift $z_b$.
A least-squares fit to our median $\ln\mathcal{R}$ using Eq.~\eqref{eq:z-bpl} gives
    \begin{equation}
    \begin{aligned}
    &\kappa_1 = \RedshiftFitBplKappaLow \\
    &\kappa_2 = \RedshiftFitBplKappaHigh \\
    &z_b = \RedshiftFitBplZBreak \\
    &\mathcal{R}_b = \RedshiftFitBplRateBreak\,{\rm Gpc}^{-3}\,{\rm yr}^{-1}.
    \end{aligned}
    \end{equation}
A fit using Eq.~\eqref{eq:sigmoid}, in turn, gives
    \begin{equation}
    \begin{aligned}
    &z_b = \RedshiftFitSigmoidZCenter \\
    & \delta z = \RedshiftFitSigmoidDeltaZ \\
    &\mathcal{R}_0 = \RedshiftFitSigmoidRateLow\,{\rm Gpc}^{-3}\,{\rm yr}^{-1} \\
    & \delta\mathcal{R} = \RedshiftFitSigmoidDeltaRate\,{\rm Gpc}^{-3}\,{\rm yr}^{-1};
    \end{aligned}
    \end{equation}
this fit is shown as a dotted curve in Fig.~\ref{fig:z}.
As our autoregressive results begin to revert to the prior above $z=1$, these fits are performed only in the restricted range $z\leq 0.8$.

Recall that the above fits describe the merger rate per $\ln m_1$ per unit $q$ evaluated at $m_1 = 20\,M_\odot$ and $q=1$, \textit{not} the fully-integrated merger rate.
If the full binary black hole merger rate, integrated over all masses, is desired, it can be fit with the same functional forms:
    \begin{equation}
    \mathcal{R}(z) = \begin{cases}
    \mathcal{R}_b \left(\dfrac{1+z}{1+z_b}\right)^{\kappa_1} & (z\leq z_b) \\[5mm]
    \mathcal{R}_b \left(\dfrac{1+z}{1+z_b}\right)^{\kappa_2} & (z > z_b),
    \end{cases}
    \label{eq:m-z-bpl}
    \end{equation}
or
    \begin{equation}
    \mathcal{R}(z) = \mathcal{R}_0 + \frac{\delta\mathcal{R}}{1 + e^{-(z-z_b)/\delta z}},
    \label{eq:m-z-sigmoid}
    \end{equation}
with parameters
    \begin{equation}
    \begin{aligned}
    &\kappa_1 = \MarginalRedshiftFitBplKappaLow \\
    &\kappa_2 = \MarginalRedshiftFitBplKappaHigh \\
    &z_b = \MarginalRedshiftFitBplZBreak \\
    &\mathcal{R}_b = \MarginalRedshiftFitBplRateBreak\,{\rm Gpc}^{-3}\,{\rm yr}^{-1}
    \end{aligned}
    \end{equation}
in Eq.~\eqref{eq:m-z-bpl} or
    \begin{equation}
    \begin{aligned}
    &z_b = \MarginalRedshiftFitSigmoidZCenter \\
    & \delta z = \MarginalRedshiftFitSigmoidDeltaZ \\
    &\mathcal{R}_0 = \MarginalRedshiftFitSigmoidRateLow\,{\rm Gpc}^{-3}\,{\rm yr}^{-1} \\
    & \delta\mathcal{R} = \MarginalRedshiftFitSigmoidDeltaRate\,{\rm Gpc}^{-3}\,{\rm yr}^{-1};
    \end{aligned}
    \end{equation}
in Eq.~\eqref{eq:m-z-sigmoid}.
    
\section{Stop Three: Component Spins}
\label{sec:spins}

Next, we turn to the distribution of spins among binary black hole systems.
A black hole binary is characterized by six spin degrees of freedom, three per component spin.
Assuming that component spins have no preferential azimuthal orientations (although see Ref.~\cite{Varma2022}), we work in a reduced four-dimensional space and fit for the distributions of component spin magnitudes, $\chi_1$ and $\chi_2$, and (cosine of the) spin-orbit tilt angles, $\cos\theta_1$ and $\cos\theta_2$.
We assume that the variation of the merger rate across spin magnitudes and tilts is described via two autoregressive processes, $\Psi(\chi)$ and $\Phi(\cos\theta)$, with the two component spins in a given binary distributed independently and identically.
As we measure $\Psi(\chi)$ and $\Phi(\cos\theta)$, we simultaneously infer the mass and redshift distributions of the binary black hole population by falling back on ordinary strongly parametrized models, assuming a primary mass and mass ratio distributions $f(m_1)$ and $p(q)$ as described in Appendix~\ref{app:strong-models} and a merger rate density that grows as $(1+z)^\kappa$.
Together, our full merger rate model is of the form
    \begin{equation}
    \begin{aligned}
    &\mathcal{R}(\ln m_1,q,\chi_1,\chi_2,\cos\theta_1,\cos\theta_2;z) \\
        &\hspace{-1mm} = r \frac{f(m_1)\,p(q)}{f(20\,M_\odot)}  \left(\frac{1+z}{1+0.2}\right)^\kappa \Big[e^{\Psi(\chi_1)}  e^{\Psi(\chi_2)} e^{\Phi(\cos\theta_1)} e^{\Phi(\cos\theta_2)}\Big].
    \end{aligned}
    \label{eq:ar-component-spins}
    \end{equation}

\begin{figure*}
    \centering
    \includegraphics[width=0.93\textwidth]{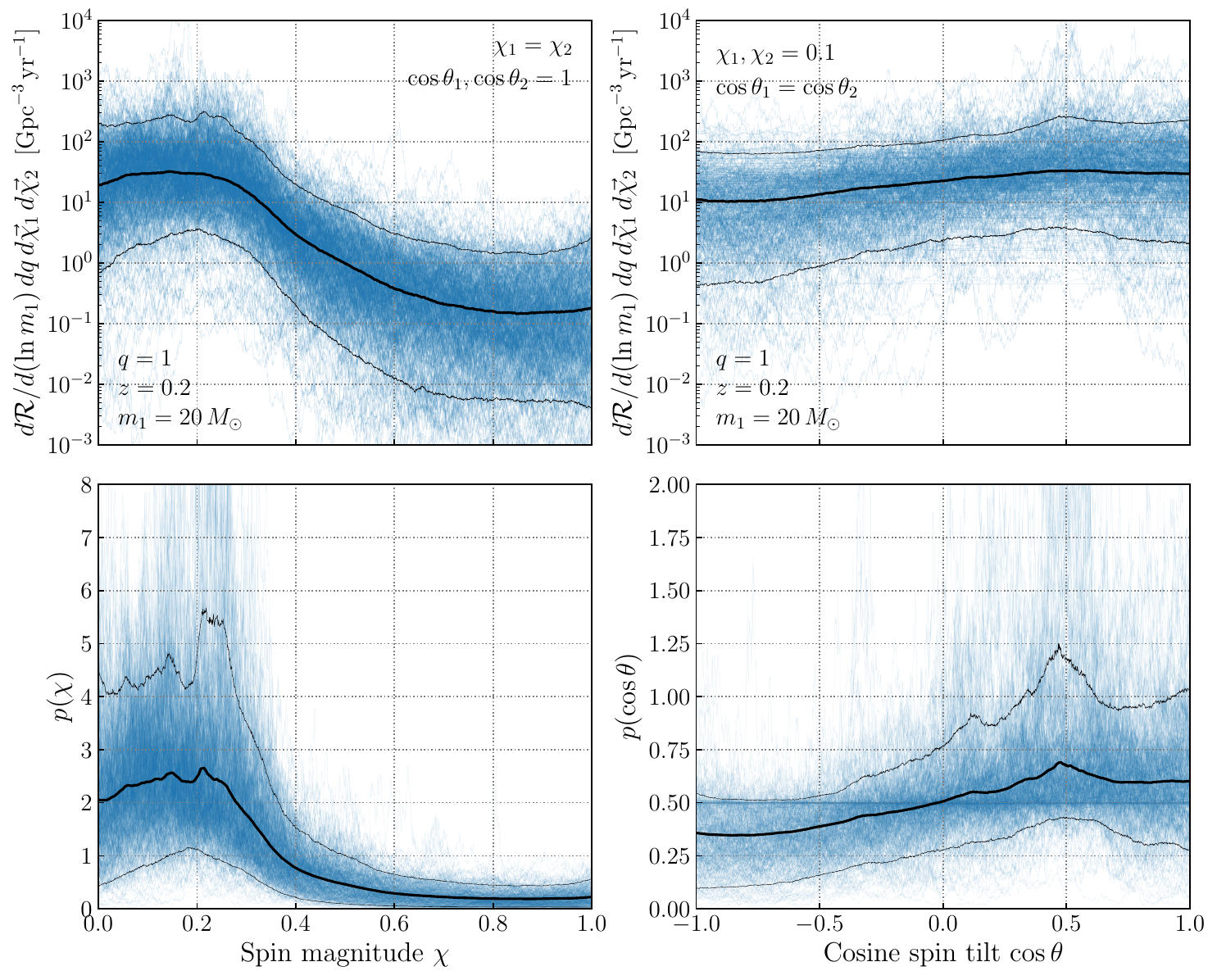}
    \caption{
    \textit{Top}: The merger rate of binary black holes as a function of component spin magnitudes (left) and spin-orbit misalignment angles (right), as inferred using our autoregressive model defined in Eq.~\eqref{eq:ar-component-spins}.
    In the axes labels, we use the shorthand $d\vec\chi_1 \equiv d\chi_1 d\cos\theta_1$ to indicate a density over both spin magnitude and cosine tilt.
    Specifically, the rates shown are that of binaries with equal component spin magnitudes ($\chi_1 = \chi_2 = \chi$; see Eq.~\eqref{eq:R-chi}) or tilts ($\cos\theta_1 = \cos\theta_2 = \cos\theta$; Eq.~\eqref{eq:R-cost}), each evaluated at fixed reference masses and redshift ($m_1 = 20\,M_\odot$, $q=1$, and $z=0.2$).
    The bottom panels show the corresponding probability distributions on component spin magnitudes and tilts among black hole binaries.
    Within each panel, the central black curve marks the mean inferred rate/probability, while outer black curves bound 90\% credible intervals.
    We see that spin magnitudes are well described by a unimodal distribution that peaks at low values, with no sign of an excess of non-spinning ($\chi=0$) or near-maximally spinning black holes.
    Meanwhile, the rate of binary mergers is non-zero across the full range of misalignment angles, with a spin-tilt distribution that is possibly (but not necessarily) isotropic.
    While there also appears to be a possible excess of black holes with $\cos\theta\approx0.4$, this feature is not statistically significant.
    }
    \label{fig:chi-cost}
\end{figure*}

\subsection{Features in the black hole spin distribution}

Figure~\ref{fig:chi-cost} shows our autoregressive measurements of the black hole spin magnitude and tilt distributions.
We plot our results in two ways.
First, the upper row shows merger rates as a function of spin magnitude and orientation.
The upper left panel shows the merger rate of binaries along the $\chi_1 = \chi_2 = \chi$ diagonal at fixed reference mass, mass ratio, redshift, and spin tilts ($m_1=20\,M_\odot$, $q=1$, $z=0.2$, and $\cos\theta_1=\cos\theta_2=1$); using Eq.~\eqref{eq:ar-component-spins}, this is given by
    \begin{equation}
    \mathcal{R}(\chi_1,\chi_2 = \chi) = r\,\left(e^{\Psi(\chi)} \right)^2 \left(e^{\Phi(1)}\right)^2.
    \label{eq:R-chi}
    \end{equation}
Similarly, the upper right panel shows the merger rate along the $\cos\theta_1 = \cos\theta_2 = \cos\theta$ diagonal at fixed spin magnitudes ($\chi_1 = \chi_2 = 0.1$) and the same reference masses and redshift:
    \begin{equation}
    \mathcal{R}(\cos\theta_1,\cos\theta_2=\cos\theta) = r\,\left(e^{\Psi(0.1)} \right)^2 \left(e^{\Phi(\cos\theta)}\right)^2.
    \label{eq:R-cost}
    \end{equation}
We choose to plot results along the $\chi_1 = \chi_2$ and $\cos\theta_1 = \cos\theta_2$ diagonals to mitigate systematic modeling uncertainties, in much the same way that we plot merger rates conditioned on specific values of other parameters rather than marginalizing over them.
The rate of black hole mergers as a function of $\chi_1$ only (marginalized over $\chi_2$), for instance, is strongly affected by assumptions regarding spin pairing which tend to differ widely across the literature.
For better comparison with other work, however, in the lower row we also show the implied probability distributions on individual component spin magnitudes and tilts.
Since we assume that component spins are independently and identically distributed, the spin magnitude and tilt probability distributions are given by
    \begin{equation}
    p(\chi) = \frac{e^{\Psi(\chi)}}{\int_0^1 e^{\Psi(\chi')} d\chi' }
    \label{eq:p-chi}
    \end{equation}
and
    \begin{equation}
    p(\cos\theta) = \frac{e^{\Phi(\cos\theta)}}{\int_{-1}^1 e^{\Phi(\cos\theta')} d\cos\theta' }.
    \label{eq:p-cost}
    \end{equation}
Note that Eqs.~\eqref{eq:R-chi} and \eqref{eq:R-cost} are proportional to the \textit{squares} of Eqs.~\eqref{eq:p-chi} and \eqref{eq:p-cost}, respectively.

From Fig.~\ref{fig:chi-cost}, we can make the following parameter-free statements regarding the binary black hole spin distribution:

\begin{figure}
    \centering
    \includegraphics[width=0.46\textwidth]{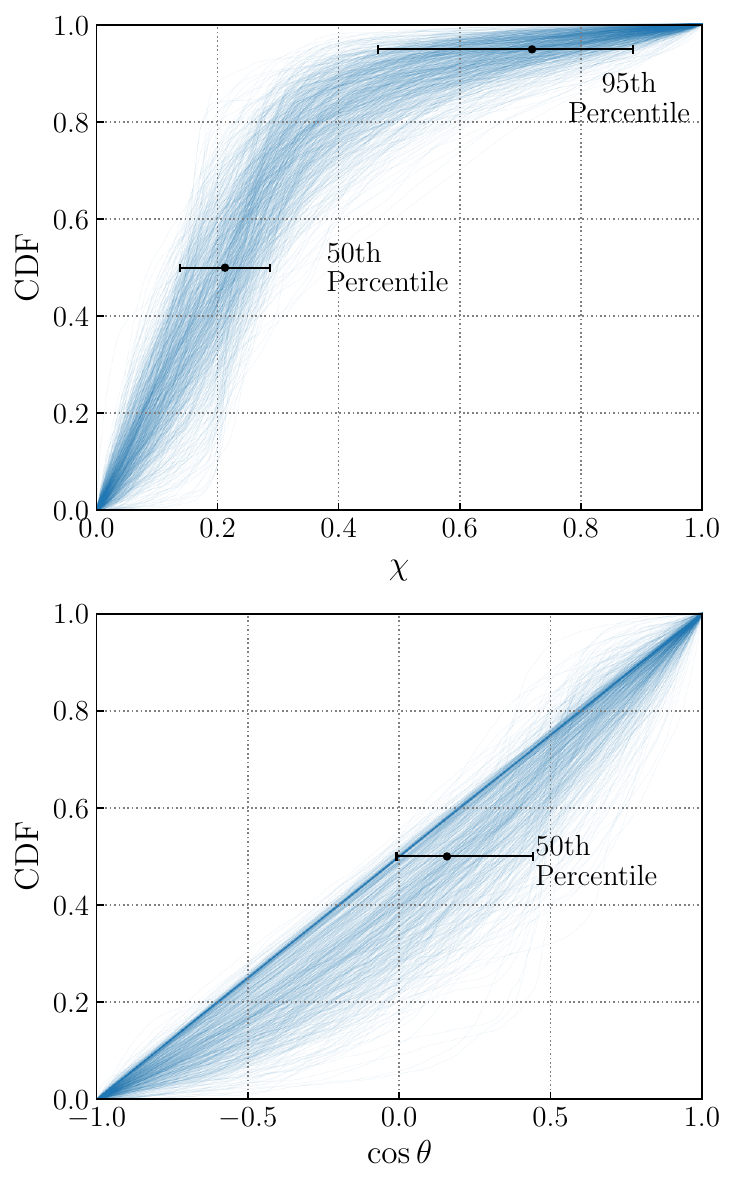}
    \caption{
    Cumulative distribution functions of binary black hole spin magnitudes (top) and cosine tilt angles (bottom), corresponding to the probability distributions shown in Fig.~\ref{fig:chi-cost-comparison}.
    For reference, we mark median estimates of the $50$th and $95$th percentiles in the spin magnitude distribution, occurring at $\chi_{50\%} = \SpinMagnitudeFifty$ and $\chi_{95\%} = \SpinMagnitudeNinetyFive$, respectively.
    We also indicate the measured $50$th percentile of the $\cos\theta$ distribution, occurring at $\cos\theta_{\,50\%} = \SpinTiltFifty$, with $\cos\theta_{\,50\%}>0$ at $\SpinTiltMedianPercentPositive$ credibility.
    }
    \label{fig:spin_cdfs}
\end{figure}

\textit{1. The binary black hole merger rate is maximized at low spin magnitudes.}
As in Sect.~\ref{sec:masses}, we can evaluate the robustness of this statement by comparing mean merger rates in  different intervals.
We find, for example, that our inferred rate of mergers with $0\leq \chi \leq 0.2$ is greater than the rate of mergers across $0.6 \leq \chi \leq 0.8$ for each of our $4500$ posterior samples on $\mathcal{R}(\chi)$.
In Fig.~\ref{fig:spin_cdfs}, we additionally show the ensemble of cumulative distribution functions corresponding to our posterior on $p(\chi)$ from Fig.~\ref{fig:chi-cost}.
We find the 50th percentile to occur at $\chi_{50\%} = \SpinMagnitudeFifty$, such that half of black holes have spin magnitudes below $\chi\lesssim0.2$.
The $p(\chi)$ distribution shown in Fig.~\ref{fig:chi-cost} furthermore suggests that the spin magnitude distribution may actually peak near $\chi\approx 0.2$; the recovered mean (shown in black) rises slightly in this region and the upper bound on $p(\chi)$ is elevated between $0.2\lesssim\chi\lesssim0.25$.
Neither of these features are statistically significant though; only $\MagnitudeSignificancePercentagePeak$ of traces give larger integrated probability in the $0.15\lesssim\chi\lesssim0.35$ interval than in the $0\lesssim\chi\lesssim0.15$ interval.
Thus the spin magnitude distribution is consistent with a peak global maximum at $\chi\approx 0$.

\textit{2. No special features required at $\chi=0$ or $\chi=1$.}
Although binary black holes exhibit a preference for small spins, the data do not require sharp or discontinuous excesses of non-spinning or maximally spinning black holes.
The possible existence of these features has been the subject of much scrutiny.
Initial work found that gravitational-wave data were consistent with two distinct subpopulations: a ``spike'' comprising the majority of the binary population and a secondary broad sub-population centered at $\chi\approx 0.5$ and possibly extending to large spins~\cite{Roulet2021}.
Later work further asserted that such features were in fact \textit{required} by the data~\cite{Galaudage2021}.\footnote{{A subsequent erratum~\cite{Galaudage2022} diminished the initial evidence in Ref.~\cite{Galaudage2021} for distinct non-spinning and spinning sub-populations, bringing their conclusions into closer agreement with those of Refs.~\cite{O3b-pop,Mould2022,Callister2022,Tong2022}}}
And both Refs.~\cite{Roulet2021,Galaudage2021} suggested that the failure by other analyses to properly model a zero-spin sub-population led to spurious identification of spin-orbit misalignment among the black hole population (to be discussed further below).
Follow-up investigations, however, have since concluded that the data remain agnostic about zero-spin or rapidly-spinning sub-populations~\cite{O3b-pop,Mould2022,Callister2022,Tong2022}.

In our Fig.~\ref{fig:chi-cost}, we see no indication of an excess of non-spinning systems, nor do we see any feature suggesting a sub-population of rapidly spinning black holes.
There may \textit{exist} a small number of rapidly spinning black holes; as illustrated in Fig.~\ref{fig:spin_cdfs}, we infer the 95th percentile of the spin magnitude distribution to occur at $\chi_{95\%} = \SpinMagnitudeNinetyFive$.
We emphasize, however, that there is no observational evidence that these systems comprise a physically distinct subpopulation and not simply an extended tail of a single, predominantly low-spin population.
Consistent results have also been found when alternatively using splines to flexibly model the black hole spin distribution~\cite{Edelman2022b,Golomb2022}.

Although an excess of zero-spin systems is not \textit{ruled out}, the current lack of discernible features at $\chi=0$ or $\chi=1$ is in possible tension with common assumptions in the population synthesis of compact binaries~\cite{Qin2018,Bavera2020,Belczynski2020}: that efficient angular momentum transport yields isolated black holes born with very small (e.g. $\chi\lesssim0.1$)~\cite{Spruit1999,Spruit2002} or vanishing ($\chi\lesssim0.01$) natal spin magnitudes~\cite{Fuller2019}.
Very small natal spin magnitudes should yield a sharp excess of low- or nonspinning systems in the binary black hole spin distribution.
Meanwhile, if some fraction of mergers arise from isolated stellar binaries, then late time tidal spin-up of the second-born black hole's progenitor can override otherwise efficient angular momentum loss, yielding a secondary sub-population of black holes with spins up to $\chi\approx 1$~\cite{Qin2018,Zaldarriaga2018,Bavera2020,Bavera2022}.
The absence of such features in current data may suggest that angular momentum transport is less efficient than usually expected.

\textit{3. The merger rate is non-zero at $\chi=0$.}
Despite the fact that there is no \textit{excess} of systems with vanishing spin, the the binary black hole merger rate is confidently non-zero at $\chi=0$.
This finding is in conflict with commonly-used parametric models that assume component spins follow non-singular Beta distributions~\cite{Talbot2017,O3a-pop,Galaudage2021,O3b-pop}, which, by definition, require that $p(\chi) = 0$ at $\chi=0$; see Fig.~\ref{fig:chi-cost-comparison} and further discussion below.\footnote{Sometimes \textit{singular} Beta distributions are also allowed. Singular Beta distributions give $p(\chi)\to\infty$ as $\chi\to 0$, which is also precluded in Fig.~\ref{fig:chi-cost}.}
The fact that the spin magnitude is non-zero at $\chi=0$ may have implications for the processes by which black holes acquire their spins.
If black holes acquire their spins via stochastic or incoherent isotropic processes (e.g., random bombardment by gravity waves soon before core collapse~\cite{Fuller2014,Fuller2015} or statistically isotropic fallback accretion), then the spin magnitude distribution should have a Maxwellian-like form $p(\chi) \propto \chi^2$ near $\chi=0$.
The fact that this is not seen suggests, instead, that black hole spins originate from longer-lived or directionally-coherent processes~\cite{Chan2020,Janka2022}.

\textit{4. Black holes exhibit a broad range of spin-orbit misalignment angles.}
As illustrated in the upper- and lower-right panels of Fig.~\ref{fig:chi-cost}, we infer a non-zero merger rate across the full range of $\cos\theta$.
Using our autoregressive constraints on $\mathcal{R}(\cos\theta)$, we estimate that $\SpinTiltPercentNegative$ of black hole spins are misaligned by more than $90^\circ$ with respect to binaries' orbital angular momenta and that the rate of mergers with at least one component spin tilted by $\theta>90^\circ$ is $\SpinTiltRateOneOrBothNegative\,\mathrm{Gpc}^{-3}\mathrm{yr}^{-1}$.
Past studies using both strongly parametrized models~\cite{O3a-pop,O3b-pop,Callister2022,Vitale2022} and flexible splines~\cite{Edelman2022b,Golomb2022} have also concluded that the binary black hole population exhibits significant spin-orbit misalignment.
The results presented here, obtained under our highly agnostic and parameter-free autoregressive model, corroborate these conclusions.

\textit{5. A perfectly isotropic distribution is moderately disfavored.}
As seen in Fig.~\ref{fig:chi-cost}, both the merger rate $\mathcal{R}(\cos\theta)$ and the probability distribution $p(\cos\theta)$ have a tendency to increase towards positive $\cos\theta$.
In Fig.~\ref{fig:spin_cdfs} we show the corresponding cumulative distribution of $\cos\theta$ and the inferred median $\cos\theta$ among the black hole population.
We find this median to be $\cos\theta_{\,50\%} = \SpinTiltFifty$, with $\cos\theta_{\,50\%}>0$ for $\SpinTiltMedianPercentPositive$ of our posterior samples (the \textit{mean} value of $\cos\theta$ is also positive at comparable credibility).
These results somewhat disfavor a purely isotropic component spin distribution, although isotropy cannot yet be ruled out.

\begin{figure}
    \centering
    \includegraphics[width=0.46\textwidth]{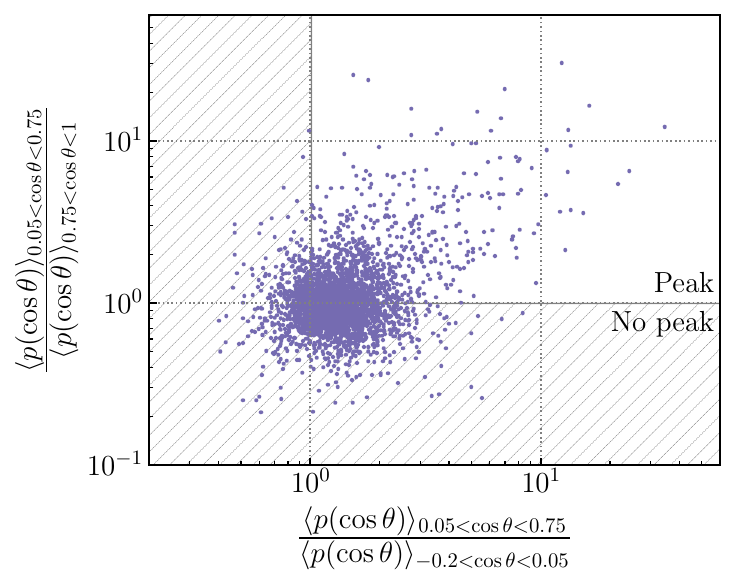}
    \caption{
    Evaluation of the significance of the $\cos\theta\approx0.4$ peak in Fig.~\ref{fig:chi-cost}.
    For each probability distribution $p(\cos\theta)$ in Fig.~\ref{fig:chi-cost}, we show the ratios between the mean probability in the window $0.05<\cos\theta < 0.75$ (centered on the possible peak) and the mean probabilities across adjacent windows at smaller and larger $\cos\theta$.
    When a peak is present, both ratios should be greater than one, corresponding to the upper right quadrant.
    We find that only $\TiltSignificancePercentagePeak$ of posterior samples fall in this quadrant, indicating that the $\cos\theta\approx0.4$ peak is not statistically significant.
    }
    \label{fig:theta_features}
\end{figure}

\begin{figure*}
    \centering
    \includegraphics[width=0.93\textwidth]{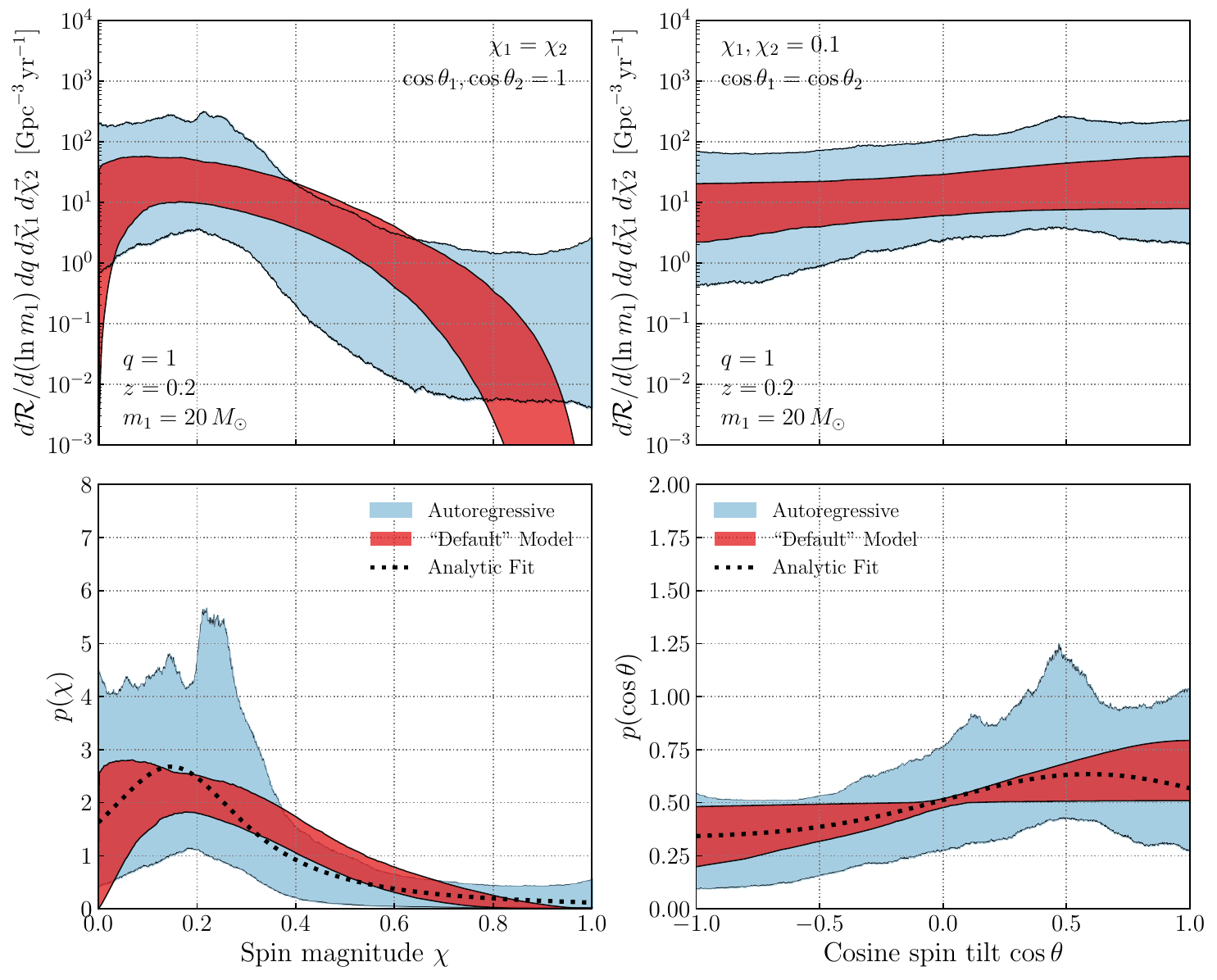}
    \caption{
    A comparison of the binary black hole spin distributions inferred using our autoregressive model (blue) and that recovered by a strongly parametrized approach (red, the \textsc{Default} model of Ref.~\cite{O3b-pop}).
    As in Fig.~\ref{fig:chi-cost}, the top row shows the binary merger rate as a function of component spin magnitude and spin-orbit tilt angle, at fixed $m_1$, $q$, and $z$, while the lower row shows the corresponding probability distributions.
    Component spins are assumed to be independently and identically distributed.
    Overall, there is good reasonable qualitative agreement between both sets of results; each recovers similar merger rates across the range of $\cos\theta$ values and for $0.1\lesssim\chi\lesssim0.3$.
    At the same time, the autoregressive results indicate that the merger rate remains finite for both smaller and larger spin magnitudes, whereas the parametric model requires \textit{a priori} that it vanishes as $\chi\to0$ and $\chi\to 1$.
    }
    \label{fig:chi-cost-comparison}
\end{figure*}

\textit{6. A possible excess of systems with $\cos\theta \approx 0.4$?}
As identified in Ref.~\cite{Vitale2022}, we also see a possible excess of black holes with $\cos\theta\approx0.4$.
We find that although this feature is \textit{possible}, it is not required by the data.
Following our procedure from Sect.~\ref{sec:masses} above, we can evaluate the significance of the $\cos\theta\approx0.4$ peak by asking what fraction of posterior samples give a higher mean probability in a window centered on the peak than in windows at both higher and lower $\cos\theta$ values.
Shown in Fig.~\ref{fig:theta_features}, only $\TiltSignificancePercentagePeak$ of samples are consistent with a peak at $\cos\theta\approx0.4$.
While the probability distribution of spin tilts is very likely to \textit{increase} between $\cos\theta\approx-0.1$ and $\cos\theta\approx0.4$, few samples exhibit the subsequent drop necessary for a peak.

\subsection{Discussion}

Figure~\ref{fig:chi-cost-comparison} compares our flexible autoregressive inference with results from the strongly parameterized \textsc{Default} model~\cite{Talbot2017} presented in Ref.~\cite{O3b-pop}.
In this model, component spin magnitudes are independently and identically drawn from a Beta distribution, while spin tilts are drawn from a mixture between isotropic and preferentially-aligned sub-populations. 
The two approaches generally yield similar conclusions, with some notable exceptions.
First, as noted above, the \textsc{Default} model is defined such that $p(\chi)$ is necessarily zero at $\chi=0$.
Our autoregressive results indicate that this is likely not the case; we infer a non-zero rate of mergers with $\chi=0$ (although no \textit{excess} of such mergers as one might expect if isolated black holes have vanishing natal spins).
Second, strongly parametrized approaches typically require the $\cos\theta$ distribution to be either isotropic or peaked at $\cos\theta=1$.
As illustrated in, e.g., the bottom right-hand panel of Fig.~\ref{fig:chi-cost-comparison}, the data tell a more complicated story, with a possible (albeit statistically insignificant) feature at intermediate $\cos\theta$ values.
See Ref.~\cite{Vitale2022} for further investigations of this feature.

Finally, it is instructive to compare the behavior of $\mathcal{R}(\cos\theta)$, in top right panel of Fig.~\ref{fig:chi-cost-comparison}, with that of $p(\cos\theta)$, in the bottom right.
Studies of the black hole spin distribution sometimes include the following seemingly inconsistent statements: (\textit{i}) that an isotropic $\cos\theta$ distribution is disfavored but cannot be ruled out, and (\textit{ii}) that our knowledge of $p(\cos\theta)$ is accurately reflected in the red band in the lower-right panel of Fig.~\ref{fig:chi-cost-comparison}.
Figure~\ref{fig:chi-cost-comparison}, though, seems to show unambiguously that $p(\cos\theta)$ is an increasing function of $\cos\theta$, in conflict with the first of the two statements above!

The resolution to this apparent paradox involves the fact that we directly measure $\mathcal{R}(\cos\theta)$, not the normalized probability distribution $p(\cos\theta)$.
Although spin isotropy is disfavored, it is evident in Fig.~\ref{fig:chi-cost-comparison} that a flat $\mathcal{R}(\cos\theta)$ cannot yet be fully ruled out.
The renormalized probability density $p(\cos\theta)$ can inadvertently obscure this fact:
Although there may exist many distinct posterior samples that yield isotropic $\mathcal{R}(\cos\theta)$ (e.g. flat traces at different vertical positions within the red or blue bands), each of these possibilities is mapped to the \textit{same} function, $p(\cos\theta) = 1/2$, upon normalization.
Thus, hidden behind the ``uncertainty bands'' in the lower-right panel of Fig.~\ref{fig:chi-cost-comparison} is a very uneven \textit{density} of possibilities, with a high number of individual draws stacked directly on $p(\cos\theta) = 1/2$.
Because the uncertainty bounds do not communicate this density, the result is a figure that appears to indicate an unambiguous measurement of anisotropy.
In order to avoid this counterintuitive behavior, we recommend that measurements of the $\cos\theta$ distribution be shown as constraints on both the probability density $p(\cos\theta)$ \textit{and} the merger rate $\mathcal{R}(\cos\theta)$.

While our autoregressive model makes minimal physical assumptions, there remain two caveats to consider when interpreting the above results.
First, we have chosen to model component spins as independently and identically distributed.
This assumption is broken in situations like tidal spin-up of field binaries.
We note that our assumption of independence and identicality is purely a choice of model, and not a limitation of the method; one could consider, instead, adopting \textit{separate} autoregressive processes for the distributions of primary and secondary spin magnitudes and tilts.

Second, our autoregressive model necessarily imposes a degree of continuity in the merger rate as a function of $\chi$ and $\cos\theta$.
This continuity could, in principle, obscure very sharp or discontinuous features in the black hole spin distributions.
It is therefore reasonable to ask if our conclusions above are being driven by continuity conditions rather than informative data.
This question is particularly critical when interpreting our conclusions regarding the lack of sharp features near $\chi\approx 0$; is our non-detection of such features significant, or do they fall outside the coverage of our model?
In Appendix~\ref{app:demo}, we conduct a mock data challenge to test the ability of our autoregressive model to recover a sharp excess of non-spinning black holes.
Although the resolution of our results is, at times, limited by the processes' finite scale length $\tau$, we find that we can successfully identify narrow excesses or bimodalities in the merger rate arising from a population of non-spinning systems, should it exist.

A related question is the degree to which we can trust extended tails appearing in our autoregressive measurements of the spin-dependent merger rate.
As also demonstrated in Appendix~\ref{app:demo}, our autoregressive process never goes completely to zero, as this would correspond to $\ln\mathcal{R}\to-\infty$.
Consequently, are the tails in Fig.~\ref{fig:chi-cost} towards large $\chi$ and negative $\cos\theta$ physically meaningful, or do they arise from our prior modeling assumptions?
Within Appendix~\ref{app:demo}, we find that, in the absence of observations, the recovered merger rates asymptotically approach a value corresponding to $N_\mathrm{exp}\lesssim 1$ total expected detections (integrated across the region of interest).
We can leverage this behavior to gauge the extent to which tails in our $\chi$ and $\cos\theta$ distributions are prior- or likelihood-dominated.
Specifically, we use our posteriors on $\mathcal{R}(\chi_i)$ and $\mathcal{R}(\cos\theta_i)$ to compute expected detection rates at large $\chi$ and small $\cos\theta$ and identify the threshold spin magnitude and tilts beyond which we expect fewer than $N=2$ component spins to arise in our sample; these values mark the boundaries beyond which our results are likely prior dominated.
This calculation is described in more detail in Appendix~\ref{app:hyperparams}, and it accounts also for the influence of selection effects on the observed distribution of binary parameters.

Our measurements of the spin magnitude distribution imply that we expect $N\leq 2$ detections with at least one component spin magnitude falling above $\chi\geq\ComponentMagTailHighTwo$, where the uncertainties reflect our uncertain recovery of the spin magnitude distribution.
Thus, our recovered spin magnitude distribution is likely prior dominated at $\chi\gtrsim 0.9$, although prior effects may also become important above $\chi\gtrsim 0.65$ under a more conservative interpretation.
Meanwhile, we find that our results, on average, predict fewer than two detections with $\cos\theta\leq\ComponentTiltTailLowTwo$, implying that our posterior on the spin tilt distribution is likelihood dominated across nearly the full range of $\cos\theta$ values.
One might choose more conservative thresholds by instead identifying values beyond which fewer than $N=4$ detections are predicted; these occur at $\chi\geq\ComponentMagTailHighFour$ and $\cos\theta\leq\ComponentTiltTailLowFour$.

\subsection{Suitable parametric models}

When a standard strongly parametrized model is required, we find that our autoregressive measurement of $p(\chi)$ is well fit by a truncated Gaussian or a truncated Lorentzian,
    \begin{equation}
    p(\chi) = \frac{\mathcal{C}}{\gamma} \left[1+\left(\frac{\chi-\chi_0}{\gamma}\right)^2\right]^{-1},
    \label{eq:lorentzian}
    \end{equation}
with normalization
    \begin{equation}
    \mathcal{C} = \left[\tan^{-1}\left(\frac{1-\chi_0}{\gamma}\right) + \tan^{-1}\left(\frac{\chi_0}{\gamma}\right)\right]^{-1},
    \end{equation}
and $p(\cos \theta)$ by a mixture between isotropic and Gaussian components,
    \begin{equation}
    p(\cos\theta) = \frac{f_\mathrm{iso}}{2} + (1-f_\mathrm{iso}) N_{[-1,1]}(\cos\theta| \mu,\sigma),
    \label{eq:variable-default}
    \end{equation}
where $N_{[-1,1]}(\cos\theta| \mu,\sigma)$ indicates a truncated Gaussian normalized on the interval $-1\leq\cos\theta\leq 1$.
Equation~\eqref{eq:variable-default} is the same as the \textsc{Default} spin-tilt distribution~\cite{Talbot2017} but with a freely varying mean as advocated in Ref.~\cite{Vitale2022}.
A least-squares fit of our results to Eqs.~\eqref{eq:lorentzian} and \eqref{eq:variable-default} yields best-fit parameters
    \begin{equation}
    \begin{aligned}
    \chi_0 &= \MagFitLorentzianMean \\
    \gamma &= \MagFitLorentzianGamma \\
    f_\mathrm{iso} &= \CosTiltFitFractionIso \\
    \mu &= \CosTiltFitMean \\
    \sigma &= \CosTiltFitStd\,.
    \end{aligned}
    \end{equation}
These fits describe marginal probability distributions and are therefore valid at any choice of $m_1$, $q$, and $z$.

\section{Stop Four: Effective Spins}
\label{sec:effective-spins}

\begin{figure*}
    \centering
    \includegraphics[width=0.92\textwidth]{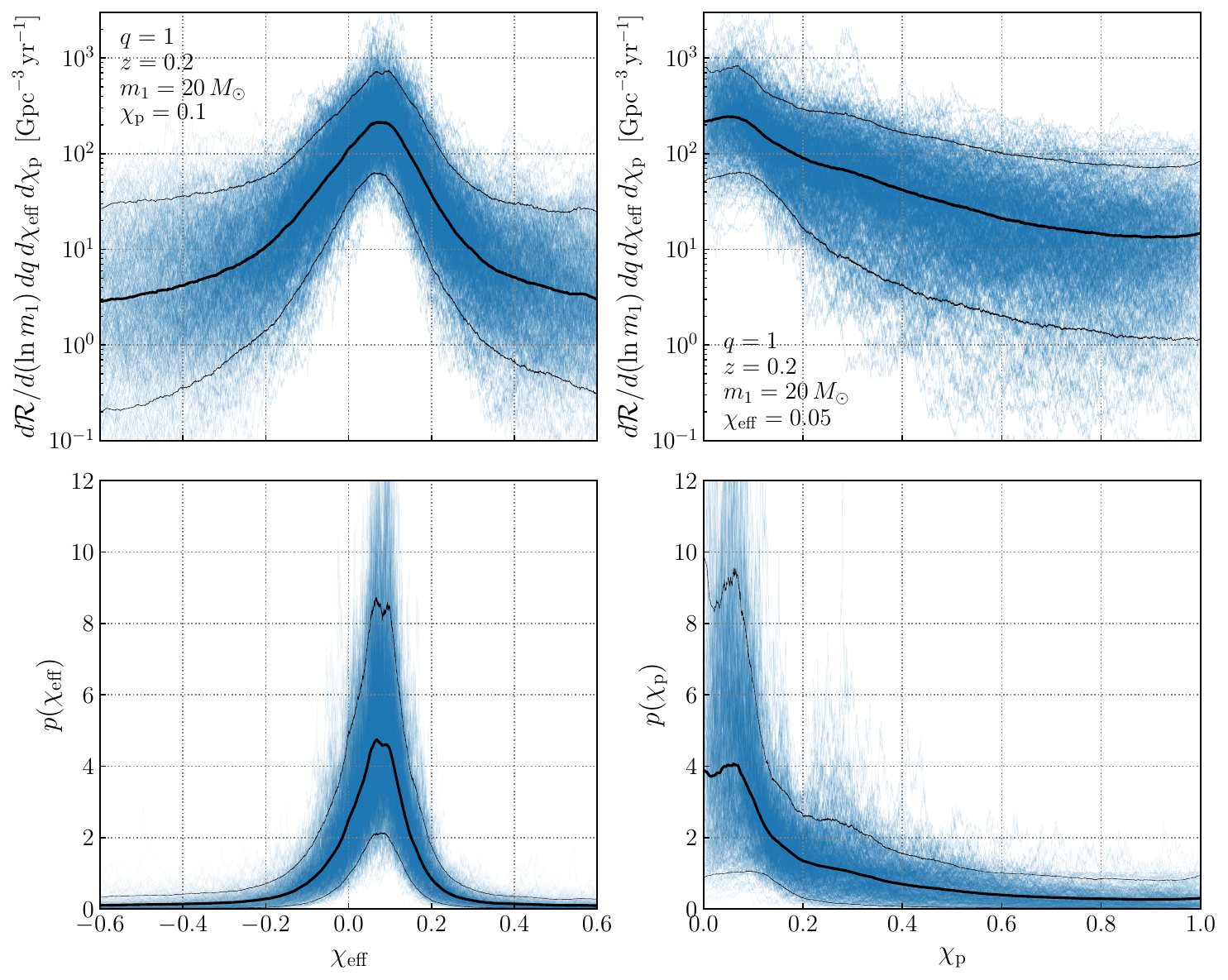}
    \caption{
    \textit{Top}: The merger rate of binary black holes as a function of effective inspiral spin ($\chi_\mathrm{eff}$, left) and effective precessing spin ($\chi_p$, right), as inferred using our autoregressive model.
    The rates shown are each evaluated at fixed reference masses and redshift ($m_1 = 20\,M_\odot$, $q=1$, and $z=0.2$).
    The bottom panels show the corresponding probability distributions on each effective spin parameter.
    Within each panel, the central black curve marks the mean inferred rate or probability densities, while outer black curves bound 90\% credible intervals.
    Effective inspiral spins exhibit a unimodal distribution.
    The center of this distribution prefers to be at positive $\chi_\mathrm{eff}$ but with a non-zero merger rate at $\chi_\mathrm{eff}<0$.
    The $\chi_\mathrm{p}$ distribution, meanwhile, preferentially peaks toward $\chi_\mathrm{p} = 0$ but with a shoulder that extends to moderate to large precessing spins.
    }
    \label{fig:xeff-xp}
\end{figure*}

\begin{figure*}
    \centering
    \includegraphics[width=0.92\textwidth]{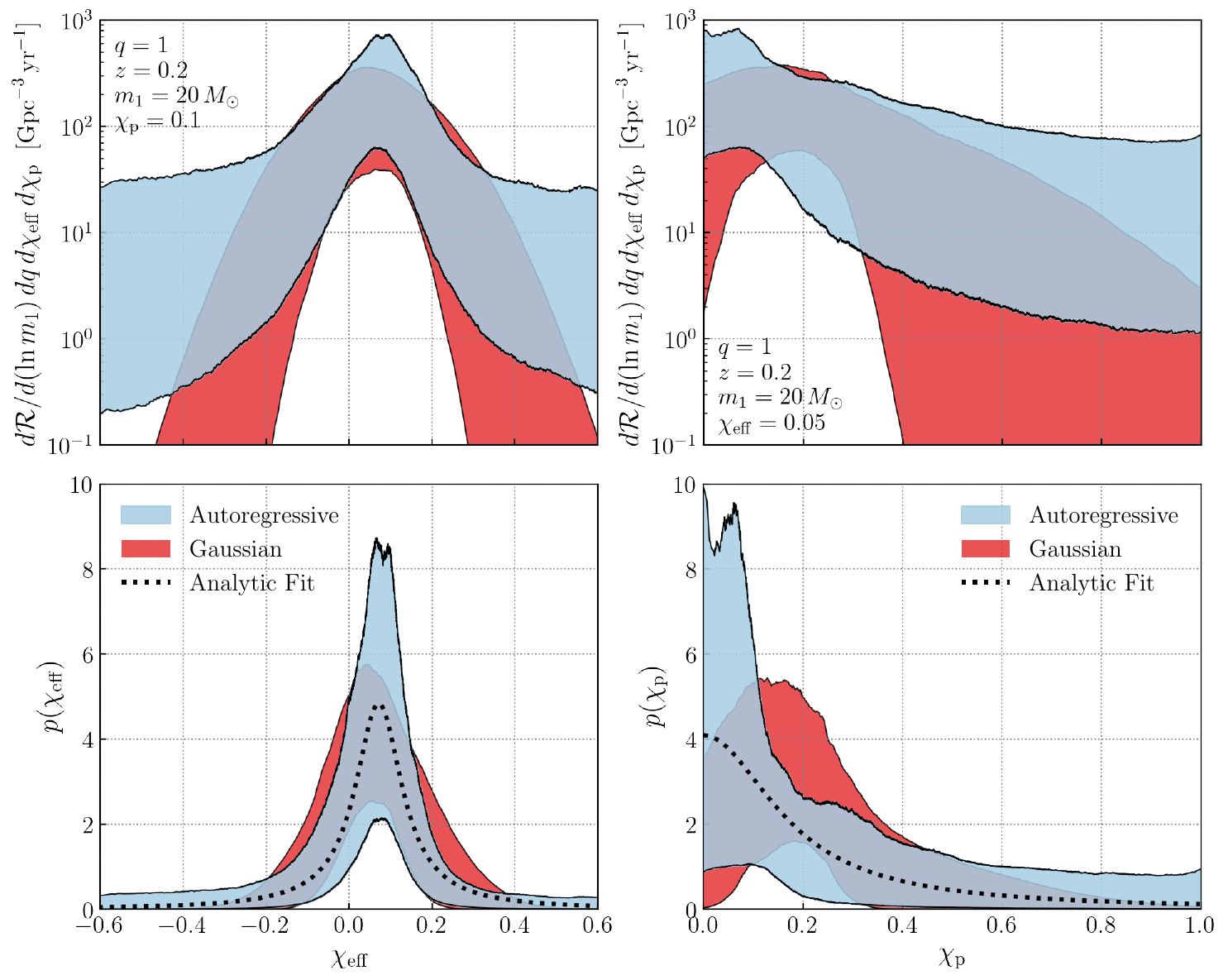}
    \caption{
    A comparison of the binary black hole effective spin distributions inferred using our autoregressive model (blue) and that recovered by a strongly parametrized approach (red, the \textsc{Gaussian} spin model of Ref.~\cite{O3b-pop}).
    As in Fig.~\ref{fig:xeff-xp}, the top row shows the binary merger rate as a function of effective spin parameters, $\chi_\mathrm{eff}$ and $\chi_\mathrm{p}$ (at fixed $m_1$, $q$, and $z$), while the lower row shows the corresponding probability distributions.
    The merger rates recovered by each approach agree well in the $-0.1 \lesssim \chi_\mathrm{eff} \lesssim 0.2$ and $0.1 \lesssim \chi_\mathrm{p} \lesssim 0.3$ ranges, beyond which the \textsc{Gaussian} rates fall to zero much more quickly than our autoregressive inference.
    }
    \label{fig:xeff-xp-comparison}
\end{figure*}

Although component spin magnitudes and spin-orbit misalignment angles have a clear physical interpretation, they are particularly difficult to measure using gravitational waves.
Easier to directly measure are various \textit{effective spins}: derived parameters that, while less physically interpretable, more directly govern a gravitational wave's morphology.
These effective parameters include the effective inspiral spin~\cite{Damour2001,Racine2008},
    \begin{equation}
    \chi_\mathrm{eff} = \frac{\chi_1\cos\theta_1 + q \chi_2\cos\theta_2}{1+q},
    \end{equation}\\
and the effective precessing spin~\cite{Schmidt2015},
    \begin{equation}
    \chi_\mathrm{p} = \mathrm{Max}\left( \chi_1\sin\theta_1, \frac{3+4q}{4+3q}q \chi_2\sin\theta_2\right).
    \end{equation}
Here, $\chi_\mathrm{eff}$ quantifies the degree of spin projected parallel to a binary's orbital angular momentum, while $\chi_\mathrm{p}$ approximately quantifies the degree of in-plane spin (and hence more directly controls the degree of spin-orbit precession).
Although $\chi_\mathrm{eff}$ and $\chi_\mathrm{p}$ are less manifestly physical than the component spin magnitudes and tilts (much like the relationship between a binary's chirp mass and component masses), they act as signposts by which to identify categorical features of the compact binary spin distribution.
Negative $\chi_\mathrm{eff}$, for example, can arise only if one or both component spins are inclined by more than $90^\circ$ with respect to their orbit.
Non-zero $\chi_\mathrm{p}$, meanwhile, can manifest only if a system has at least some in-plane spin, such that $\sin\theta>0$.

Just as we have applied our autoregressive model to non-parametrically infer the component spin magnitude and tilt distributions, we can use our autoregressive model to measure the distribution of these spin parameters.
If $\Psi(\chi_\mathrm{eff})$ and $\Phi(\chi_\mathrm{p})$ are autoregressive functions of $\chi_\mathrm{eff}$ and $\chi_\mathrm{p}$, respectively, then our merger rate model will be of the form
    \begin{equation}
    \begin{aligned}
    &\mathcal{R}(\ln m_1,q,\chi_\mathrm{eff},\chi_p;z) \\
        &\quad = r \frac{f(m_1) p(q)}{f(20\,M_\odot)}  \left(\frac{1+z}{1+0.2}\right)^\kappa \Big[e^{\Psi(\chi_\mathrm{eff})}  e^{\Phi(\chi_\mathrm{p})}\Big],
    \end{aligned}
    \label{eq:ar-effective-spins}
    \end{equation}
where we again fall back on parametric models for the dependence of the merger rate on binary masses and redshift.
Note that, while we are describing a binary's spin configuration in terms of $\chi_\mathrm{eff}$ and $\chi_\mathrm{p}$, binary spin is fundamentally six-dimensional.
Our choice to work in a reduced two-dimensional space requires that we assume some distribution for the remaining four degrees of freedom, even if that assumption is implicit.
In defining Eq.~\eqref{eq:ar-effective-spins}, we indirectly assume that the remaining spin degrees of freedom follow their default parameter estimation priors (uniform spin magnitudes and isotropic directions), conditioned on $\chi_\mathrm{eff}$ and $\chi_\mathrm{p}$.

\subsection{Features in the black hole effective spin distribution}

Figure~\ref{fig:xeff-xp} shows our inference of the $\chi_\mathrm{eff}$ and $\chi_\mathrm{p}$ distributions of binary black holes.
As in Fig.~\ref{fig:chi-cost} above, the upper row shows the inferred merger rate as a function of $\chi_\mathrm{eff}$ (with fixed $\chi_\mathrm{p}$; left) and $\chi_\mathrm{p}$ (with fixed $\chi_\mathrm{eff}$; right).
Both rates are evaluated at a fixed reference primary mass, mass ratio, and redshift.
The bottom row, meanwhile, shows the corresponding normalized probability distributions of each effective spin parameter.
From Fig.~\ref{fig:xeff-xp}, we draw the following conclusions:

\textit{1. The merger rate is non-zero for $\chi_\mathrm{eff}<0$}.
Consistent with the results of Sect.~\ref{sec:spins}, we find a non-zero merger rate for binaries with $\chi_\mathrm{eff}<0$, suggesting the presence of component spins misaligned by more than $90^\circ$ with respect to their orbital angular momenta.
We find that $\ChiEffPercentNegative$ of binary black holes have negative $\chi_\mathrm{eff}$ and that the integrated merger rate of binaries with negative effective spin is $\ChiEffRateNegative\,\mathrm{Gpc}^{-3}\mathrm{yr}^{-1}$.
These estimates are comparable to those presented in Ref.~\cite{O3b-pop}, which concluded, using a strongly-parameterized model, that $29^{+15}_{-13}\%$ of binaries exhibit negative $\chi_\mathrm{eff}$.

\textit{2. The $\chi_\mathrm{eff}$ distribution peaks at positive values.}
Despite the presence of binaries with negative effective spin, we find that the $\chi_\mathrm{eff}$ distribution is not symmetric about $\chi_\mathrm{eff}=0$ but instead prefers to peak at small but positive values.
This preference is significant;
among our posterior samples on $\mathcal{R}(\chi_\mathrm{eff})$, $\ChiEffAsymmetricPercentage$ have a larger integrated merger rate in the range $0 \leq \chi_\mathrm{eff} \leq 0.1$ than between $-0.1 \leq \chi_\mathrm{eff} \leq 0$.
Similarly, the median $\chi_\mathrm{eff}$ is inferred to be positive for $\ChiEffMedianPercentPositive$ of samples.~\footnote{
At the same time, the \textit{mean} $\chi_\mathrm{eff}$ among the binary population, found to be $\mu = \ChiEffMean$, remains consistent with zero.}

\textit{3. The binary black hole distribution exhibits non-zero $\chi_\mathrm{p}$}.
Consistent with the measurement of a range of $\cos\theta$ values, we find that the black hole merger rate extends across a wide range of $\chi_\mathrm{p}$.
The percentage of binaries with $\chi_\mathrm{p}>0.2$, for example, is $\ChiPPercentLarge$.

\begin{figure*}
    \centering
    \includegraphics[width=0.92\textwidth]{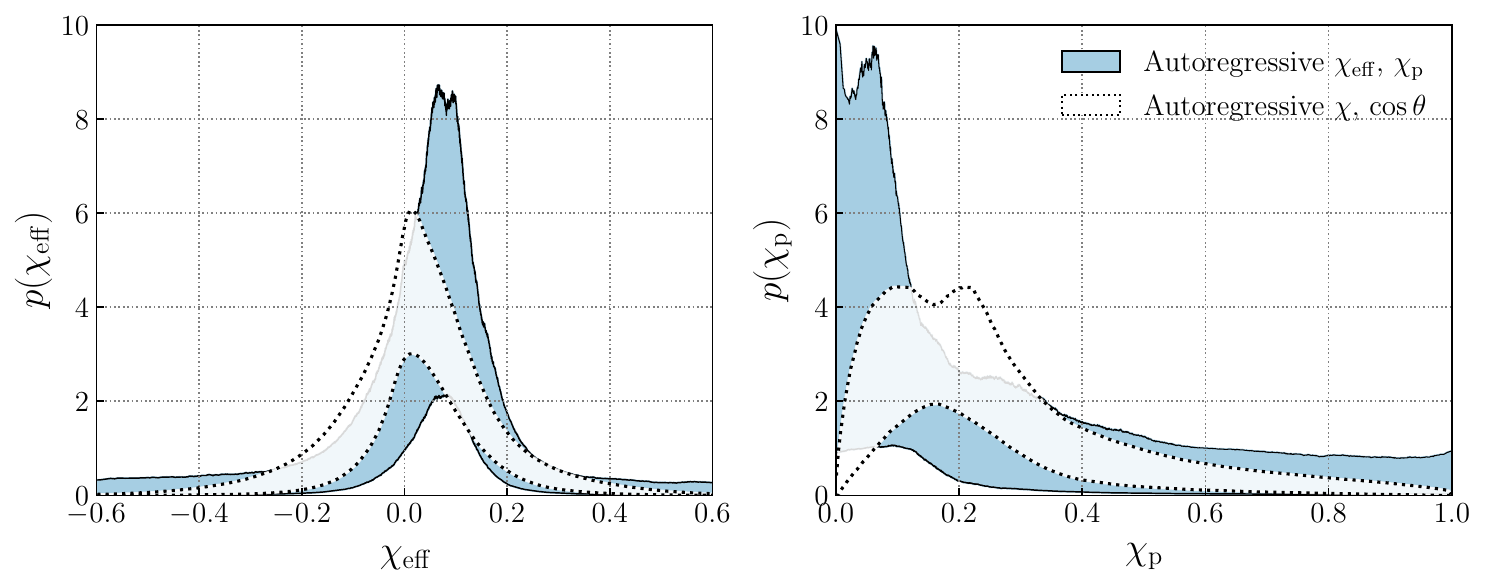}
    \caption{
    A comparison between the effective spin distributions obtained through direct autoregressive fits (blue; Eq.~\eqref{eq:ar-effective-spins} in Sect.~\ref{sec:effective-spins}) and the effective spin distributions obtained by instead fitting autoregressive models to the underlying \textit{component spin} magnitudes and tilts (white; Eq.~\eqref{eq:ar-component-spins} in Sect.~\ref{sec:spins}).
    There is not perfect agreement between results.
    Compared to the direct autoregressive measurement of $p(\chi_{\rm eff})$ and $p(\chi_{\rm p})$, fits to the underlying component spins give an effective inspiral spin distribution that is likely maximized closer to $\chi_{\rm eff} \approx 0$ with a broader tail to positive $\chi_{\rm eff}$, and $p(\chi_{\rm p})$ is meanwhile maximized at larger values of $\chi_{\rm p}$.
    These differences are expected;
    each model makes different physical assumptions, and the coordinate transformation from component to effective spins imprints further structure in the derived effective spin distributions (such as the requirement that $p(\chi_{\rm p}) = 0$ at $\chi_{\rm p} = 0$).
    At the same time, both sets of results share qualitative features in common, such as the presence of events with negative effective spins.
    That both results exhibit this feature suggests that it is robust, rather than a spurious result arising from continuity conditions imposed on our autoregressive measurement of $p(\chi_{\rm eff})$.
    }
    \label{fig:model-comparison}
\end{figure*}

\subsection{Discussion}

We compare our autoregressive measurements to previous strongly parametrized population measurements in Fig.~\ref{fig:xeff-xp-comparison}.
Blue bands show central 90\% credible intervals on the rates and probability distributions of $\chi_\mathrm{eff}$ and $\chi_\mathrm{p}$ under our autoregressive model, while red bands show results obtained when modeling the $\chi_\mathrm{eff}-\chi_\mathrm{p}$ as a bivariate Gaussian in order to measure the mean and standard deviation of each quantity.
Our autoregressive $\mathcal{R}(\chi_\mathrm{eff})$ measurement is, in fact, in reasonable agreement with a Gaussian model, although with extended tails to $\chi_\mathrm{eff}\gtrsim 0.4$ and $\chi_\mathrm{eff}\lesssim-0.4$.
Both the autoregressive and Gaussian $\chi_\mathrm{p}$ models, in turn, yield similar merger rates at $\chi_\mathrm{p}\approx 0.1$, although the Gaussian model appears to vanish too quickly as $\chi_\mathrm{p}\to 0$ or $1$.

As in the component spin case above, it is valuable to explore whether these extended tails in our $\chi_\mathrm{eff}$ and $\chi_\mathrm{p}$ distribution are due to informative data or to the continuity imposed by our autoregressive model.
We will once again estimate the regions in which our results are prior dominated by identifying the threshold $\chi_\mathrm{eff}$ and $\chi_\mathrm{p}$ values beyond which our posteriors predict fewer than $N=2$ detections.
We find that fewer than two detections are expected at $\chi_\mathrm{eff}\leq \EffectiveSpinTailLowTwo$, at $\chi_\mathrm{eff}\geq\EffectiveSpinTailHighTwo$, and at $\chi_\mathrm{p}\geq\PrecessingSpinTailHighTwo$.
This suggests that our results are prior-dominated beyond these regions, such that the apparent tension between the \textsc{Gaussian} and autoregressive models at very positive and negative $\chi_\mathrm{eff}$ and very large $\chi_p$ is consistent with differing prior assumptions.
Furthermore, the fact that we expect $N=2$ detections with effective spins $\chi_{\rm eff} \lesssim -0.3$ suggests that the inferred presence of negative $\chi_{\rm eff}$ systems is due to informative data, rather than a consequence of prior continuity conditions imposed on our model.
However, based on the uncertainties quoted above, we cannot rule out an empirical distribution function that instead only reaches $N=2$ detections above $\chi_{\rm eff} = -0.09$, such that prior effects set in at $\chi_{\rm eff} \approx -0.1$.
In this more conservative approach, is it possible that continuity conditions are ``fooling'' our model into inferring the presence of negative effective spins where none in fact exist?

If all binary black holes had spin-orbit misalignment angles below $\theta = 90^\circ$ degrees (and hence purely positive effective spins), this could manifest as a sharp truncation in the effective spin distribution at $\chi_{\rm eff} = 0$.
In this case, our autoregressive model would indeed struggle to fit such a discontinuity, possibly leading us to incorrectly conclude the existence of systems with negative effective spins.
There is, however, a simple remedy to this problem.
We should simply move to coordinates in which there are no such discontinuities: component spin magnitudes and tilts.
In Sec.~\ref{sec:spins} above, autoregressive modeling of the spin-tilt distribution implied that a significant fraction of black holes have misalignment angles greater than $90^\circ$, consistent with the need for negative $\chi_{\rm eff}$ identified in this section.

Figure~\ref{fig:model-comparison} compares our autoregressive measurements of the effective spin distributions in this section (blue) with the effective spin distributions implied by our \textit{component spin} measurements in Sect.~\ref{sec:spins} (white).
In order to compute these implied distributions, we again assume that component spins are independently and identically distributed.
The white and blue distributions are not identical.
This is expected; the two measurements each correspond to fundamentally distinct models, and further structure is necessarily imposed by the coordinate transformation from component to effective spins.~\footnote{For example, the coordinate transformation from component spins to $\chi_{\rm p}$ forces $p(\chi_{\rm p})$ to zero at $\chi_{\rm p} = 0$.}
At the same time, we see the same qualitative features in both sets of results: a $\chi_{\rm eff}$ distribution extending to negative values and a broad $\chi_{\rm p}$ distribution.
That these features emerge whether we choose to describe the black hole population via its component spins or effective spins suggests that these conclusions are robust, and not due to modeling systematics.

\subsection{Suitable parametric models}

When a strongly parametrized model is needed for $p(\chi_\mathrm{eff})$, we find our autoregressive result to be well approximated by a truncated Gaussian with mean and standard deviation
    \begin{equation}
    \begin{aligned}
    \mu &= \ChiEffFitGaussianMean \\
    \sigma &= \ChiEffFitGaussianStd
    \end{aligned}
    \end{equation}
or a truncated Lorentzian (see Eq.~\eqref{eq:lorentzian}) with
    \begin{equation}
    \begin{aligned}
    \chi_0 &= \ChiEffFitLorentzianMean \\
    \gamma &= \ChiEffFitLorentzianGamma;
    \end{aligned}
    \end{equation}
this latter fit is shown as a dotted line in the lower-right panel of Fig.~\ref{fig:xeff-xp-comparison}.
Similarly $p(\chi_\mathrm{p})$ can be approximated by either a truncated Gaussian or Lorentzian, with best-fit parameters 
    \begin{equation}
    \begin{aligned}
    \mu &= \ChiPFitGaussianMean \\
    \sigma &= \ChiPFitGaussianStd
    \end{aligned}
    \end{equation}
and
    \begin{equation}
    \begin{aligned}
    \chi_0 &= \ChiPFitLorentzianMean \\
    \gamma &= \ChiPFitLorentzianGamma,
    \end{aligned}
    \end{equation}
respectively, the latter of which is shown in Fig.~\ref{fig:xeff-xp-comparison}.
As in previous sections, these fits are valid at any choice of $m_1$, $q$, or $z$.

\section{Conclusion}
\label{sec:conclusion}

In this paper, we have developed and demonstrated a novel means of measuring the population properties of merging binary black holes.
By describing the black hole merger rate as a stochastic process, we hierarchically inferred the black hole mass, redshift, and spin distributions without resorting to strongly parametrized models that \textit{a priori} assume some particular structure.
The advantage of highly flexible models like autoregressive processes is two-fold.
They allow us to agnostically study the ``known unknowns,'' like theoretically-predicted features in the black hole population, but also reveal the ``unknown unknowns'' -- unexpected and impactful features that may otherwise be missed by standard strongly parametrized approaches.

We accordingly searched for expected and unexpected features alike in the distributions of binary black hole masses, redshifts, and spins.
Our results reiterated known features in the black hole mass spectrum (peaks at approximately $10$ and $35\,M_\odot$), but also revealed more nuanced structure like an additional steepening of $\mathcal{R}(m_1)$ towards high masses.
We found signs of unexpected structure in the redshift distribution of binary black holes, recovering a merger rate that prefers to remain flat at low redshifts followed by steeper growth at $z\gtrsim0.5$.
And our autoregressive results offered a direct and model agnostic look at the black hole spin distribution, revealing features like severe spin-orbit misalignment and a unimodal spin magnitude distribution that have previously been controversial.

A challenge that arises when using flexible models is how exactly to translate results (e.g. our posterior on $\mathcal{R}(m_1)$) into statements about physical features and their significances.
We find it useful to conceptually distinguish between two steps: (\textit{i}) data fitting and (\textit{ii}) feature extraction.
When performing hierarchical inference with strongly parametrized models, these two steps are accomplished simultaneously.
A clear example is the \textsc{Power Law+Peak} model for $\mathcal{R}(m_1)$, whose parameters directly encode the location, width, and height of a possible Gaussian peak.
Fitting the \textsc{Power Law+Peak} model to data, therefore, automatically extracts information about the feature of interest.
When using highly flexible models, on the other hand, data fitting and feature extraction are necessarily distinct.
Although hierarchically fitting our autoregressive model yields, for instance, the mass spectrum shown in Fig.~\ref{fig:lnm1}, this result offers no immediate information about the presence and/or significance of possible features.
Instead, we need to visually inspect our results and devise further tests or summary statistics to make any quantitative statements about the features we see.
A major focus of ours has accordingly been the use of parameter-free summary statistics, like the ratios of merger rates in adjacent bins, to identify and characterize the features summarized above.
These parameter-free techniques for feature extraction can be employed for any model, and additionally offer a means of directly comparing results obtained under two or more different models (strongly-parameterized or not).

While highly flexible models like ours enable a very agnostic exploration of the compact binary population, we do not necessarily advocate for replacing standard strongly-parameterized models.
Instead, we envision using both strongly-parameterized and flexible models in a cyclic development process:
flexible models enable the identification of possible new features, which are followed up and characterized using targeted strongly-parameterized models, whose validity is finally re-checked with flexible models as new data become available.
In the spirit of this cyclic development, in each section above we have offered refined strongly-parameterized models that capture the range of features identified in our autoregressive results.

One limitation of the autoregressive model employed here is the fact that it is fundamentally one dimensional.
Although we can simultaneously measure the dependence of the merger rate on different binary parameters, each with its own autoregressive process, this approach cannot capture any intrinsic \textit{correlations} among parameters.
As strongly-parameterized models begin to identify possible correlations between binary parameters~\cite{Callister2021,Biscoveanu2022}, flexible population models that can operate in higher dimensions will be critical in following up these results and agnostically identifying new correlations.
Some alternative approaches, like spline-based~\cite{Edelman2022a,Edelman2022b} or binned~\cite{Mandel2017,Farr2018} models, can be very easily extended to more than $n=1$ dimension but likely become computationally infeasible when $n$ becomes large.
Future work will involve the exploration of multi-dimensional stochastic processes as tools with which to measure the merger rate across the complete higher-dimensional space of binary black hole parameters. 

\begin{acknowledgements}

We thank Tom Dent, Bruce Edelman, Amanda Farah,  Salvatore Vitale, and others within the LIGO, Virgo, and KAGRA collaborations for their numerous conversations and valuable comments.
We also thank our anonymous referees, whose feedback has significantly improved the quality of this work.
This project is supported by the Eric and Wendy Schmidt AI in Science Postdoctoral Fellowship, a Schmidt Futures program.
This research has made use of data, software and/or web tools obtained from the Gravitational Wave Open Science Center (https://www.gw-openscience.org), a service of LIGO Laboratory, the LIGO Scientific Collaboration and the Virgo Collaboration.
Virgo is funded by the French Centre National de Recherche Scientifique (CNRS), the Italian Istituto Nazionale della Fisica Nucleare (INFN) and the Dutch Nikhef, with contributions by Polish and Hungarian institutes.
This material is based upon work supported by NSF's LIGO Laboratory which is a major facility fully funded by the National Science Foundation.\\

\noindent \textit{Data and code availability}:
The code used for this study is hosted on GitHub at \url{https://github.com/tcallister/autoregressive-bbh-inference/}, and data produced by our analyses can be download from Zenodo at
 \url{https://doi.org/10.5281/zenodo.7616096}.
\\

\noindent \textit{Software}: {\tt astropy}~\cite{astropy1,astropy2}, {\tt h5py}~\cite{h5py}, {\tt jax}~\cite{jax}, {\tt matplotlib}~\cite{Hunter:2007}, {\tt numpy}~\cite{numpy}, {\tt numpyro}~\cite{numpyro1,numpyro2}, {\tt scipy}~\cite{scipy}.

\end{acknowledgements}

\appendix
\section{More on Autoregressive Models}
\label{app:ar-appendix}

As discussed in the main body, in this work we agnostically model the rate density of compact binaries as an autoregressive process.
Given a merger rate $\ln\mathcal{R}_i$ at some mass $\ln m_i$, Eq.~\eqref{eq:mass-ar1} offers a prescription with which to randomly propose a merger rate $\ln\mathcal{R}_{i+1}$ at the next mass $\ln m_{i+1}$ of interest.
We still need to \textit{initialize} this process, though, picking some initial rate $\ln\mathcal{R}_1$ at the smallest mass $\ln m_1$ considered in our sample.
This initial value is randomly drawn via
    \begin{equation}
    \label{eq:initial-point}
    \ln\mathcal{R}_1 \sim N(\ln r,\sigma).
    \end{equation}

The exact form of Eq.~\eqref{eq:mass-ar1} and the definitions of $c_i$ and $w_i$ (Eqs.~\eqref{eq:c} and \eqref{eq:w}, respectively) are chosen to guarantee that all subsequent rates $\ln \mathcal{R}_i$ have the same marginal prior as $\ln \mathcal{R}_1$, such that the autoregressive process is stationary.
For example, from Eq.~\eqref{eq:mass-ar1} the prior expectation value of $\ln \mathcal{R}_i$ is given by
    \begin{equation}
    \langle \ln \mathcal{R}_i \rangle
        = \ln r + c_i \left[ \langle \ln\mathcal{R}_{i-1}\rangle - \ln r \right],
    \end{equation}
where we have used the fact that $\langle w_i \rangle = 0$.
From above, though, we know that our initial point satisfies $\langle \ln \mathcal{R}_1\rangle = \ln r$, implying
    \begin{equation}
    \langle \ln \mathcal{R}_i \rangle = \ln r
    \end{equation}
for all $i$.
Similarly, the variance of $\ln\mathcal{R}_i$ is
    \begin{equation}
    \label{eq:variance-ar1}
    \begin{aligned}
    \mathrm{Var}&\left(\ln\mathcal{R}_i\right) \\
        &= c_i^2\,\mathrm{Var}\left(\ln\mathcal{R}_{i-1}\right)
            + \mathrm{Var}(w_i) \\
        &= e^{-2\Delta_i/\tau} \mathrm{Var}\left(\ln\mathcal{R}_{i-1}\right) + \sigma^2 \left(1-e^{-2\Delta_i/\tau}\right).
    \end{aligned}
    \end{equation}
Consider the $i=2$ case.
From Eq.~\eqref{eq:initial-point} we know that the variance of $\ln\mathcal{R}_1$ is $\sigma^2$, giving
    \begin{equation}
    \begin{aligned}
    \mathrm{Var}\left(\ln\mathcal{R}_2\right)
        &= e^{-2\Delta_i/\tau} \sigma^2 + \sigma^2 \left(1-e^{-2\Delta_i/\tau}\right) \\
        &= \sigma^2.
    \end{aligned}
    \end{equation}
By induction,
    \begin{equation}
    \mathrm{Var}\left(\ln\mathcal{R}_i\right) = \sigma^2
    \label{eq:ar-var}
    \end{equation}
for all subsequent $i$.
We can finally consider the covariance between the rates $\ln\mathcal{R}_{i+n}$ and $\ln\mathcal{R}_i$ at two different locations:
    \begin{equation}
    \begin{aligned}
    &\mathrm{Cov}\left(\ln\mathcal{R}_{i+n},\ln\mathcal{R}_i\right) \\
        &= \left\langle \ln\mathcal{R}_{i+n} \ln\mathcal{R}_i \right\rangle - \langle \ln\mathcal{R}_{i+n}\rangle \langle \ln\mathcal{R}_i \rangle  \\
        &= \left\langle
            \left[ \ln r + c_{i+n}(\ln \mathcal{R}_{i+n-1} - \ln r) + w_{i+n}\right] \ln \mathcal{R}_i
            \right\rangle - \left(\ln r\right)^2 \\
        &= c_{i+n} \langle \ln \mathcal{R}_{i+n-1} \ln \mathcal{R}_i \rangle
            - c_{i+n} \left(\ln r\right)^2 \\
        &= c_{i+n} \mathrm{Cov} \left(\ln\mathcal{R}_{i+n-1},\ln\mathcal{R}_i\right)
    \end{aligned}
    \label{eq:ar-cov}
    \end{equation}
To obtain the third line, we used the definition of our autoregressive process to write $\ln\mathcal{R}_{i+n}$ in terms of $\ln\mathcal{R}_{i+n-1}$.
To move to the fourth line, we then used the facts that $w_{i+n}$ and $\ln \mathcal{R}_i$ are uncorrelated, that $\langle w_{i+n} \rangle = 0$, and that $\langle \ln\mathcal{R}_i \rangle = \ln r$.
Continuing to iterate in this fashion gives
    \begin{equation}
    \begin{aligned}
    \mathrm{Cov}&\left(\ln\mathcal{R}_{i+n},\ln\mathcal{R}_i\right) \\
        &= \left(\prod_{j=i+1}^{i+n} c_j\right) \mathrm{Cov}(\ln \mathcal{R}_i,\ln \mathcal{R}_i) \\
        &= e^{-\left(\Delta_{i+n} + \Delta_{i+n-1} + ... + \Delta_{i+1}\right)/\tau} \,\mathrm{Var}(\ln \mathcal{R}_i) \\
        &= \sigma^2 e^{-\left(\ln m_{i+n} - \ln m_i\right)/\tau}
    \end{aligned}
    \end{equation}
So $\tau$ is indeed the scale over which the autoregressive process retains significant autocorrelation.

Implementing our hierarchical likelihood model in \texttt{numpyro}~\cite{numpyro1,numpyro2} and \texttt{jax}~\cite{jax} necessitates efficient proposals of new autoregressive processes drawn from our prior.
To make our discussion more concrete, consider the proposal of an autoregressive process $\Psi$ over (log) black hole masses $\ln m$.
As a preprocessing step, define $\llbracket \ln m\rrbracket$ to be the \textit{sorted union} of all of our posterior samples and injections,
     \begin{equation}
     \llbracket \ln m\rrbracket = \mathrm{Sort}\Big(\{\ln m_{1}\} \cup ... \cup\{\ln m_{N_\mathrm{obs}}\} \cup \{\ln m_{\mathrm{inj}}\}\Big),
     \label{eq:sort}
     \end{equation}
where $\{\ln m_i\}$ is the set of posterior samples associated with event $i$ and $\{\ln m_\mathrm{inj}\}$ is the set of masses corresponding to successfully recovered injections.
Define $n_\mathrm{pts}$ to be the length of $\llbracket \ln m \rrbracket$.
Also precompute the set of differences $\llbracket \Delta \rrbracket$ between each adjacent pair in $\llbracket \ln m\rrbracket$,
    \begin{equation}
    \llbracket \Delta \rrbracket = \mathrm{Diff}\, \llbracket \ln m\rrbracket,
    \end{equation}
where $\llbracket \Delta \rrbracket$ is of length $n_\mathrm{pts}-1$.
Given priors $p(\sigma)$, $p(\tau)$, and $p(\ln r)$ on the standard deviation, length scale, and mean of the autoregressive process $\Psi$, realizations of $\Psi$ can then be proposed as follows.
\begin{enumerate}
    \item Draw $\sigma \sim p(\sigma)$, $\tau \sim p(\tau)$, and $\ln r \sim p(\ln r)$ from their respective priors.
    \item Draw a set $\llbracket n_i \rrbracket_{i=1}^{n_\mathrm{pts}} \sim N(0,1)$ of values from a unit normal distribution.
    \item Initialize the autoregressive process by defining $\tilde \Psi_1 = \sigma n_1$.
    \item Compute sets $\llbracket w_i \rrbracket = \llbracket \sigma (1-e^{-2\Delta_i/\tau}) n_i\rrbracket$ and $\llbracket c_i \rrbracket = \llbracket e^{-\Delta_i/\tau}\rrbracket$ for $i \in (2,n_\mathrm{pts})$.
    \item Compute all $\tilde \Psi_{i>1}$ via iterating $\tilde \Psi_i = c_i \tilde \Psi_{i-1} + w_i$.
    \item Finally, apply the mean: $\llbracket \Psi_i \rrbracket = \llbracket \tilde \Psi_i + \ln r\rrbracket$
\end{enumerate}
To enable efficient sampling, the $\llbracket \Psi_i\rrbracket$, $\llbracket w_i\rrbracket$, and $\llbracket c_i\rrbracket$ are generated following non-centered approaches; we directly sample in $\llbracket \tilde \Psi_i \rrbracket$ and $\llbracket n_i \rrbracket$ and then transform to the actual parameters of interest.
Once the complete set $\llbracket \Psi_i \rrbracket$ of log-merger rates is generated, the sorting performed to obtain Eq.~\eqref{eq:sort} can be reversed to repartition $\llbracket \Psi_i \rrbracket$ back into the merger rates across individual events' posterior samples and found injections.

In some cases, the merger rate $\ln\mathcal{R}(\theta)$ is not well measured at the lowest $\theta$ in our set of samples but at some intermediate value.
The merger rate as a function of mass, for example, is much better constrained near $m \approx 25\,M_\odot$ than at the very lowest masses $m\lesssim 5\,M_\odot$.
In this case, sampling efficiency is maximized by not initializing our autoregressive process at its left-most point (as in Step 3 above), but instead initializing the process in the \textit{middle} of our parameter range, near the best-measured rate.
In this case, Steps 4-6 above are just repeated twice, once to generate forward steps to the right of our reference point, and once to generate backward steps to the left of the reference point.

\section{Further Inference Details and Prior Constraints}
\label{app:inference-details}

In this appendix, we give additional information about the exact data used in this paper and further details regarding our implementation and inference of the autoregressive population model.

In our analyses, we include binary black holes in the GWTC-3 catalog~\cite{GWTC3} detected with false alarm rates below $1\,\mathrm{yr}^{-1}$.
GWTC-3 contains two events, GW190814~\cite{GW190814} and GW190917, that are likely binary black holes but which are known to be outliers with respect to the bulk binary population~\cite{O3b-pop}; we exclude these two events, leaving 69 binary black holes to be included in our analysis.
We use publicly-available parameter estimation samples hosted by the Gravitational-Wave Open Science Center\footnote{https://www.gw-openscience.org/}~\cite{losc,gwosc} and/or Zenodo.
Each binary typically has several distinct sets of associated posterior samples.
For events first identified in GWTC-1~\cite{GWTC1}, we use the ``\texttt{Overall\_posterior}'' samples.\footnote{Available at https://dcc.ligo.org/LIGO-P1800370/public}
For events announced in GWTC-2~\cite{GWTC2}, we use the ``\texttt{PrecessingSpinIMRHM}'' samples,\footnote{Available at https://dcc.ligo.org/LIGO-P2000223/public} and for new events in GWTC-3~\cite{GWTC3} we use the ``\texttt{C01:Mixed}'' samples.\footnote{Available at https://zenodo.org/record/5546663}
Each of these sets corresponds to a union of parameter estimation samples from various waveform families. 
All waveforms include the physical effects of spin precession, although parameter estimation accounting for higher-order modes is only available for GWTC-2 and GWTC-3.

Our hierarchical inference relies on the use of injected signals to characterize search selection effects; see Eq.~\eqref{eq:Nexp-mc}.
We use the injection set discussed in Ref.~\cite{O3b-pop},\footnote{https://zenodo.org/record/5636816} characterizing injections as ``found'' if they are recovered with false-alarm rates below $1\,\mathrm{yr}^{-1}$ in at least one search pipeline.
Note that the subset of injections performed for the O1 and O2 observing runs does not have associated false-alarm rates, only network signal-to-noise ratios $\rho$.
For these events, we consider them ``found'' if $\rho \geq 10$.

\begin{figure}
    \centering
    \includegraphics[width=0.49\textwidth]{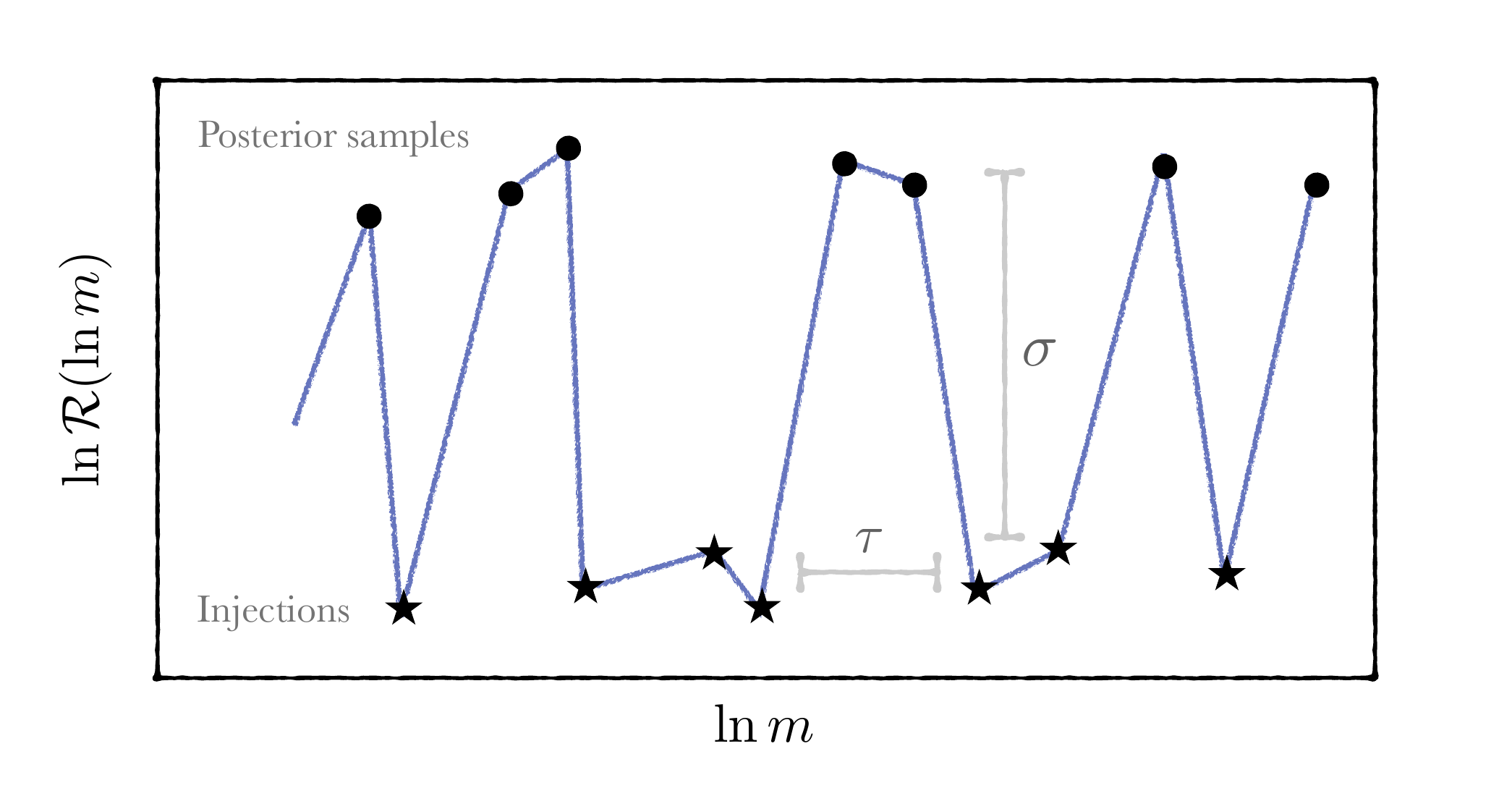}
    \caption{
    Illustration of the runaway situation described in Appendix~\ref{app:inference-details}.
    When performing inference with an autoregressive process population model, one can encounter a runaway instability occurring at small scale lengths $\tau$ and large variances $\sigma^2$.
    In this regime, the likelihood (Eq.~\eqref{eq:pop-likelihood-mc}) can be made arbitrarily large by allowing the merger rate at the locations of posterior samples (denoted by circles) to approach $\ln \mathcal{R}\to\infty$ while sending the merger rate at the locations of found injections (stars) to $\ln\mathcal{R}\to-\infty$.
    This behavior can be combated with additional regularization that penalizes very large ratios $\sigma/\sqrt{\tau}$, as in Eq.~\eqref{eq:ratio-prior}.
    }
    \label{fig:runaway-cartoon}
\end{figure}

Ensuring convergence of our inference sometimes requires careful regularization of $\sigma$ and $\tau$.
In particular, when performing inference with an autoregressive prior, we find that a common instability is a runaway towards $\sigma \to \infty$ and $\tau \to 0$.
The cause of this behaviour can be seen from the hierarchical likelihood defined in Eqs.~\eqref{eq:pop-likelihood-mc} and \eqref{eq:Nexp-mc}.
In particular, the likelihood is maximized if the merger rates at each posterior sample can grow to $\mathcal{R}(\lambda_{I,j})\to\infty$ while sending the merger rates at each injection to $\mathcal{R}(\lambda_{\mathrm{inj},i})\to 0$ (in turn sending $N_\mathrm{exp} \to 0$).
This situation is sketched in Fig.~\ref{fig:runaway-cartoon}, in which posterior samples are denoted as filled circles while injections are marked with stars.
Strongly-parametrized models usually impose strict constraints on the continuity and smoothness of the merger rate, preventing this behavior.
Our autoregressive model and similar approaches, in contrast, are explicitly designed to allow for rapid variations in the merger rate, and so are subject to this instability.
In particular, our autoregressive inference is most susceptible to runaway oscillatory behavior when the processes' variance becomes large and the autocorrelation length becomes small.

This instability can be regulated by placing suitable priors on the parameters governing the autoregressive process.
To motivate physically meaningful priors, it is useful to think about how an autoregressive process is allowed to vary between two points.
Consider an autoregressive process $\Psi(x)$ with zero mean, variance $\sigma^2$, and autocorrelation length $\tau$ defined across some parameter $x$.
Also let $\delta\Psi = \Psi_2 - \Psi_1$ be the \textit{difference} in the process' values between two points $x_1$ and $x_2$ (separated by $\delta x = x_2-x_1$).
The expectation value of $\delta \Psi$ is
    \begin{equation}
    \begin{aligned}
    \langle \delta \Psi\rangle
        &= \langle \Psi_2 \rangle - \langle \Psi_1 \rangle \\
        &= 0,
    \end{aligned}
    \end{equation}
since, by definition, the process has zero mean.
The expectation value of $\delta \Psi^2$, though, is nonzero:
    \begin{equation}
    \begin{aligned}
    \langle \delta \Psi^2 \rangle
        &= \langle \left(\Psi_2 - \Psi_1\right)^2 \rangle \\
        &= \langle \Psi_2^2 \rangle + \langle \Psi_1^2 \rangle - 2 \,\mathrm{Cov}(\Psi_2,\Psi_1) \\
        &= 2\sigma^2 - 2 \sigma^2 e^{-\delta x/\tau} \\
        &= 2\sigma^2\left(1-e^{-\delta x/\tau}\right),
    \end{aligned}
    \label{eq:diff-var}
    \end{equation}
using Eqs.~\eqref{eq:ar-var} and \eqref{eq:ar-cov} for the variance and covariance of an autoregressive process.
It is helpful to consider Eq.~\eqref{eq:diff-var} in two different limits.
In the limit $\delta x \ll \tau$,
    \begin{equation}
    \lim_{\delta x \ll \tau} \langle \delta \Psi^2 \rangle = 2\, \delta x \frac{\sigma^2}{\tau}.
    \label{eq:dpsi-limit-low}
    \end{equation}
In the opposite limit,
    \begin{equation}
    \lim_{\delta x \gg \tau} \langle \delta \Psi^2 \rangle = 2 \sigma^2.
    \label{eq:dpsi-limit-high}
    \end{equation}
    
More generally, we can show that in each of the above limits, $\delta \Psi^2$ is $\chi^2$-distributed with $k=1$ degrees of freedom.
Recall that $\Psi_1$ and $\Psi_2$ are related by
    \begin{equation}
    \Psi_2 = e^{-\delta x/\tau} \Psi_1 + \sigma \left(1-e^{-2\delta x/\tau}\right)^{1/2} n,
    \end{equation}
where $n\sim N(0,1)$ is drawn from a unit normal distribution.
Then,
    \begin{equation}
    \delta \Psi^2
        = \left[ \left(e^{-\delta x/\tau} - 1\right)\Psi_1 + \sigma \left(1-e^{-2\delta x/\tau}\right)^{1/2} n \right]^2.
    \label{eq:full-squared-diff}
    \end{equation}
First consider the $\delta x \ll \tau$ limit.
Expanding to lowest order in $\delta x/\tau$,
    \begin{equation}
    \begin{aligned}
    \delta \Psi^2
        &\approx \left( -\frac{\delta x}{\tau} \Psi_1 + \sigma n \sqrt{\frac{2\delta x}{\tau}}  \right)^2 \\
        &\approx \sigma^2 n^2 \left(\frac{2\delta x}{\tau}\right),
    \end{aligned}
    \end{equation}
where the first term is subdominant to the second in the limit of small $\delta x/\tau$.
We therefore have
    \begin{equation}
    \frac{\delta \Psi^2/\delta x}{2 \sigma^2/\tau} \approx n^2
    \end{equation}
such that, by definition, this quantity is chi-squared distributed with one degree of freedom:
    \begin{equation}
    \frac{\delta \Psi^2/\delta x}{2 \sigma^2/\tau} \sim \chi^2(1).
    \end{equation}
Its corresponding expectation value is
    \begin{equation}
    \left\langle \frac{\delta \Psi^2/\delta x}{2 \sigma^2/\tau} \right\rangle
        = \frac{\langle\delta \Psi^2\rangle /\delta x}{2 \sigma^2/\tau} 
        = 1;
    \end{equation}
compare to Eq.~\eqref{eq:dpsi-limit-low} above.
In the opposite limit, where $\delta x \gg \tau$, Eq.~\eqref{eq:full-squared-diff} becomes
    \begin{equation}
    \delta \Psi^2 \approx \left( -\Psi_1 + \sigma n\right)^2.
    \end{equation}
Note that $\Psi_1$ is itself drawn from a  normal distribution (see Appendix~\ref{app:ar-appendix}).
We can therefore write
    \begin{equation}
    \delta \Psi^2 \approx \sigma^2 (n-m)^2,
    \end{equation}
where $m\sim N(0,1)$ is another normally distributed random variable.
The difference $n-m$ is itself a Gaussian random variable with zero mean and standard deviation $\sqrt{2}$, and so
    \begin{equation}
    \delta \Psi^2 = 2 \sigma^2 \overline n^2
    \end{equation}
where $\overline n = (n-m)/\sqrt{2} \sim N(0,1)$ now follows a unit normal distribution.
Hence
    \begin{equation}
    \frac{\delta \Psi^2}{2\sigma^2} \sim \chi^2(1)
    \end{equation}
is also chi-squared distributed with one degree of freedom and with mean
    \begin{equation}
    \frac{\langle \delta \Psi^2\rangle}{2\sigma^2} = 1;
    \end{equation}
compare to Eq.~\eqref{eq:dpsi-limit-high}.
All together,
    \begin{equation}
    \begin{cases}
    \dfrac{\delta \Psi^2/\delta x}{2 \sigma^2/\tau} \sim \chi^2(1) & \left(\delta x\ll \tau\right) \\[10pt]
    \dfrac{\delta \Psi^2}{2\sigma^2} \sim \chi^2(1) & \left(\delta x \gg \tau\right).
    \end{cases}
    \label{eq:dpsi-cases}
    \end{equation}

We use Eq.~\eqref{eq:dpsi-cases} to motivate physical priors on $\sigma^2$ and $\tau$.
First, we might expect the log merger rate to vary by no more than $\delta \Psi_\mathrm{max}$ over the full parameter space.
Let $\Delta x$ be the full extent of the parameter space and assume that $\Delta x \gtrsim \tau$, such that we are in the second case in Eq.~\eqref{eq:dpsi-cases}.
Then our expectation is that 
    \begin{equation}
    q < \frac{\delta \Psi_\mathrm{max}^2}{2\sigma^2}
    \label{eq:q-large-delta}
    \end{equation}
for a chi-squared distributed random variable $q\sim \chi^2(1)$.
In particular, we might assert that $\delta \Psi_\mathrm{max}^2/2\sigma^2$ occurs at the 99th percentile $q_{99}$ of the chi-squared distribution.
The cumulative distribution of a $\chi^2(1)$ distribution is the regularized gamma function $P(1/2;q/2)$, and so $q_{99}$ occurs at
    \begin{equation}
    q_{99} = P^{-1}(1/2;0.99) \approx 3.32.
    \end{equation}
Inserting into Eq.~\eqref{eq:q-large-delta} and solving for $\sigma$, we have
    \begin{equation}
    \sigma < \frac{\delta \Psi_\mathrm{max}}{\sqrt{2\,q_{99}}}.
    \label{eq:sigma-inequality}
    \end{equation}
We therefore choose a prior on $\sigma$ to enforce Eq.~\eqref{eq:sigma-inequality}.
Specifically, we adopt a half-Gaussian prior
    \begin{equation}
    p(\sigma|\Sigma_\sigma) \propto \exp\left(-\frac{\sigma^2}{2\Sigma_\sigma^2}\right)
    \label{eq:sig-prior}
    \end{equation}
on the range $0\leq \sigma < \infty$.
The scale $\Sigma_\sigma^2$ of this prior is chosen so that 99\% of our prior weight occurs below the threshold set by Eq.~\eqref{eq:sigma-inequality}.
Specifically,
    \begin{equation}
    \Sigma_\sigma = \frac{\delta\Psi_\mathrm{max}}{2 \sqrt{q_{99}}\, \mathrm{Erf}^{-1}(0.99)}
    \label{eq:sig-prior-var}
    \end{equation}

\begin{table*}[]
    \setlength{\tabcolsep}{8pt}
    \renewcommand{\arraystretch}{1.1}
    \centering
    \caption{
    Priors governing the various autoregressive process models used in this paper.
    For each physical parameter, we give our choices for the scale values characterizing the priors defined in Appendix~\ref{app:inference-details}:
    the domain width $\delta x$, the maximum variation $\delta \Psi_\mathrm{max}$ in the log merger rate, the maximum inter-event variation $\delta \Psi_\mathrm{event}$ in the log rate, and the number $N$ of measurements considered.
    We specifically give $e^{\delta \Psi_\mathrm{max}}$ and $e^{\delta \Psi_\mathrm{event}}$, so that that the quantities listed in the table are directly interpretable as merger rate variations.
    Also note that while the majority of priors use $N=69$ (the number of events in our sample), we take $N=138$ when analyzing the component spin magnitude and tilt distributions, as each binary contributes two component spins to our sample.
    Given these parameter choices, we also show the derived quantities directly appearing in Eqs.~\eqref{eq:sig-prior}, \eqref{eq:ratio-prior}, and \eqref{eq:tau-prior}.
    }
    \begin{tabular}{l c| r r r r | r r r r r }
    \hline \hline
    Param. & Section & $\Delta x$ & $e^{\delta\Psi_\mathrm{max}}$ & $e^{\delta\Psi_\mathrm{event}}$ & $N$ & $\Sigma_\sigma$ & $\ln(\Delta x/2)$ & $\Sigma_{\ln\tau}$  & $\Sigma_r$ &  $\delta x_\mathrm{min}$ \\
    \hline
    $\ln m_1$ & 
        Sect.~\ref{sec:masses} & 
        $4$ &
        $10^2$ &
        $2$ &
        $69$ &
        $0.91$ &
        $0.69$ &
        $4.72$ &
        $0.57$ & 
        $8.5\times10^{-4}$ \\
    $q$ & 
        Sect.~\ref{sec:masses} & 
        $1$ &
        $10^2$ &
        $2$ &
        $69$ &
        $0.91$ &
        $-0.69$ &
        $4.72$ &
        $1.14$ & 
        $2.1\times10^{-4}$ \\
    $z$ & 
        Sect.~\ref{sec:redshifts} & 
        $2$ & 
        $10^2$ &
        $2$ &
        $69$ &
        $0.91$ &
        $0.0$ &
        $4.72$ &
        $0.81$ &
        $4.2\times10^{-4}$ \\
    $\chi_1, \chi_2$ &
        Sect.\ref{sec:spins} & 
        $1$ &
        $10^2$ &
        $1.5$ &
        $138$ &
        $0.91$ &
        $0.69$ &
        $5.57$ &
        $0.94$ & 
        $5.3\times10^{-5}$ \\
    $\cos\theta_1, \cos\theta_2$ & 
        Sect.\ref{sec:spins} & 
        $2$ &
        $10^2$ &
        $1.5$ &
        $138$ &
        $0.91$ &
        $0.0$ &
        $5.57$ &
        $0.67$ & 
        $1.1\times10^{-4}$ \\
    $\chi_\mathrm{eff}$ & 
        Sect.\ref{sec:effective-spins} &
        $2$ &
        $10^2$ &
        $2$ &
        $69$ &
        $0.91$ &
        $0.0$ &
        $4.72$ &
        $0.81$ & 
        $4.2\times10^{-4}$ \\
    $\chi_\mathrm{p}$ & 
        Sect.\ref{sec:effective-spins} & 
        $1$ &
        $10^2$ &
        $2$ &
        $69$ &
        $0.91$ &
        $-0.69$ &
        $4.72$ &
        $1.14$ & 
        $2.1\times10^{-4}$ \\
    \hline
    $\chi$ & 
        Appendix~\ref{app:demo} & 
        $2$ &
        $10^2$ &
        $2$ &
        $69$ &
        $0.91$ &
        $0.0$ &
        $4.72$ &
        $0.81$ & 
        $4.2\times10^{-4}$ \\
    \hline
    \hline
    \end{tabular}
    \label{tab:ar-priors}
\end{table*}

Next, consider how we expect our autoregressive process to vary on small scales.
In particular, let $\delta\Psi_\mathrm{event}$ be the maximum variation we expect in the log rate on the typical inter-event distance scale $\delta x \approx \Delta x/N$, where $N$ is the number of events in our catalog.
This distance scale is likely smaller than the autoregressive process' correlation length $\tau$, and so we now use the first inequality in Eq.~\eqref{eq:dpsi-cases}, demanding that
    \begin{equation}
    q < \frac{N\delta \Psi_\mathrm{event}^2/\Delta x}{2\sigma^2/\tau},
    \label{eq:q-constraint}
    \end{equation}
for $q\sim \chi^2(1)$.
We proceed as above, using Eq.~\eqref{eq:q-constraint} to define the 99th percentile of $q$ and rearranging to obtain the limit
    \begin{equation}
    \frac{\sigma}{\sqrt{\tau}} < \delta \Psi_\mathrm{event} \sqrt{\frac{N}{2\, q_{99} \,\Delta x}}.
    \label{eq:ratio-limit}
    \end{equation}
We impose this limit by adopting another half-Gaussian prior on the ratio $\sigma/\sqrt{\tau}$,
    \begin{equation}
    p(\sigma/\sqrt{\tau}\,|\,\Sigma_r) \propto \mathrm{exp}\left(-\frac{(\sigma/\sqrt{\tau})^2}{2\Sigma^2_r}\right),
    \label{eq:ratio-prior}
    \end{equation}
with $\Sigma_r$ chosen such that $99\%$ of the prior weight occurs below the limit set by Eq.~\eqref{eq:ratio-limit}:
    \begin{equation}
    \Sigma_r = \frac{\delta\Psi_\mathrm{event}}{2\,\mathrm{Erf}^{-1}(0.99)} \sqrt{\frac{N}{q_{99}\,\Delta x}}.
    \label{eq:ratio-prior-var}
    \end{equation}

We can also place limits on the expected length scale $\tau$ of our autoregressive processes.
We generally expect the merger rate to be smoothly varying across the parameter space of interest, but also want a prior that will nevertheless allow for rapid, small-scale variations should they be demanded by the data.
We accordingly place an unbounded Gaussian prior on $\ln \tau$:
    \begin{equation}
    p(\ln\tau|\Sigma_{\ln\tau}) \propto \mathrm{exp}\left(-\frac{(\ln\tau - \ln(\Delta x/2))^2}{2\Sigma_{\ln\tau}^2}\right).
    \label{eq:tau-prior}
    \end{equation}
This prior is centered at $\ln(\Delta x/2)$, and we set $\Sigma_{\ln\tau}$ by considering the minimum length scale that can be meaningfully constrained by $N$ detections.
In particular, the data contain no information about features on scales smaller than the \textit{minimum spacing} between events.
If we consider randomly placing $N$ events across an interval of width $\Delta x$, the spacing $\delta x$ between events will be exponentially distributed:
    \begin{equation}
    p(\delta x) = \frac{(N/\Delta x) e^{-N \delta x/\Delta x}}{1-e^{-N}}
    \end{equation}
with a cumulative distribution
    \begin{equation}
    F(\delta x) = \frac{1-e^{-N \delta x/\Delta X}}{1-e^{-N}},
    \label{eq:cumulative-exp}
    \end{equation}
normalizing over the range $0 \leq \delta x \leq \Delta x$.
With $N$ events, we expect the minimum spacing $\delta x_\mathrm{min}$ in our sample to probe the $1/N$ quantile of this exponential distribution.
Setting $F=1/N$ and inverting Eq.~\eqref{eq:cumulative-exp} then gives
    \begin{equation}
    \delta x_\mathrm{min} = -\frac{\Delta x}{N} \ln\left[ 1 - \frac{1}{N}\left(1-e^{-N}\right)\right];
    \end{equation}
this is the minimum length scale we expect to probe with $N$ events.
We choose $\Sigma_{\ln\tau}$ such that 95\% of our prior lies above $\ln\delta x_\mathrm{min}$:
    \begin{equation}
    \Sigma_{\ln\tau} = \frac{\ln\delta x_\mathrm{min} - \ln(\Delta x/2)}{\sqrt{2}\mathrm{Erf}^{-1}(1-2\times0.95)}.
    \label{eq:tau-prior-var}
    \end{equation}

So far, all our discussion has concerned priors on the variance and lengthscale of an autoregressive process.
We also need to consider a prior on the \textit{mean} of the process, denoted above as $\ln r$.
For all analyses, we place a logarithmically uniform prior on $r$,
    \begin{equation}
    p(r) = \frac{1}{r},
    \label{eq:rate-prior}
    \end{equation}
across the interval $10^{-6} \leq (r/\mathrm{Gpc}^{-3}\,\mathrm{yr}^{-1}) \leq 10^{5}$, with $p(r) = 0$ outside this interval.

Together, the product of Eqs.~\eqref{eq:sig-prior}, \eqref{eq:ratio-prior}, \eqref{eq:tau-prior}, and \eqref{eq:rate-prior} comprise the priors placed on our autoregressive models, with variances defined by Eqs.~\eqref{eq:sig-prior-var}, \eqref{eq:ratio-prior-var}, and \eqref{eq:tau-prior-var}, respectively.
These still depend on several yet-undefined choices for $\Delta x$, $\delta \Psi_\mathrm{max}$, and $\delta \Psi_\mathrm{event}$; these are listed in Table~\ref{tab:ar-priors} along with several other derived quantities characterizing the above priors.

In Sect.~\ref{sec:method}, we discussed how the likelihood is approximated via a weighted Monte Carlo average over ensembles of posterior samples; see Eq.~\eqref{eq:pop-likelihood-mc}.
Similarly, the expected number of detections was evaluated by a Monte Carlo average over a set of successfully found injections; see Eq.~\eqref{eq:Nexp-mc}.
Both approximations break down if these averages become dominated by a very small number of posterior samples or found injections.
One metric for gauging the health of Monte Carlo averaging is the effective sample number.
Define $w_i(\Lambda) = R_d(\lambda_{\mathrm{inj},i};\Lambda)/p_\mathrm{pe}(\lambda_{\mathrm{inj},i})$ to be the weights appearing in the calculation of $N_\mathrm{exp}(\Lambda)$ in Eq.~\eqref{eq:Nexp-mc}.
The effective number of samples informing this calculation is given by
    \begin{equation}
    N_\mathrm{eff}^\mathrm{inj}(\Lambda) = \frac{(\sum_i w_i(\Lambda))^2}{\sum_j(w_j(\Lambda))^2}.
    \end{equation}
In order for systematic uncertainty in $N_\mathrm{exp}=(\Lambda)$ to remain a subdominant effect in our hierarchical analysis, it is necessary that $N_\mathrm{eff}^\mathrm{inj}(\Lambda)\gtrsim 4 N_\mathrm{obs}$~\cite{Farr2019,Talbot2023}.
Similarly, define $w_{I,j}(\Lambda) = R_d(\lambda_{I,j};\Lambda)/p_\mathrm{pe}(\lambda_{I,j})$ to be the weights defined in Eq.~\eqref{eq:pop-likelihood-mc} over the posterior samples $j$ of each event $I$.
The number of effective posterior samples informing each event's likelihood is then
    \begin{equation}
    N_{\mathrm{eff},I}^\mathrm{samp}(\Lambda) = \frac{(\sum_j w_{I,j}(\Lambda))^2}{\sum_k(w_{I,k}(\Lambda))^2}.
    \end{equation}
In particular, a useful metric is the minimum number of effective posterior samples, $\min_I N_{\mathrm{eff},I}^\mathrm{samp}(\Lambda)$, taken over events $I$.
Healthy inference generally requires $\min_I N_{\mathrm{eff},I}^\mathrm{samp}(\Lambda) \gg 1$~\cite{Essick2022}.

For each of our analyses, we monitor $N_\mathrm{eff}^\mathrm{inj}$ and $\min N_\mathrm{eff}^\mathrm{samp}$ for signs of poor effective sample counts; see Figs.~\ref{fig:mass-corner}-\ref{fig:effective-spin-corner} in Appendix~\ref{app:hyperparams}.
Although potential population models with low effective sample counts are largely discouraged by the above priors on $\sigma$, $\ln\tau$, and $\sigma/\sqrt{\tau}$, we further prevent our inference from exploring models with pathologically low effective samples by severely penalizing proposed populations with $N_\mathrm{eff}^\mathrm{inj}<4 N_\mathrm{obs}$ and/or $\min \log_{10} N_\mathrm{eff}^\mathrm{samp}<0.6$.
Specifically, we define the function
    \begin{equation}
    \mathcal{S}(x) = \left(\frac{1}{1+x^{-30}}\right);
    \end{equation}
this asymptotes to unity when $x$ is large and falls to zero when $x$ approaches zero.
We then add the terms $\ln\mathcal{S}(N_\mathrm{eff}^\mathrm{inj}/4 N_\mathrm{obs}) + \ln\mathcal{S}(\min \log_{10} N_\mathrm{eff}^\mathrm{samp}/0.6)$ to the log-likelihood implemented in \texttt{numpyro}.
These send the log-likelihood towards $-\infty$ when either of the above conditions are violated.

\section{Demonstrations on Known Populations}
\label{app:demo}

We demonstrate the machinery developed in Appendices~\ref{app:ar-appendix} and \ref{app:inference-details} by injecting and recovering a set of known distributions.
This is useful in verifying that our methodology works as expected, and in diagnosing any limitations in the performance of our autoregressive model or interpretation of our results.
We consider four toy populations:
\begin{enumerate}
    \item A Gaussian distribution
        \begin{equation}
        p(\chi) = \frac{1}{\sqrt{2\pi\sigma^2}} e^{-(\chi-\mu)^2/2\sigma^2}
        \end{equation}
        with $\mu=0.05$ and $\sigma=1$.
    \item A delta function at zero,
        \begin{equation}
        p(\chi) = \delta(\chi).
        \end{equation}
    \item A mixture between a delta function and a broad Gaussian,
        \begin{equation}
        p(\chi) = \frac{f_g}{\sqrt{2\pi\sigma^2}} e^{-(\chi-\mu)^2/2\sigma^2} + (1-f_g)\delta(\chi)\,,
        \end{equation}
    with $\mu = 0.5$, $\sigma=0.3$, and $f_p=0.3$.
    \item A half-normal distribution truncated to purely positive values,
        \begin{equation}
        p(\chi) = \begin{cases}
            \sqrt{\frac{2}{\pi\sigma^2}} e^{-(\chi-\mu)^2/2\sigma^2} & (\chi\geq 0) \\
            0 & (\chi<0)\,
        \end{cases}
        \end{equation}
    with $\mu=0$ and $\sigma=0.35$.
\end{enumerate}
All four are chosen as plausible distributions of the effective inspiral spin $\chi_\mathrm{eff}$ among binary black hole mergers.
Specifically, the first model corresponds approximately to the measured effective spin distribution reported in Ref.~\cite{O3b-pop}.
The second and third correspond to cases in which some or all black hole spins are zero, as predicted in Ref.~\cite{Fuller2019} and initially claimed in Ref.~\cite{Galaudage2021} (but disfavored by Refs.~\cite{Callister2022,Mould2022,Tong2022}).
The fourth population corresponds to a situation in which there is no excess of zero-spin black holes, but in which all spins are preferentially aligned and hence give purely positive $\chi_\mathrm{eff}$

\begin{figure*}
    \centering
    \includegraphics[width=0.75\textwidth]{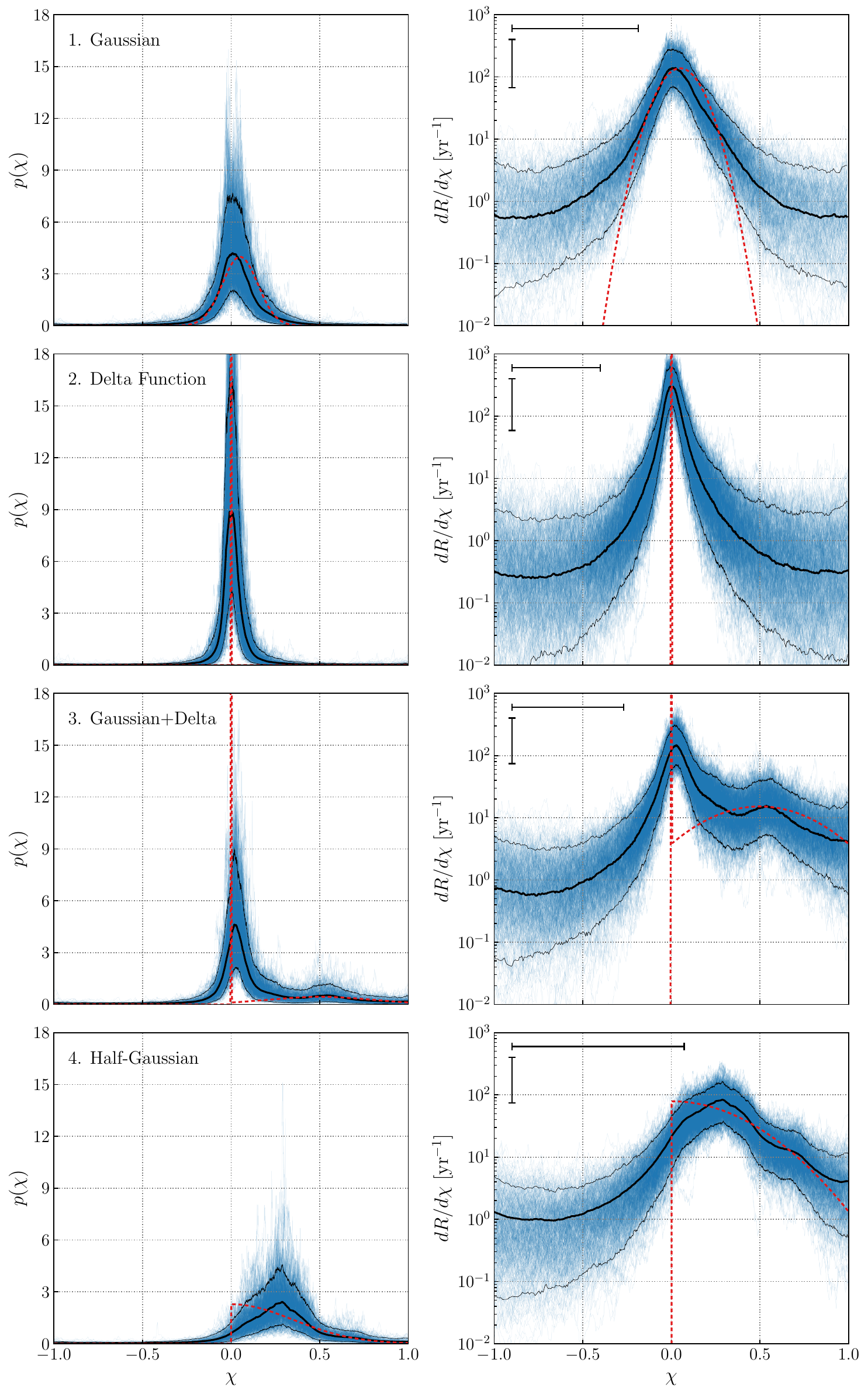}
    \caption{
    Demonstrated autoregressive inference of four simulated populations, as described in Appendix~\ref{app:demo}.
    The left column indicates the normalized probability distributions recovered in each case, while the right column shows inferred rate densities.
    In the right-hand column, the median recovered autocorrelation length $\tau$ and standard deviation $e^\sigma$ of the process are shown via the horizontal and vertical error bars (recall that $\sigma^2$ defines the variance of the \textit{logarithmic} number density).
    In each case, true distributions are indicated as dashed red curves.
    }
    \label{fig:spin_injections}
\end{figure*}

We draw 69 simulated events from each population, matching the size of the GWTC-3 catalog, and assume that these events are obtained over $T=1\,\mathrm{yr}$ of observation.
For each simulated event $i$, we draw a ``true'' value $\chi_{{\rm true},i} \sim p(\chi)$ and then a random ``observed'' maximum likelihood value.
To obtain an observed value, we first draw a random measurement uncertainty according to $\log_{10}\sigma_i \sim N(-0.9,0.3)$, chosen to match the distribution of $\chi_\mathrm{eff}$ uncertainties among events in GWTC-3.
We then randomly draw an observed value $\chi_{{\rm obs},i} \sim N(\chi_{{\rm true},i},\sigma_i)$.
Finally, we randomly draw $1000$ mock posterior samples per event, assuming these are Gaussian-distributed about $\chi_{{\rm obs},i}$ with standard deviation $\sigma_i$.
For simplicity, we assume a flat selection function.
To most accurately test the behavior of our method, however, we still compute the expected number of detections $N_{\rm exp}$ using a Monte Carlo average over a set of $10^4$ discrete samples, as in Eq.~\eqref{eq:Nexp-mc}.
In particular, this means that our inference can still experience the instability discussed above in Appendix~\ref{app:inference-details}.
The final line in Table~\ref{tab:ar-priors} defines the prior placed on our autoregressive model in each of the four cases.

Our results are shown in Fig.~\ref{fig:spin_injections}.
For each injected population, the left-hand column shows our recovery of the normalized probability distribution, while the right-hand column shows inferred rate densities $dR/d\chi$.
In all cases, the underlying distributions from which mock events are drawn are shown via dashed red curves.
Case 1 shows faithful recovery of a Gaussian distribution, with the reconstructed probability distribution and rate density closely matching the injected population within the range $-0.5\lesssim \chi \lesssim 0.5$.
Beyond this range, we see that, although our recovered rate density is able to fall by approximately three orders from the center to the edge of the domain, it is unable to go strictly to zero.
Instead, the autoregressive process asymptotes to $dR/d\chi \approx 0.5$ in regions with no data.
Note that this asymptotic limit corresponds to the threshold below which we expect no additional events on top of our 69 simulated detections after $T=1\,\mathrm{yr}$ of observation.
As our catalog increases in size, the continued non-observation of events at $|\chi|\gtrsim 0.5$ will further suppress tails in the recovered rate density.

Case 2 behaves similarly.
Although our autoregressive model cannot truly produce a delta function distribution, we recover a narrow distribution centered at $\chi=0$.
Once more, we see that the inferred rate falls by three orders of magnitude between $\chi=0$ and $|\chi|\gtrsim 0.5$, with asymptotic tails in regions with no data.
In Case 4, illustrating recovery of a broad half-Gaussian distribution, we correctly recover the injected rate density from its peak of $dR/d\chi\approx 10^2$ at $\chi=0$ down to $dR/d\chi \approx 1$ at $\chi = 0$.
At negative $\chi$, Case 4 shows the same asymptotic tail exhibited by Cases 1 and 2.

One limitation of our autoregressive model is illustrated in Case 3, in which our injected population is a mixture between a delta function at $\chi=0$ and a broad Gaussian at positive $\chi$.
We see that both features, a primary peak at $\chi=0$ and a secondary peak at $\chi=0.5$, are correctly and confidently identified by our inference despite the order-of-magnitude difference in the prominence of these features.
The rate density of events arising from the half-Gaussian is, in particular, well-recovered even out to $\chi=1$.
At the same time, our autoregressive model struggles to produce an extremely sharp feature at $\chi=0$ while \textit{also} fitting the broad Gaussian.
Recovery of the Gaussian serves to fix the autocorrelation length $\tau$ of the process to reasonably large values, in turn limiting the resolution with which the delta function can be resolved.
Compare, for example, the recovery of the delta function in Case 3 to the much sharper recovery of an isolated delta function in Case 2.

\section{Strongly-Parametrized Models}
\label{app:strong-models}

\begin{table}[]
    \setlength{\tabcolsep}{6pt}
    \renewcommand{\arraystretch}{1.1}
    \centering
    \caption{Parameters characterizing the ordinary strongly-parameterized models discussed in Appendix~\ref{app:strong-models}.
    For each parameter, we indicate its defining equation, the paper sections in which it is used, and its associated prior.
    We use $N(a,b)$ to denote a Gaussian prior with mean $a$ and standard deviation $b$, $\mathrm{U}(a,b)$ to indicate a uniform prior between $a$ and $b$, and $\mathrm{LU}(a,b)$  to indicate a logarithmically-uniform prior between the given bounds.
    }
    \begin{tabular}{l l l l}
    \hline \hline
    Param. & Defined & Sections & Prior \\
    \hline
    $\lambda$ & 
        Eq.~\eqref{eq:plpeak} & 
        \ref{sec:redshifts},\ref{sec:spins},\ref{sec:effective-spins} & 
        $\mathrm{N}(0,10)$ \\
    $\mu_m$ &
        Eq.~\eqref{eq:plpeak} & 
        \ref{sec:redshifts},\ref{sec:spins},\ref{sec:effective-spins} &  
        $\mathrm{U}(20\,M_\odot,50\,M_\odot)$ \\
    $\sigma_m$ & 
        Eq.~\eqref{eq:plpeak} & 
        \ref{sec:redshifts},\ref{sec:spins},\ref{sec:effective-spins} & 
        $\mathrm{U}(2\,M_\odot,15\,M_\odot)$ \\
    $f_p$ & 
        Eq.~\eqref{eq:plpeak} & 
        \ref{sec:redshifts},\ref{sec:spins},\ref{sec:effective-spins} & 
        $\mathrm{LU}(10^{-3},1)$ \\
    $m_\mathrm{low}$ &
        Eq.~\eqref{eq:cutoff} & 
        \ref{sec:redshifts},\ref{sec:spins},\ref{sec:effective-spins} & 
        $\mathrm{U}(5\,M_\odot,15\,M_\odot)$ \\
    $m_\mathrm{high}$ &
        Eq.~\eqref{eq:cutoff} & 
        \ref{sec:redshifts},\ref{sec:spins},\ref{sec:effective-spins} & 
        $\mathrm{U}(50\,M_\odot,100\,M_\odot)$ \\
    $\delta m_\mathrm{low}$ &
        Eq.~\eqref{eq:cutoff} & 
        \ref{sec:redshifts},\ref{sec:spins},\ref{sec:effective-spins} &
        $\mathrm{LU}(0.1\,M_\odot,10\,M_\odot)$ \\
    $\delta m_\mathrm{high}$ &
        Eq.~\eqref{eq:cutoff} & 
        \ref{sec:redshifts},\ref{sec:spins},\ref{sec:effective-spins} &
        $\mathrm{LU}(3\,M_\odot,30\,M_\odot)$ \\
    $\beta_q$ &
        Eq.~\eqref{eq:mass-ratio} & 
        \ref{sec:redshifts},\ref{sec:spins},\ref{sec:effective-spins} & 
        $\mathrm{N}(0,4)$ \\
    \hline
    $\mu_\chi$ & 
        Eq.~\eqref{eq:default-mag} & 
        \ref{sec:masses},\ref{sec:redshifts} & 
        $\mathrm{U}(0,1)$ \\
    $\sigma_\chi$ &
        Eq.~\eqref{eq:default-mag} &
        \ref{sec:masses},\ref{sec:redshifts} &
        $\mathrm{LU}(0.1,1)$ \\
    $\sigma_u$ &
        Eq.~\eqref{eq:default-tilt} &
        \ref{sec:masses},\ref{sec:redshifts} & 
        $\mathrm{U}(0.3,2)$ \\
    \hline
    $\kappa$ &
        Eq.~\eqref{eq:kappa-eqn} & 
        \ref{sec:masses},\ref{sec:spins},\ref{sec:effective-spins} &
        $\mathrm{N}(0,5)$  \\
    \hline
    \hline
    \end{tabular}
    \label{tab:param-priors}
\end{table}

In the main text, we studied the distributions of black hole masses (Sect.~\ref{sec:masses}), spins (Sects.~\ref{sec:spins} and~\ref{sec:effective-spins}), and redshifts (Sect.~\ref{sec:redshifts}), modeling each set of distributions in turn as autoregressive processes.
To accurately measure the population distribution of any one parameter, it is generally necessary to simultaneously fit for the distributions of other parameters.
Therefore, wherever we focused on modeling a specific subset of parameters using autoregressive models, we concurrently fit the remaining parameters using simple strongly-parametrized models.
The priors used for each of the following models are given in Table~\ref{tab:param-priors}.

Our parametric mass model assumes that primary masses are drawn from a mixture between a power law and a Gaussian peak, with possible tapering at low and high masses.
This is a variant of the \textsc{Power Law+Peak} model first defined in Ref.~\cite{Talbot2018} and used in depth in Refs.~\cite{O3a-pop,O3b-pop}.
Specifically, we define
    \begin{equation}
    \begin{aligned}
    \phi(m_1) &= 
    \frac{f_p}{\sqrt{2\pi\sigma_m^2}}\, \mathrm{exp}\left(-\frac{(m_1-\mu_m)^2}{2\sigma_m^2}\right) \\
    &+(1-f_p) \left(\frac{1+\lambda}{(100\,M_\odot)^{1+\lambda} - (2\,M_\odot)^{1+\lambda}}\right)\,m_1^\lambda.
    \end{aligned}
    \label{eq:plpeak}
    \end{equation}
to be the superposition of a power law and Gaussian; the former is normalized between $2\,M_\odot \leq m_1 \leq 100\,M_\odot$ with spectral index $\lambda$, while the latter is centered at mean $\mu_m$ with standard deviation $\sigma_m$.
The parameter $f_p$ controls the relative contribution of each component.
Our complete primary mass distribution is of the shape
    \begin{equation}
    f(m_1) = \begin{cases}
    \phi(m_1)\, \mathrm{exp}\left[\frac{-(m_1-m_\mathrm{low})^2}{2\delta m_\mathrm{low}^2}\right] & (m_1 < m_\mathrm{low}) \\
    \phi(m_1) & (m_\mathrm{low} \leq m_1 \leq m_\mathrm{high}) \\
    \phi(m_1)\, \mathrm{exp}\left[\frac{-(m_1-m_\mathrm{high})^2}{2\delta m_\mathrm{high}^2}\right] & (m_\mathrm{high} < m_1),
    \end{cases}
    \label{eq:cutoff}
    \end{equation}
with squared exponentials that taper $f(m_1)$ towards zero above and below $m_\mathrm{low}$ and $m_\mathrm{high}$, respectively.
The tapering scales $\delta m_\mathrm{low}$ and $\delta m_\mathrm{high}$ are additional free parameters inferred from the data.
Mass ratios, in turn, are assumed to follow a power-law distribution, with
    \begin{equation}
    p(q|m_1) = \left(\frac{1+\beta_q}{m_1^{1+\beta_q} - (2M_\odot)^{1+\beta_q}}\right) m_2^{\beta_q}
    \label{eq:mass-ratio}
    \end{equation}
This parametric mass model is used in Sects.~\ref{sec:redshifts}-\ref{sec:effective-spins}, and  when focusing on autoregressive modeling of black hole spins and redshifts.
    
When non-parametrically exploring the black hole mass and redshift distributions, we revert to a parametric spin model in which component spin magnitudes and spin-orbit tilt angles are independently and identically distributed as truncated normal distributions.
Each component spin magnitude in a given binary has a probability distribution
    \begin{equation}
    p(\chi_i) = \sqrt{\frac{2}{\pi\sigma_\chi^2}}
    \frac{e^{-(\chi_i - \mu_\chi)^2/2\sigma_\chi^2}}
    {\mathrm{Erf}\left(\frac{1-\mu_\chi}{\sqrt{2\sigma_\chi^2}}\right) + \mathrm{Erf}\left(\frac{\mu_\chi}{\sqrt{2\sigma_\chi^2}}\right)},
    \label{eq:default-mag}
    \end{equation}
with a mean $\mu_\chi$ and standard deviation $\sigma_\chi$ that are inferred from the data.
The cosines of component spin tilt angles, meanwhile, are independently distributed as
    \begin{equation}
    p(\cos\theta_i) = \sqrt{\frac{2}{\pi\sigma_u^2}}
    \frac{e^{-(\cos\theta_i - 1)^2/2\sigma_u^2}}
    {\mathrm{Erf}\left(\frac{-2}{\sqrt{2\sigma_u^2}}\right)},
    \label{eq:default-tilt}
    \end{equation}
with a mean fixed to $1$ but a standard deviation $\sigma_u$ measured from the data.
This parameteric model is used in Sects.~\ref{sec:masses} and \ref{sec:redshifts} when targeting binary masses and redshifts with our autoregressive prior.
    
Finally, when targeting binary masses or spins, we revert to a standard parametric redshift model in which the comoving merger rate density grows as
    \begin{equation}
    \mathcal{R}(\theta; z) \propto (1+z)^\kappa.
    \label{eq:kappa-eqn}
    \end{equation}
The observed detector-frame merger rate per unit redshift correspondingly grows as
    \begin{equation}
    R(z) \propto (1+z)^{\kappa-1} \left(\frac{dV_c}{dz}\right),
    \end{equation}
where the additional factor of $(1+z)^{-1}$ converts between source- and detector-frame rates.
This parametric model is adopted in Sects.~\ref{sec:masses}, \ref{sec:spins}, and \ref{sec:effective-spins} when non-parametrically measuring the black hole mass and spin distributions.

\section{Additional Posteriors and Predictive Checks}
\label{app:hyperparams}

\begin{figure*}
    \centering
    \includegraphics[width=0.85\textwidth]{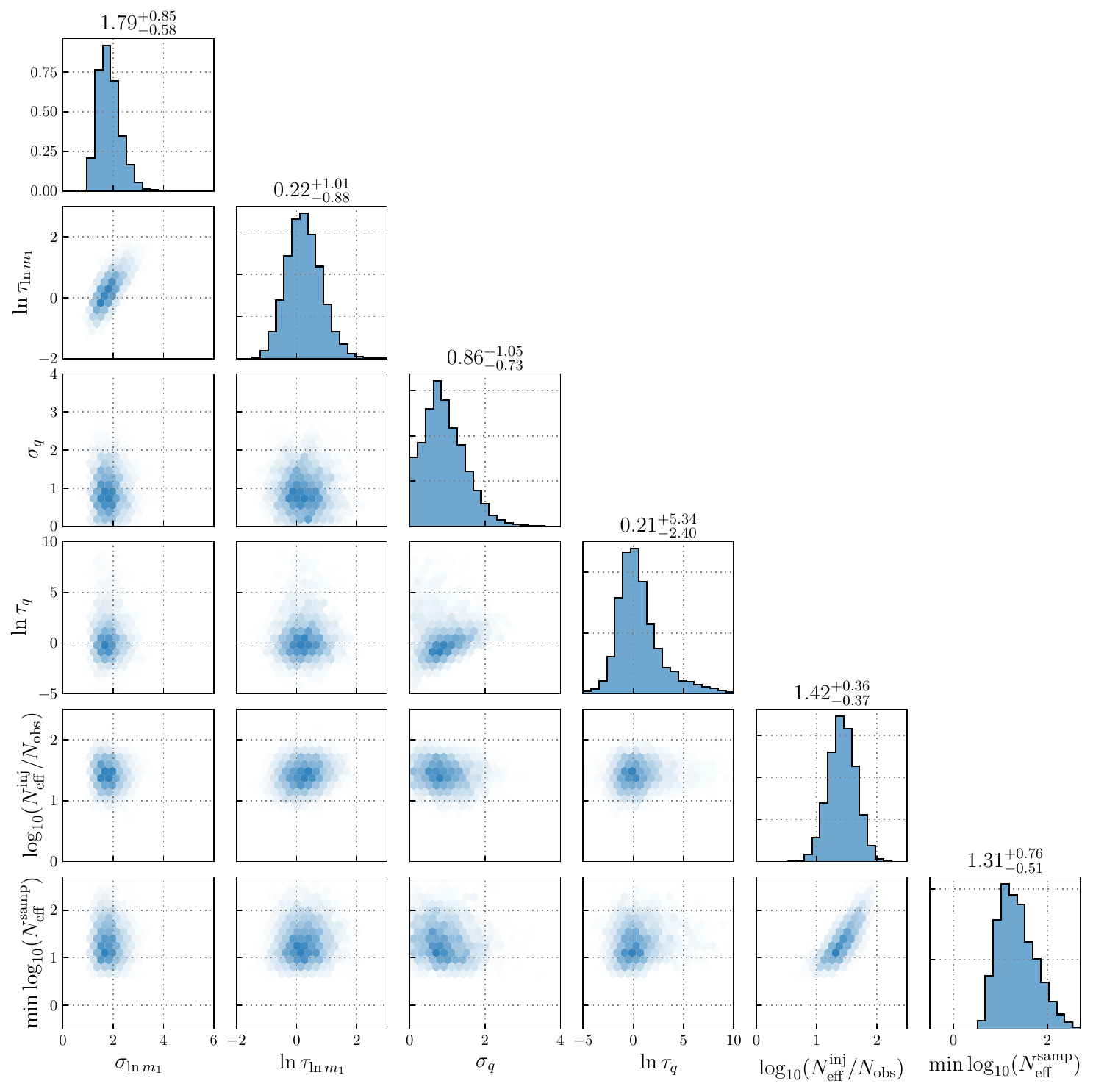}
    \caption{
    Posteriors on the hyperparameters characterizing the primary mass and mass ratio distributions presented in Sect.~\ref{sec:masses}.
    As a diagnostic, we also include the effective number of injections per observed event and the effective number of posterior samples informing the per-event likelihood, minimized over the 69 events in our sample.
    }
    \label{fig:mass-corner}
\end{figure*}

\begin{figure*}
    \centering
    \includegraphics[width=0.6\textwidth]{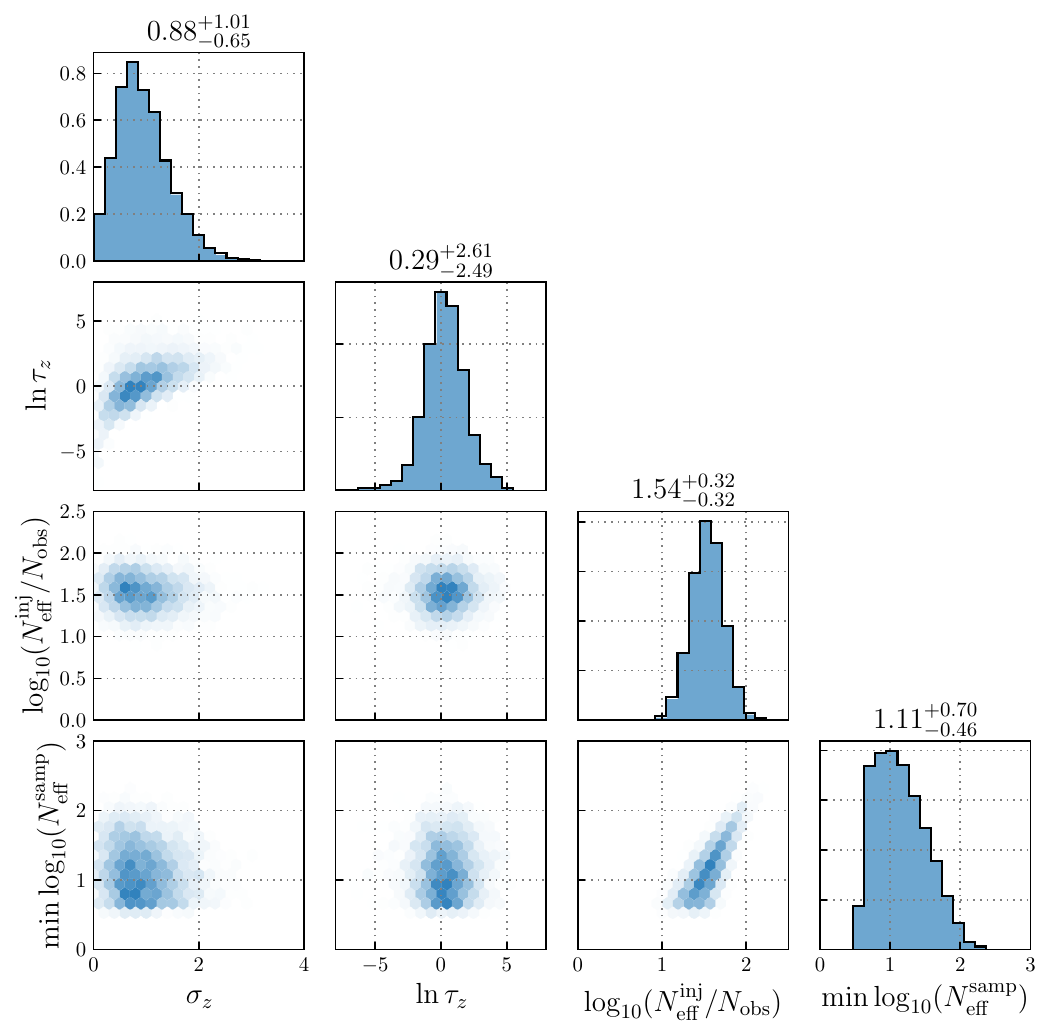}
    \caption{
    As in Fig.~\ref{fig:mass-corner}, but for the redshift evolution of the merger rate shown in Sect.~\ref{sec:redshifts}.
    }
    \label{fig:z-corner}
\end{figure*}

\begin{figure*}
    \centering
    \includegraphics[width=0.85\textwidth]{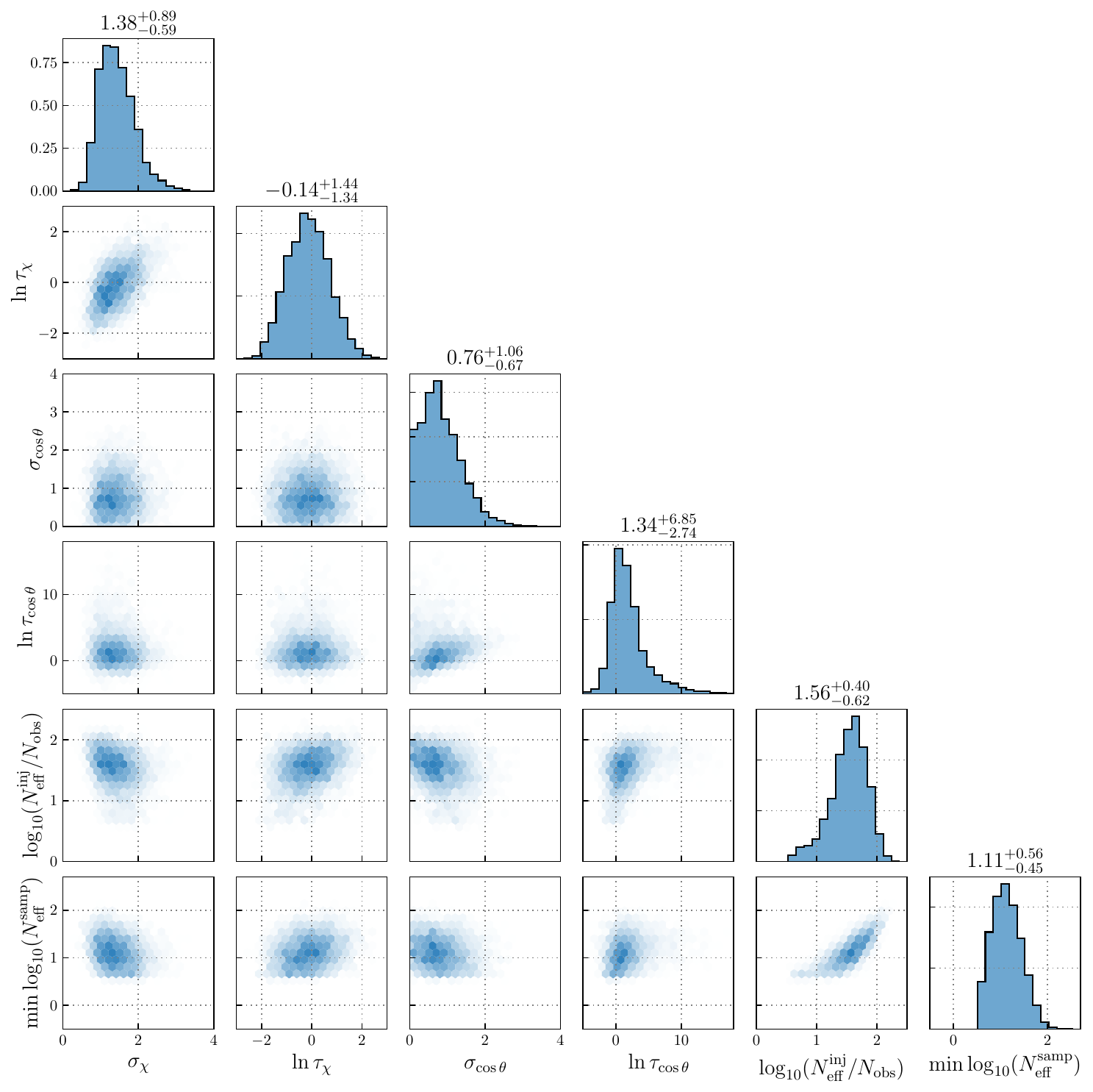}
    \caption{
    As in Fig.~\ref{fig:mass-corner}, but for the component spin distributions shown in Sect.~\ref{sec:spins}.
    }
    \label{fig:spin-corner}
\end{figure*}

\begin{figure*}
    \centering
    \includegraphics[width=0.85\textwidth]{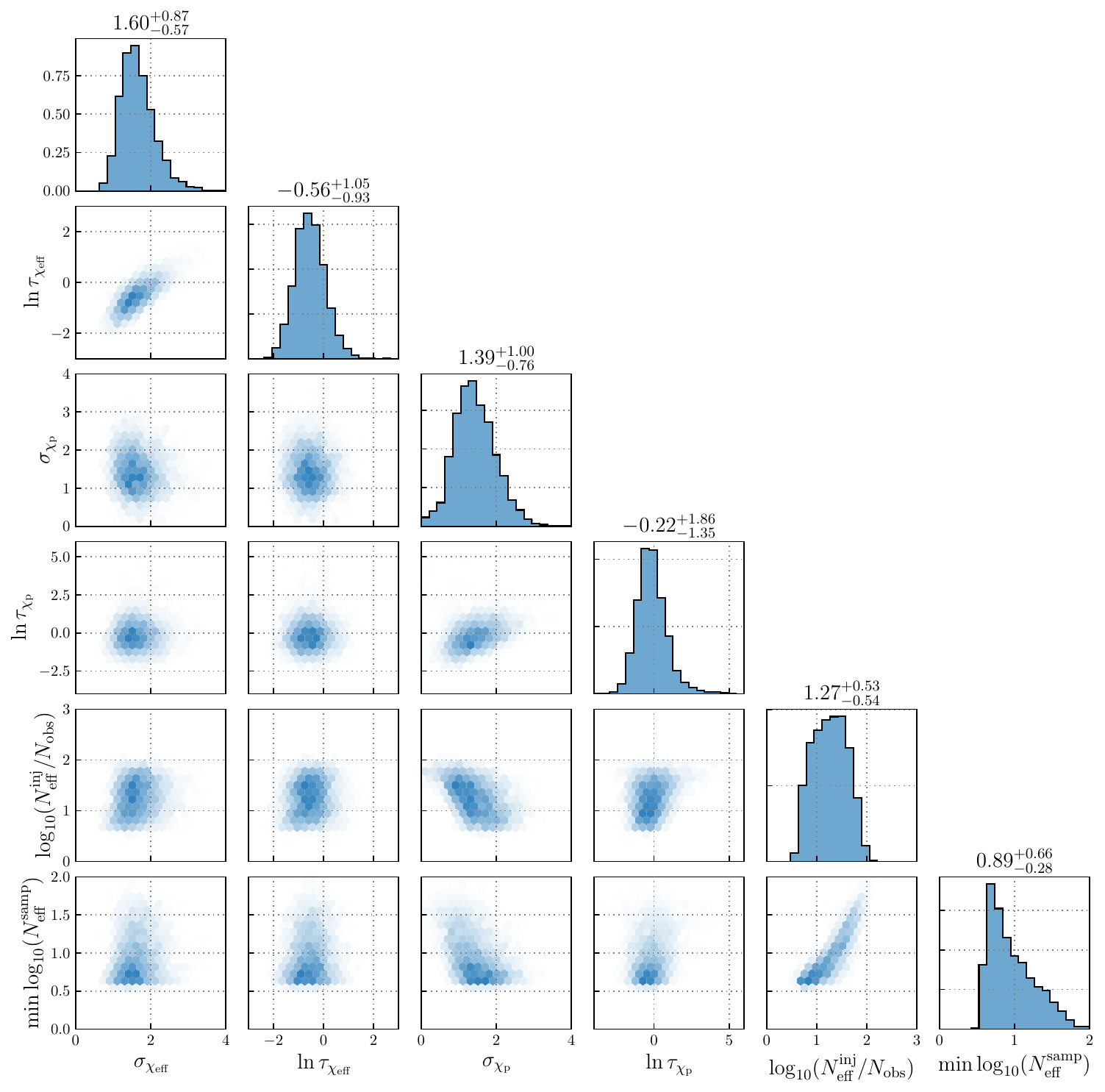}
    \caption{
    As in Fig.~\ref{fig:mass-corner}, but for the effective spin distributions shown in Sect.~\ref{sec:effective-spins}.
    }
    \label{fig:effective-spin-corner}
\end{figure*}

\begin{figure*}
    \centering
    \includegraphics[width=0.75\textwidth]{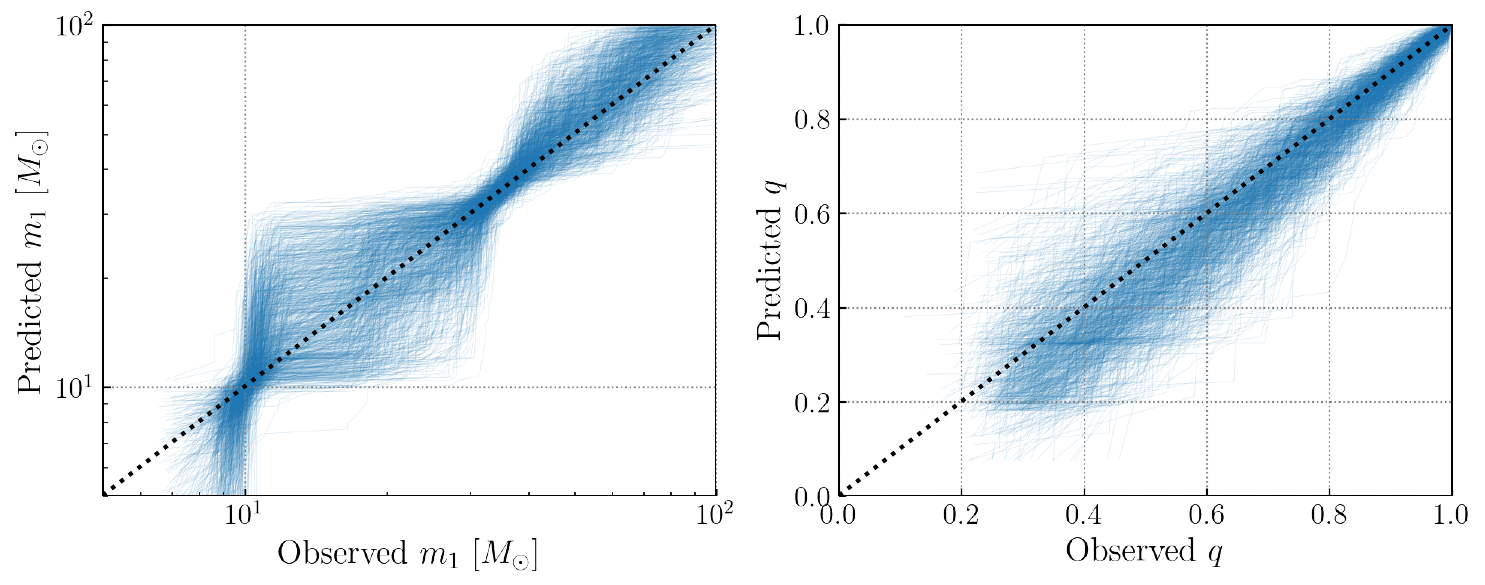}
    \caption{
    Posterior predictive check on primary masses and mass ratios, according to the autoregressive model over $\ln m_1$ and $q$ presented in Sect.~\ref{sec:masses}.
    Each trace corresponds to an individual pass through the algorithm described in Appendix~\ref{app:hyperparams}.
    In both cases, traces are centered along the diagonal (marked with a dotted black line), indicating no tension between the fitted population model and our observed data.
    }
    \label{fig:mass-ppc}
\end{figure*}

\begin{figure*}
    \centering
    \includegraphics[width=0.4\textwidth]{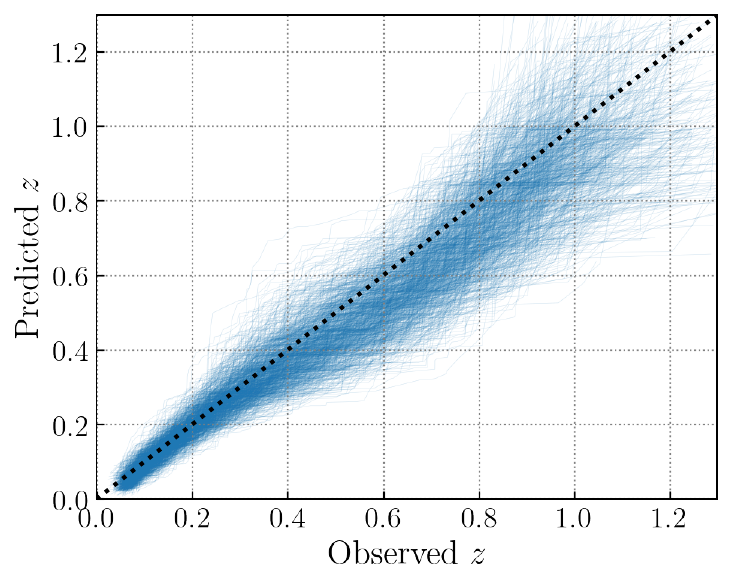}
    \caption{
    As in Fig.~\ref{fig:mass-ppc}, but for the redshift distribution modeled as an autoregressive process in Sect.~\ref{sec:redshifts}.
    }
    \label{fig:z-ppc}
\end{figure*}

\begin{figure*}
    \centering
    \includegraphics[width=0.75\textwidth]{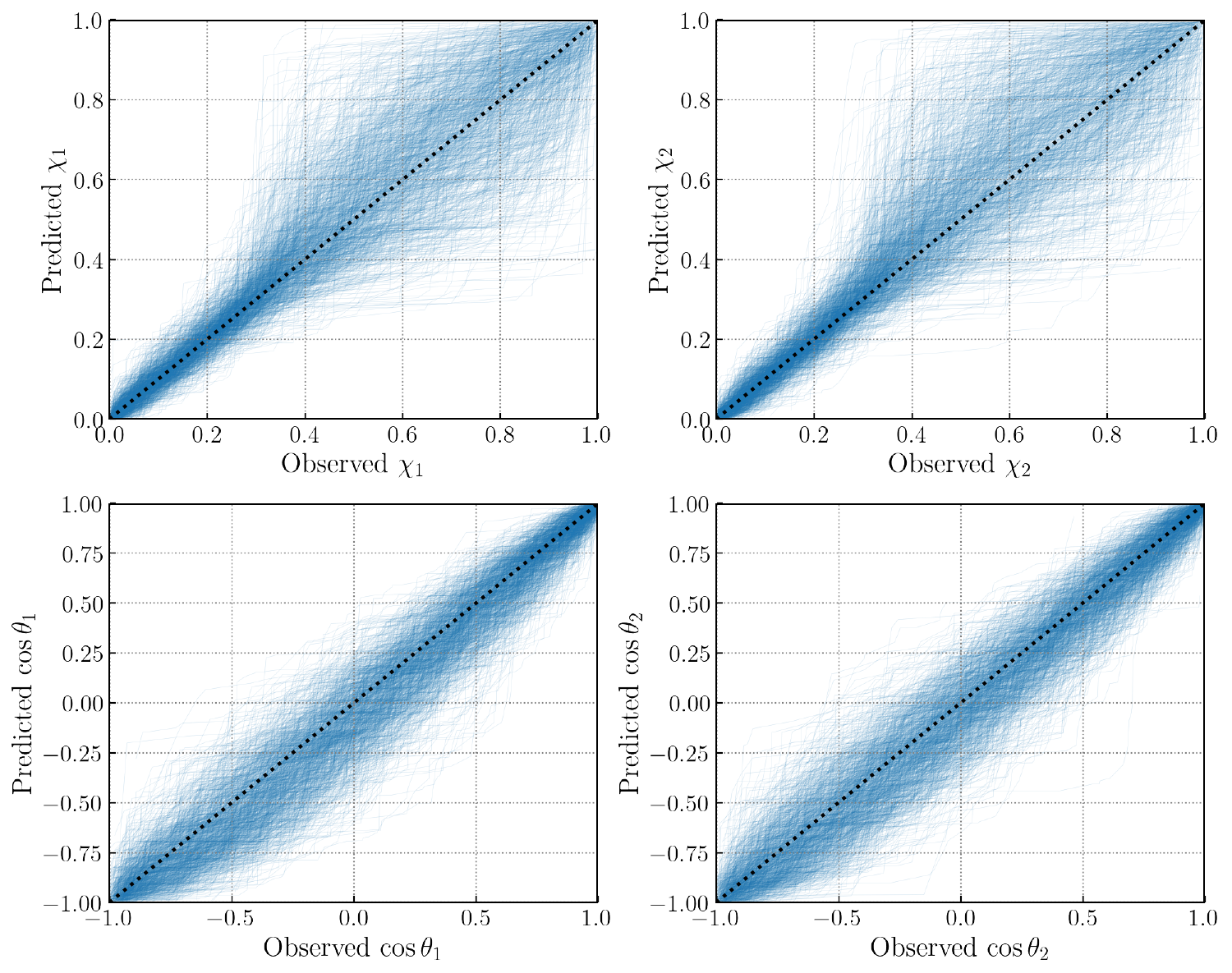}
    \caption{
    As in Fig.~\ref{fig:mass-ppc}, but for the component spin magnitude and tilt distributions modeled as autoregressive processes in Sect.~\ref{sec:spins}.
    We show predictive checks on both component spins to gauge any inconsistency with our assumptions regarding component spin independence; see Eq.~\eqref{eq:ar-component-spins}.
    }
    \label{fig:spin-ppc}
\end{figure*}

\begin{figure*}
    \centering
    \includegraphics[width=0.75\textwidth]{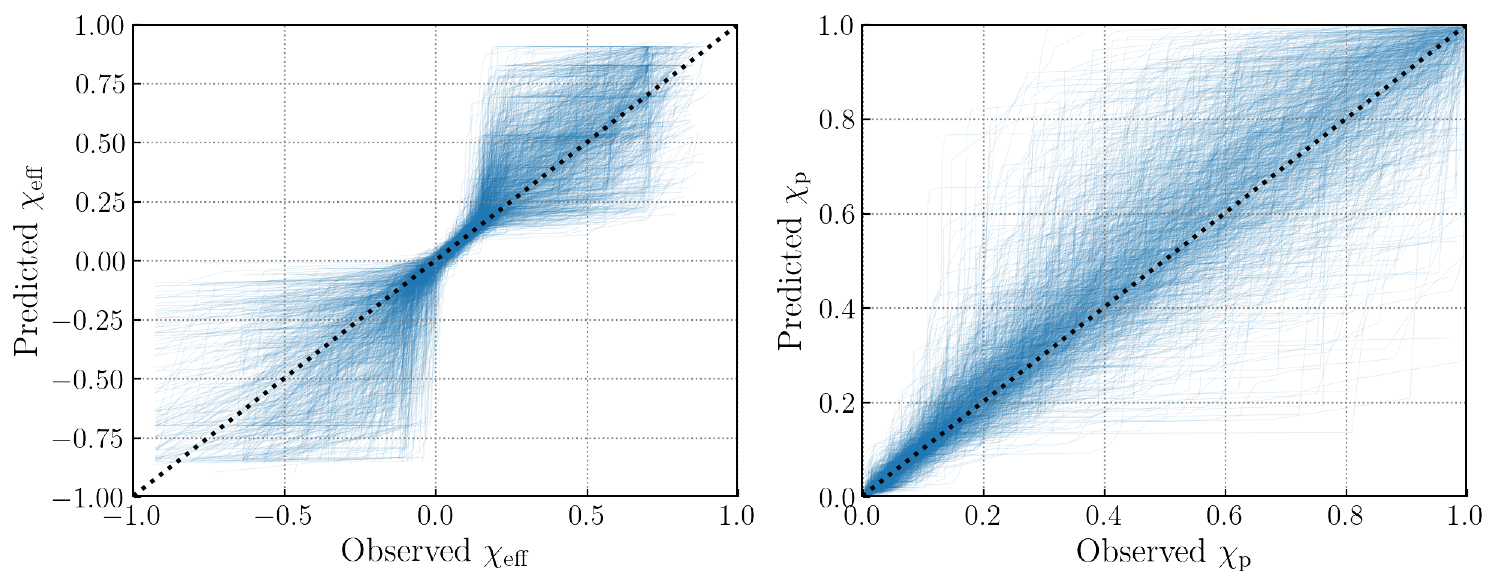}
    \caption{
    As in Fig.~\ref{fig:mass-ppc}, but for the effective spin distributions modeled as an autoregressive process in Sect.~\ref{sec:effective-spins}.
    }
    \label{fig:effective-spin-ppc}
\end{figure*}

In this appendix, we show a few more supplemental results that may be useful in assessing the behavior and performance of our population inference with an autoregressive model.

As discussed elsewhere, our autoregressive process models are characterized by hyperparameters $\sigma$ and $\tau$ controlling the variance and autocorrelation length of the (log) merger rate.
Figures~\ref{fig:mass-corner}-\ref{fig:effective-spin-corner} show the posteriors obtained on these parameters for each autoregressive model used in the main text.
For diagnostic purposes, these figures also show quantities related to the effective number of samples informing our inference.
Included in each corner plot is the distribution of effective injections per observation, $N_\mathrm{eff}^\mathrm{inj}/N_\mathrm{obs}$ and the distribution of effective posterior sample counts, minimized across events.
See Appendix~\ref{app:inference-details} for further details regarding these diagnostics.

We furthermore assess the validity of our results using posterior predictive checks, comparing distributions of observed binary parameters to distributions predicted by our fitted models.
Catalogs of ``observed'' and ``predicted'' parameters are generated by resampling individual event posteriors as well as the set of injections used elsewhere to compute $N_\mathrm{exp}(\Lambda)$.
This process proceeds as follows~\cite{Miller2020}:
    \begin{enumerate}
    \item Randomly draw a sample $\Lambda_i$ from our posterior on the population parameters $\Lambda$ (this includes the variables characterizing strongly-parametrized models, as well as latent parameters defining any autoregressive processes in use).
    \item For each observed event, with posterior samples $\{\lambda_{j}\}$, compute weights $w_{j} = R_d(\lambda_{j}|\Lambda_i)/p_\mathrm{pe}(\lambda_{j})$, the ratios between the detector-frame distribution defined by $\Lambda_i$ and the parameter estimation prior $p_\mathrm{pe}(\lambda)$.
    \item From each observed event, draw a single sample $\lambda_{j}$ with draw probabilities proportional to the weights $\{w_j\}$.
    The resulting set $\{\lambda\}$ defines a single catalog of ``observed'' parameters consistent with our data.
    \item For each successfully recovered injection, with parameters $\{\lambda_{\mathrm{inj},j}\}$, compute weights $w_{j} = R_d(\lambda_{\mathrm{inj},j}|\Lambda_i)/p_\mathrm{inj}(\lambda_{\mathrm{inj},j})$, where $p_\mathrm{inj}(\lambda)$ is the distribution from which injections were drawn.
    \item Draw $N_\mathrm{obs}$ values $\lambda_{\mathrm{inj},j}$ with draw probabilities proportional to the above weights.
    The resulting set $\{\lambda_\mathrm{inj}\}$ defines a single catalog of ``predicted'' parameters consistent with the proposed population $\Lambda_i$ and with appropriate selection effects.
    \item Sort the ``observed'' and ``predicted'' catalogs, and plot against one another.
    \item Repeat.
    \end{enumerate}

Figure~\ref{fig:mass-ppc} shows the result of this algorithm using the population model adopted in Sect.~\ref{sec:masses}, in which the primary mass and mass ratio distributions of binary black holes are described as autoregressive processes.
For convenience, the dotted black lines mark the diagonals along which ``observed'' and ``predicted'' values are equal.
A systematic deviation from this diagonal would indicate  tension between our model (and its corresponding predictions) and our observed data.
In Fig.~\ref{fig:mass-ppc}, all traces are systematically clustered around the diagonal, with no indication of systematic mismodeling.
Figures~\ref{fig:z-ppc}-\ref{fig:effective-spin-ppc} similarly show predictive checks on the autoregressive models used in Sects.~\ref{sec:redshifts}-\ref{sec:effective-spins} to describe distributions of binary redshifts, component spins, and effective spins; these also indicate good agreement between our autoregressive models and observed data.

In addition to checking model validity, the catalogs of ``predicted'' detections enable other related calculations regarding the locations in which detections are expected (or not expected) to arise.
In particular, we saw in Appendix~\ref{app:demo} that autoregressive processes can exhibit prior-dominated tails in regions with no data, and that the onset of these tails can be characterized by identifying regions in which fewer than one detection is predicted by the fitted model.
We can accordingly use the ensembles of ``predicted'' detections appearing in Figs.~\ref{fig:mass-ppc}-\ref{fig:effective-spin-ppc} to assess the statistical significances of tails appearing in our autoregressive results, calculating threshold values above or below which fewer than $N$ detections are predicted by our fitted models, on average.
The results of this calculation are quoted in Sects.~\ref{sec:spins} and \ref{sec:effective-spins} in the main text, when discussing the robustness of our recovered spin distributions.    

\bibliography{References}

\end{document}